\documentclass[aps,prb,twocolumn,amsmath,amssymb,nofootinbib,eqsecnum,superscriptaddress,floatfix]{revtex4-1}
\usepackage{amsmath}
\usepackage{amssymb}
\usepackage{amsthm}
\usepackage[dvips]{color} 
\usepackage{graphicx}
\usepackage{dcolumn} 
\usepackage{bm} 
\usepackage[hypertex]{hyperref}
\usepackage{longtable}
\usepackage{ulem}   
\begin{document}

\title{ 
Global phase diagram of two-dimensional
Dirac fermions in random potentials
      }

\author{S.\ Ryu} 
\affiliation{
Department of Physics, University of Illinois at Urbana-Champaign, 
1110 W. Green Street, 
Urbana, Illinois 61801-3080, USA
            } 
\author{C.\ Mudry} 
\affiliation{
  Condensed matter theory group, 
  Paul Scherrer Institute, CH-5232 Villigen PSI,
  Switzerland            } 
\author{A.\ W.\ W.\ Ludwig
            } 
\affiliation{
  Department of Physics, 
  University of California, Santa Barbara, CA 93106, USA}
\author{A.\ Furusaki
            } 
\affiliation{
  Condensed Matter Theory Laboratory,
  RIKEN, Wako, Saitama 351-0198, Japan
            }

\date{\today}

\begin{abstract}
Anderson localization
is studied for two flavors of 
massless Dirac fermions in two-dimensional space
perturbed by static disorder 
that is invariant under a chiral symmetry (chS)
and a time-reversal symmetry (TRS) operation 
which, when squared, is equal either to plus or minus the identity.
The former TRS 
(symmetry class BDI)
can for example be
realized when the Dirac fermions emerge from
spinless fermions hopping on a two-dimensional lattice
with a linear energy dispersion
such as the honeycomb lattice (graphene) or 
the square lattice with $\pi$-flux per plaquette.
The latter TRS is realized by
the surface states of three-dimensional
$\mathbb{Z}^{\ }_{2}$-topological 
band insulators in symmetry class CII.
In the phase diagram parametrized by the disorder strengths,
there is an infrared stable line of critical points 
for both symmetry classes BDI and CII.
Here we discuss a ``global phase diagram''
in which disordered Dirac fermion systems
in all three chiral symmetry classes, AIII, CII, and BDI,
occur in 4 quadrants, sharing one corner which
represents the clean Dirac fermion limit.
This phase diagram also includes symmetry classes AII
[e.g., appearing at the surface of a disordered three-dimensional 
$\mathbb{Z}^{\ }_2$-topological band insulator
in the spin-orbit (symplectic) symmetry class]
and D (e.g., the random bond Ising model in two dimensions)
as boundaries separating regions of the phase diagram belonging
to the three chS classes AIII, BDI, and CII.
Moreover, we argue that 
physics of Anderson localization in 
the CII phase can be presented in terms of a 
non-linear-sigma model (NL$\sigma$M) 
with a $\mathbb{Z}^{\ }_{2}$-topological term. 
We thereby complete the derivation of topological or
Wess-Zumino-Novikov-Witten terms in the NL$\sigma$M description of
disordered fermionic models in all 10 symmetry classes relevant to 
Anderson localization in two spatial dimensions.
\end{abstract}

\maketitle

\section{Introduction}
\label{sec: intro}

\subsection{Dirac fermions in condensed matter physics}

Massless Dirac fermions emerge 
quite naturally
from non-interacting and bipartite tight-binding Hamiltonians
at low energies and long wave-lengths
when
the fermion spectrum of energy eigenvalues
is symmetric about the band center
and the Fermi surface reduces to a finite
number of discrete Fermi points at the band center. 
This situation is generic for non-interacting
electrons hopping with a uniform
nearest-neighbor amplitude $t$ along a one-dimensional chain.
For non-interacting
electrons hopping on higher dimensional lattices,
this situation is the exception rather than the rule, 
for it is only fulfilled when
the hopping amplitudes are fine-tuned to the lattice.

\begin{figure}
\begin{center}
(a)
\includegraphics[height=4.5cm,clip]{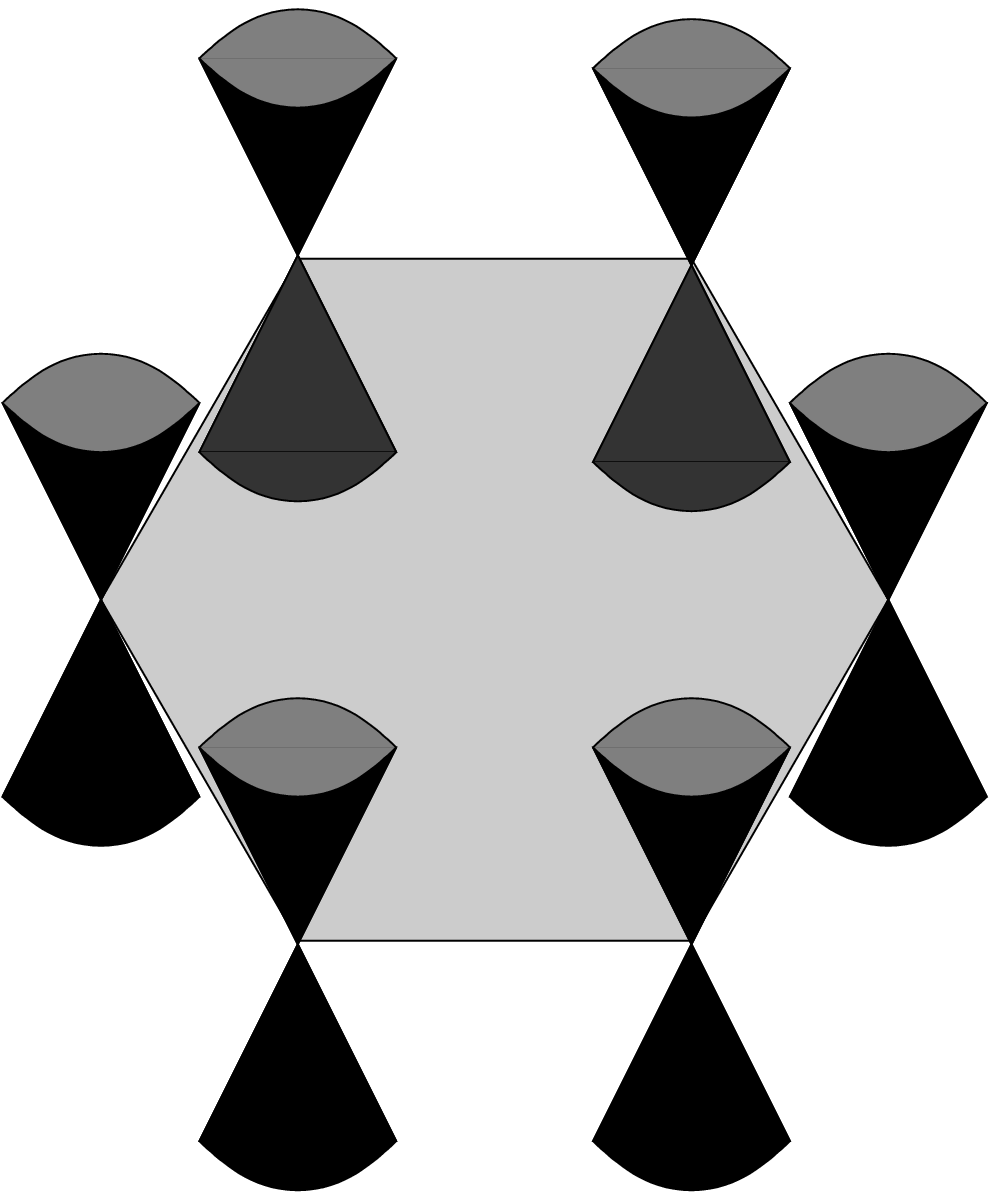}
(b)
\includegraphics[height=4.5cm,clip]{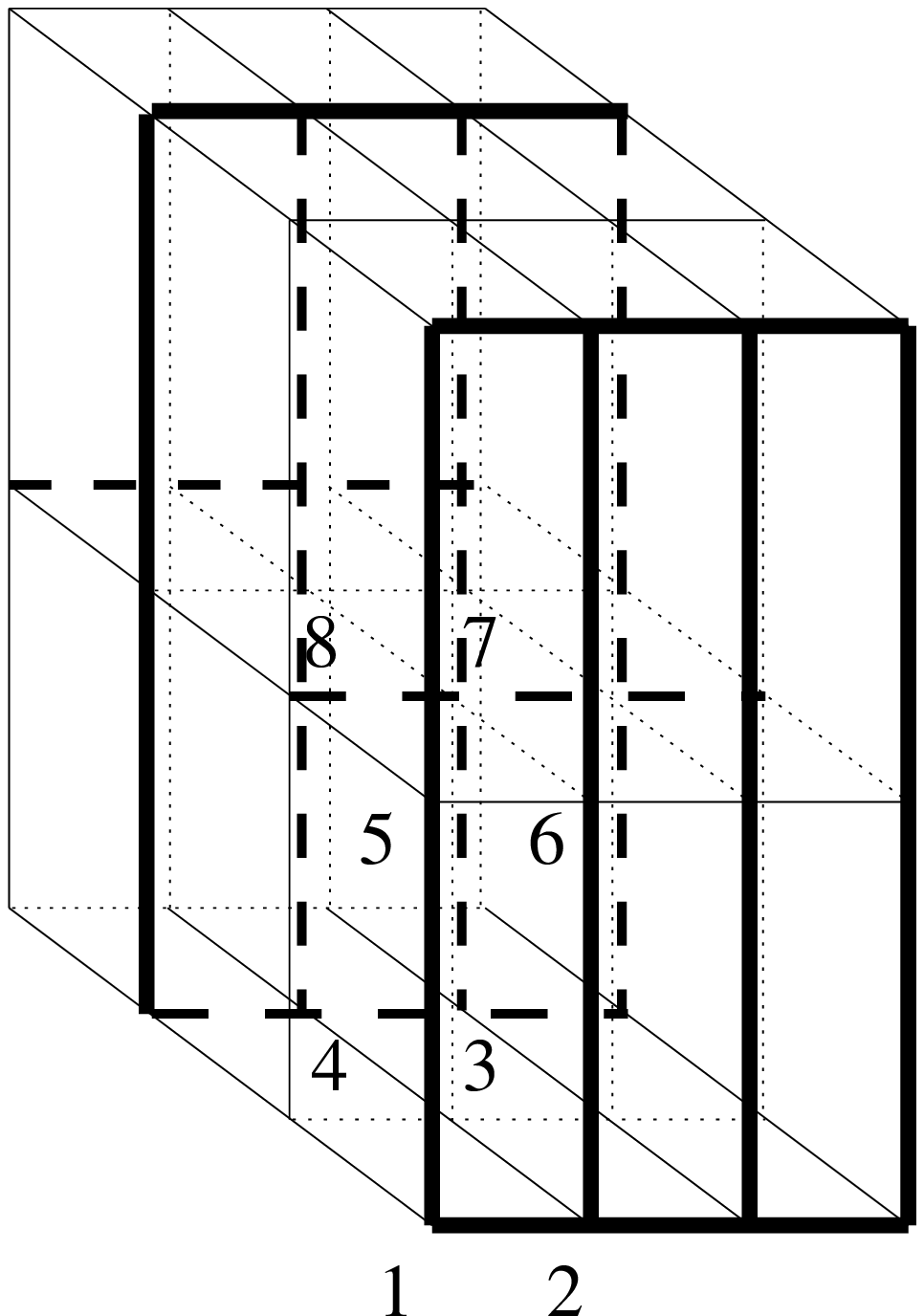}
\caption{
(a) Hexagonal Brillouin zone of graphene with 
the conduction and valence bands touching at the zone corners
in the linear approximation. There are $6/3=2$
inequivalent Fermi points (Dirac cones).
(b) The $\pi$-flux phase for the cubic lattice 
assigns the nearest-neighbor hopping amplitudes 
$+t$ for the thin bonds and $-t$ for the thick bonds
with $t$ a real number. There are $8=2^{3}$ 
inequivalent sublattices labeled 1 to 8.
        }
\label{fig: graphene + pi flux phase}
\end{center}
\end{figure}

In the case of graphene, when described by the uniform hopping
amplitude $t$ between the nearest-neighbor sites 
of the honeycomb lattice, 
there are two bands in the Brillouin zone
of the underlying triangular Bravais lattice
that touch at the 6 corners of the Brillouin zone
[see Fig.~\ref{fig: graphene + pi flux phase}(a)].%
~\cite{Wallace47}
Because the unit cell contains two sites
and because the number
of inequivalent Fermi points is two, these Dirac fermions
realize a four-dimensional representation of the Dirac equation
in two-dimensional space if we ignore the spin degrees of freedom.

For non-interacting spinless electrons
hopping on the square and (hyper-)cubic lattices, 
Dirac fermions emerge in the vicinity of the band center
whenever the translation invariance of the lattice
is broken by choosing the sign of the nearest-neighbor
hopping amplitudes of uniform magnitude $t$
in such a way that their products
along any elementary closed path (a plaquette) is $-t^{4}$
[see Fig.~\ref{fig: graphene + pi flux phase}(b)].
This pattern of nearest-neighbor hopping amplitudes
preserves time-reversal symmetry. It amounts to threading each plaquette 
by a magnetic flux of $\pi$ or, equivalently, $-\pi$ 
in appropriate units and is thus called the $\pi$ 
flux phase. In the $\pi$-flux phase for the $d$-dimensional
hypercubic lattice, there are $2^{d}$ non-equivalent 
sublattices.
Correspondingly there are $2^{d}$ Fermi points and the
emerging Dirac Hamiltonian in the vicinity of these Fermi points
is $2^{d}$ dimensional. Because the minimal irreducible
representation of the Dirac equation in $d$ dimensions
is $2^{[(d+1)/2]}$ dimensional 
($[x]$ denotes the largest
integer smaller than or equal to $x$), 
the $\pi$-flux phase yields a representation of the
Dirac equation larger than the minimal one in 
all dimensions except for $d=1$. This is called the fermion
doubling problem, for it prevents a lattice regularization
of the standard model 
of Elementary Particle Physics
that represents its particle content (quarks, leptons).
~\cite{Wilson75}

The fact that the fermion-doubling problem affects both
graphene and the $\pi$-flux phase in two dimensions is
not a coincidence. The fermion-doubling problem
is a generic property of non-interacting
local tight-binding Hamiltonians
with time-reversal symmetry.%
~\cite{Nielsen81}

It is possible to circumvent the fermion-doubling problem
in the following way. 

\begin{figure}
\begin{center}
(a)
\includegraphics[height=3cm,clip]{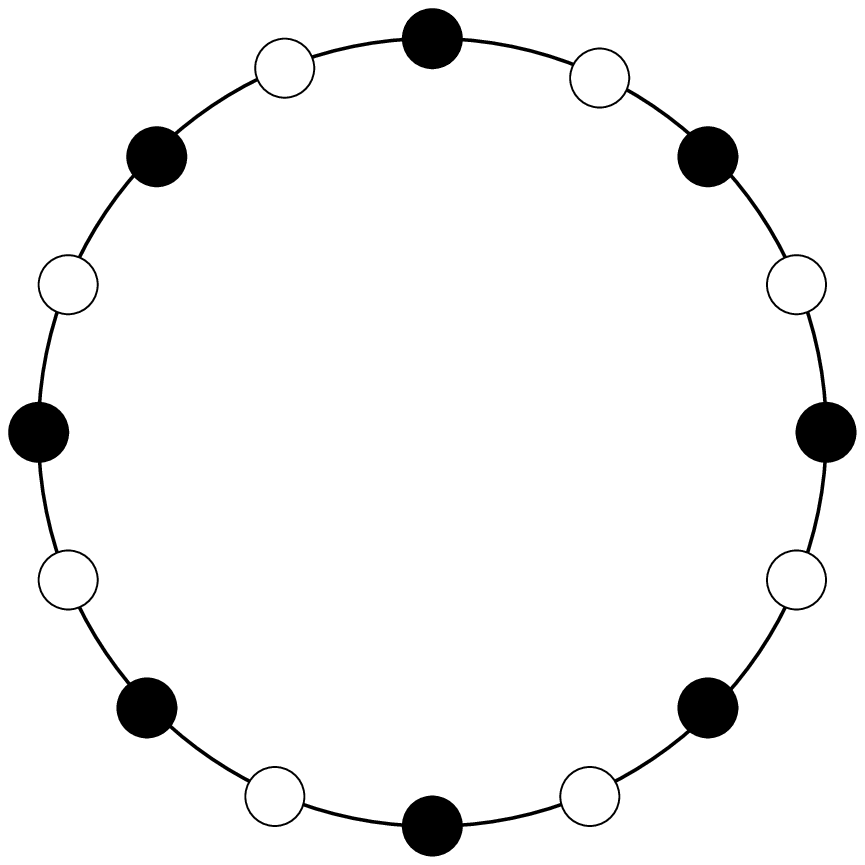}
(b)
\includegraphics[height=3cm,clip]{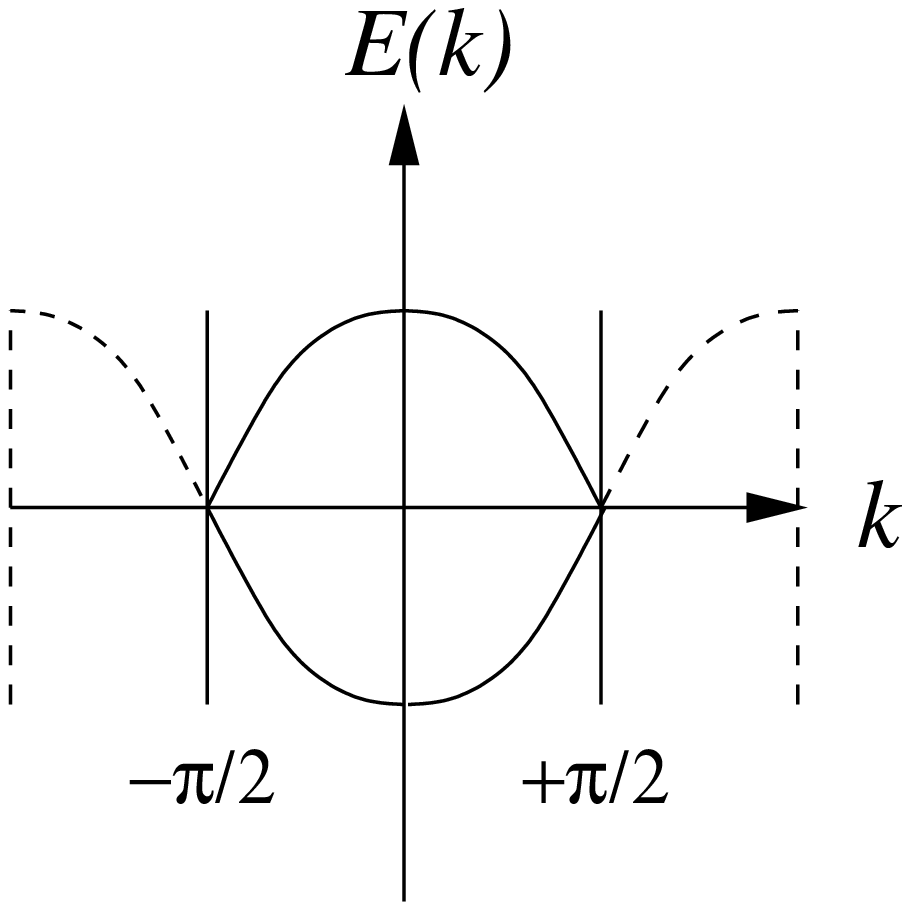}
\\ \vskip 10 true pt
(c)
\includegraphics[height=3cm,clip]{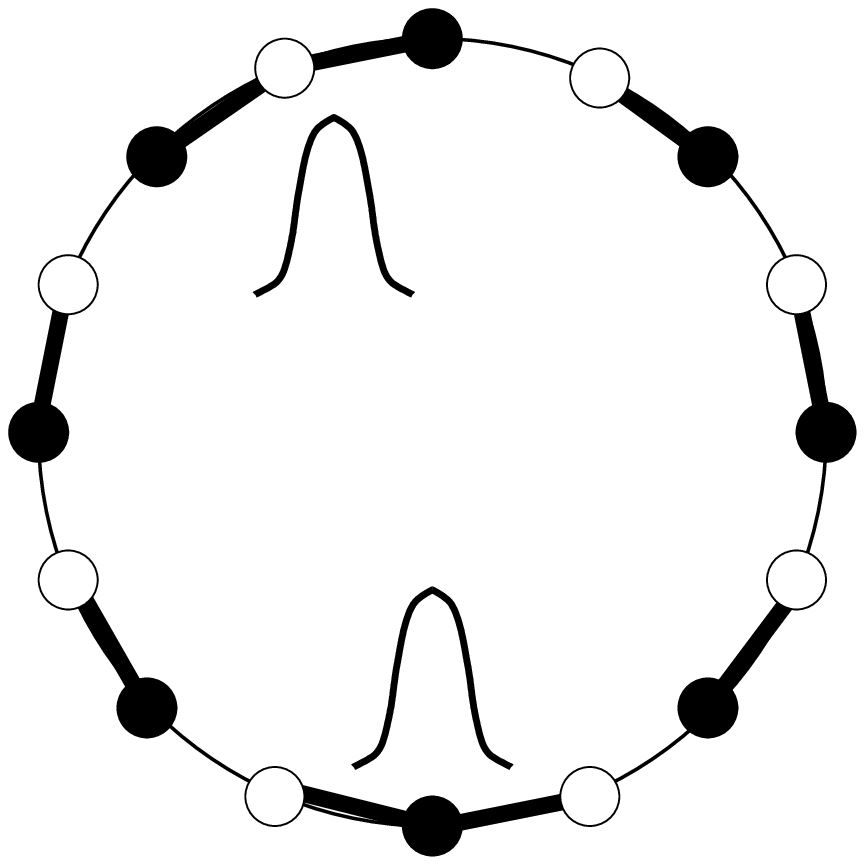}
(d)
\includegraphics[height=3cm,clip]{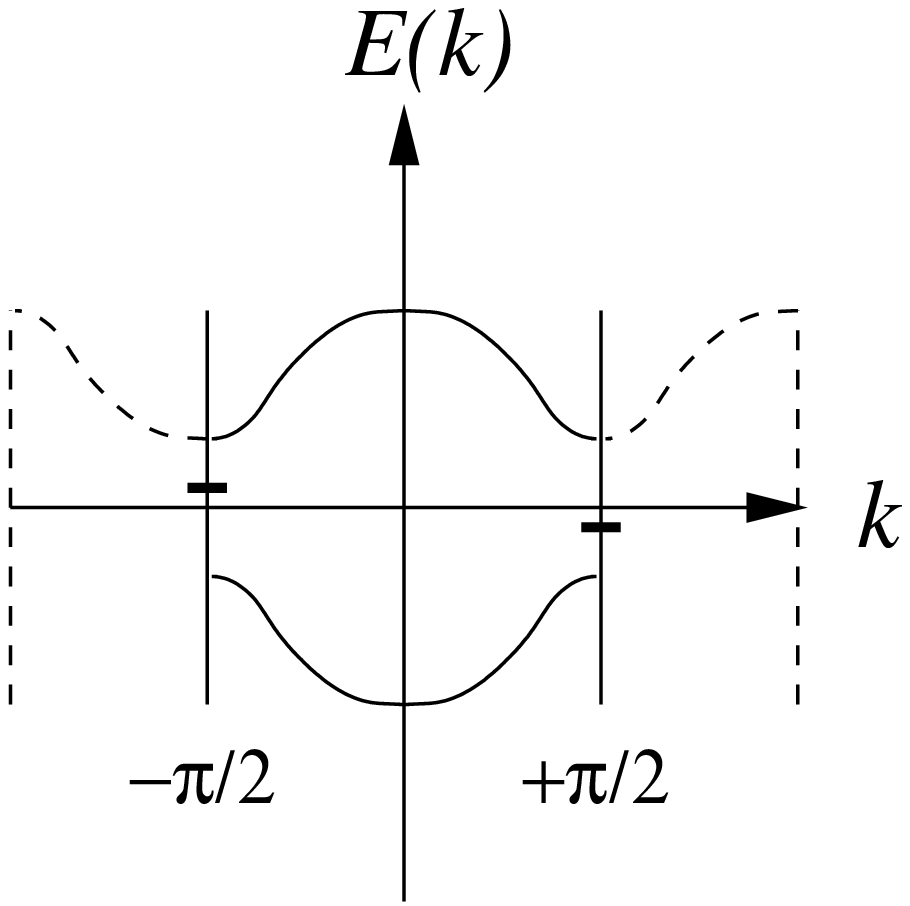}
\\ \vskip 10 true pt
(e)
\includegraphics[height=4cm,clip]{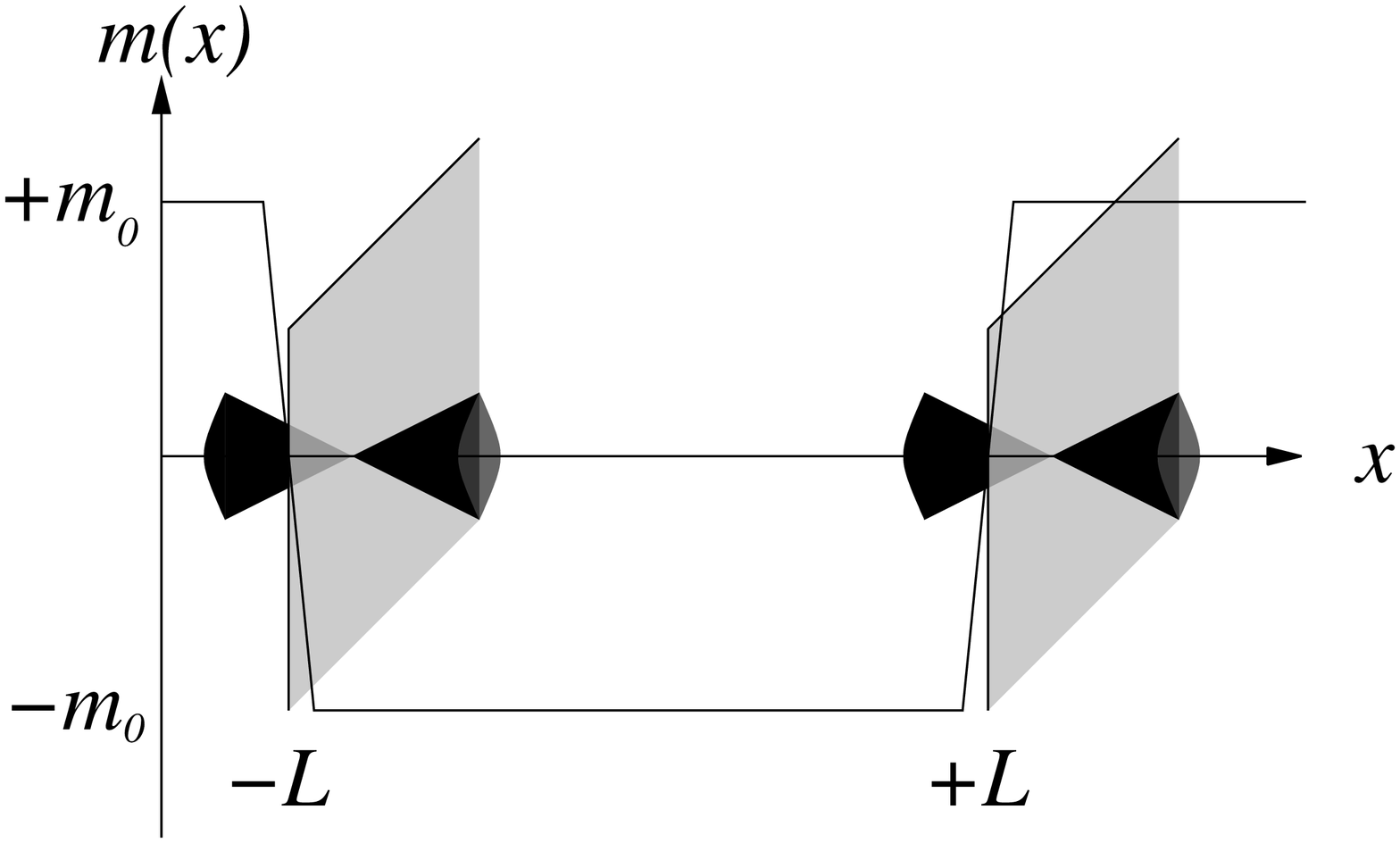}
\caption{
(a) Ring along which a spinless electron hops between nearest-neighbor sites
shown as circles with the uniform real-valued amplitude $t$. 
The lattice sites are colored in black on one sublattice and 
white on the other sublattice.
(b) Electronic dispersion corresponding to (a) after folding
the Brillouin zone.
(c) Ring along which a spinless electron hops between nearest-neighbor sites
with the dimerized real-valued amplitude $t\pm\delta t$. There are
two defective sites belonging to opposite sublattices
at which two strong bonds $t+\delta t$ meet.
(d) The breaking of translation invariance in (c) has opened a
gap at the reduced zone boundaries and localized two bound states
around the two defective sites.
(e) A generalization of (c) and (d) in three dimensions can be
achieved with the help of a suitable dimerization of the 
$\pi$-flux phase depicted in Fig.~\ref{fig: graphene + pi flux phase}(b)
for spinless electrons. The continuum approximation yields a
massive $8\times8$ Dirac equation.
Two-dimensional defective surfaces normal to the direction $x$, say,
occur when the mass changes sign. One mid-gap state is bound to each 
of the two-dimensional defective surfaces. Each midgap state
obeys a $4\times 4$ two-dimensional massless Dirac equation 
as depicted by a Dirac cone. The two mid-gap states have opposite 
chiralities.
        }
\label{fig: defects and mid-gap states}
\end{center}
\end{figure}

We consider first a 
one-dimensional chain along which a spinless electron
hops with the uniform nearest-neighbor amplitude $t$.
We also impose periodic boundary conditions
[see Fig.~\ref{fig: defects and mid-gap states}(a)].
We fold the spinless electron's dispersion 
on half of its Brillouin zone
and open a gap at the folded zone boundaries by
dimerization of the hopping amplitude,
$t\to t\pm\delta t$,
as it occurs for example through its interaction with an optical phonon
within a Born-Oppenheimer approximation.
At low energies, the effective fermionic Hamiltonian is the
one-dimensional massive Dirac equation with the mass set by the 
dimensionless parameter $\delta t/t$ assumed to be 
smaller than unity.
Imagine now that the dimerization pattern is defective at
two sites that are far apart relative to the characteristic
length scale $(t/\delta t)\mathfrak{a}$ where 
$\mathfrak{a}$ is the lattice spacing
[see Fig.~\ref{fig: defects and mid-gap states}(c)].
At the level of the effective Dirac equation, this means
that the mass term 
changes sign twice, once at each defective site.
Two bound (i.e., normalizable) states 
appear in the spectrum 
[see Fig.~\ref{fig: defects and mid-gap states}(d)]
with the remarkable property that they have opposite helicity (chirality)
and an exponentially small overlap or, equivalently, energy splitting,
for they are exponentially localized with the localization
length of order $(t/\delta t)\mathfrak{a}$
around their respective defective sites.%
~\cite{Jackiw76,Su79}

The same mechanism applies in any $d$-dimensional space, 
be it for the massive Dirac equation,%
~\cite{Callan85}
or for tight-binding Hamiltonians with sublattice symmetry
[see Fig.~\ref{fig: defects and mid-gap states}(e)],%
~\cite{Fradkin86-,Brouwer02}
and has been used in lattice gauge theory 
as a means to overcome
the fermion doubling problem.\cite{Kaplan92,Jansen96}
For example, the massive Dirac equation in odd $d$-dimensional space
supports massless boundary states with a common helicity (chirality) 
along each even $(d-1)$-dimensional boundary where the mass term vanishes.
A complete classification of
all such two-dimensional boundary states was part of the
classification of topological insulators in spatial dimensions $d=1,2,3$
given in Ref.~\onlinecite{Schnyder08}
in terms of the generic symmetry classes
arising from the antiunitary operations of time reversal
and particle-hole symmetry
[underlying the work of Altland and Zirnbauer
on random matrix theory (RMT)].%
~\cite{Zirnbauer96,Altland97,Heinzner05}
A systematic regularity (periodicity) of the classification
as the dimensionality is varied, in general dimension,
was discovered upon the use of K-Theory by Kitaev%
~\cite{Kitaev09} (see also Ref.~\onlinecite{Stone11}).
As shown in Refs.~\onlinecite{Schnyder09b} and \onlinecite{Ryu10b},
this can, alternatively, 
be understood in terms of the lack of Anderson localization
at the boundaries. More recently, an understanding of this classification
of topological insulators in terms of quantum anomalies was developed.%
~\cite{Ryu11}

\begin{table*} 
\caption{
\label{table: Fendley classification}
Table of topological terms that can be added to the replicated fermionic
non-linear-sigma model (NL$\sigma$M)  
describing Anderson localization in two dimensions
and the classification\cite{Schnyder08} of topological insulators
(superconductors) in three dimensions.
Symmetry classes indicated by the ``Cartan label''
are classified according to the
presence or absence of 
time-reversal, 
particle-hole, and
``sublattice''
symmetries
which we abbreviate as TRS, PHS, and SLS, respectively.
The presence of TRS and PHS is denoted by ``$+1$'' or ``$-1$,''
depending on whether the square of the (antiunitary) operator
implementing the symmetries equals $+1$ (identity) or $-1$,
whereas the presence of SLS is denoted by ``1''.
The absence of these symmetries is denoted by ``0''.
The SLS is a product of TRS and PHS.
For historical reasons, 
the first three rows of the table are also referred to
as the orthogonal, unitary, and symplectic symmetry classes.
When the disorder respects a sublattice symmetry 
as in the next three rows, the terminology chiral is also used. 
Finally, the last four rows
can be realized as
random Bogoliubov-de-Gennes (BdG) Hamiltonians.
Target spaces for fermionic replicated NL$\sigma$M 
($\mathsf{N}$ is the replica index and the limit $\mathsf{N}\to0$ 
is understood)
are given in the fifth column.
The penultimate column lists the nature of the topological term
compatible with the target
and two-dimensional base spaces.
The symbols $\mathbb{Z}$ and $\mathbb{Z}_2$ in the last column
indicate that the topologically distinct phases within a given
symmetry class of topological insulators or superconductors
in three spatial dimensions
are characterized by an integer topological invariant ($\mathbb{Z}$)
or a $\mathbb{Z}_2$ quantity.
The symbol ``0'' denotes the case when there exists no topological insulator
(superconductor).
        }
\begin{tabular}{lcccccc}
\hline 
\hline 
Cartan label
&
TRS
& 
PHS
& 
SLS
& 
Target space
&
Topological term \quad
&
3d-TI/TSC
\\ 
\hline
AI (orthogonal)
& 
$+1$
& 
0
& 
0
&
$
\mathrm{Sp}(4\mathsf{N})/
\mathrm{Sp}(2\mathsf{N})
\times
\mathrm{Sp}(2\mathsf{N})
$
&
--
&
0
\\
A (unitary)
& 
0
& 
0
& 
0
&
$
\mathrm{U}(2\mathsf{N})/
\mathrm{U}(\mathsf{N})
\times
\mathrm{U}(\mathsf{N})
$
&
$\theta$ term
&
0
\\
AII (symplectic)
& 
$-1$
& 
0
& 
0
&
$
\mathrm{O}(2\mathsf{N})/
\mathrm{O}(\mathsf{N})
\times
\mathrm{O}(\mathsf{N})
$
& 
$\mathbb{Z}^{\ }_{2}$ term
&
$\mathbb{Z}_2$
\\ 
\hline
BDI (chiral orthogonal)
& 
$+1$
& 
$+1$
& 
1
& 
$
\mathrm{U}(2\mathsf{N})/
\mathrm{Sp}(2\mathsf{N})
$
& 
--
&
0
\\
AIII (chiral unitary)
&
0
& 
0
& 
1
& 
$
\mathrm{U}(\mathsf{N})
\times
\mathrm{U}(\mathsf{N})/
\mathrm{U}(\mathsf{N})
$
&
WZNW term
&
$\mathbb{Z}$
\\
CII (chiral symplectic)
& 
$-1$
& 
0
& 
1
&
$
\mathrm{U}(\mathsf{N})/
\mathrm{O}(\mathsf{N})
$
& 
$\mathbb{Z}^{\ }_{2}$ term
&
$\mathbb{Z}_2$
\\ 
\hline
CI (BdG)
& 
$+1$
& 
$-1$
& 
1
& 
$
\mathrm{Sp}(2\mathsf{N})
\times
\mathrm{Sp}(2\mathsf{N})/
\mathrm{Sp}(2\mathsf{N})
$
&
WZNW term
&
$\mathbb{Z}$
\\
C (BdG)
& 
0
& 
$-1$
& 
0
&
$ 
\mathrm{Sp}(2\mathsf{N})/
\mathrm{U}(\mathsf{N})
$
&
$\theta$ term
&
0
\\
DIII (BdG)
& 
$-1$
& 
$+1$
& 
1
& 
$ 
\mathrm{O}(\mathsf{N})
\times
\mathrm{O}(\mathsf{N})/
\mathrm{O}(\mathsf{N})
$
&
WZNW term
&
$\mathbb{Z}$
\\
D (BdG)
& 
0
& 
$+1$
& 
0
& 
$
\mathrm{O}(2\mathsf{N})/
\mathrm{U}(\mathsf{N})
$
&
$\theta$ term
&
0
\\
\hline
\hline
\end{tabular}
\end{table*} 

\subsection{
Anderson localization for Dirac fermions in two dimensions
           }

Anderson localization%
~\cite{Evers08} 
for non-interacting
two-dimensional Dirac fermions was 
first studied in narrow gap semiconductors by Fradkin in 1986.%
~\cite{Fradkin86}
This work was followed up in the 90's with 
non-perturbative results motivated by the physics
of the integer quantum Hall effect (IQHE), 
the random bond Ising model, and dirty $d$-wave superconductors.%
\cite{Ludwig94,Nersesyan95,Chamon96a,Chamon96b,Mudry96,Bocquet00,%
Bhaseen01,Ludwig00,Altland02}
With the recently available
transport measurements in mesoscopic samples of graphene,
as well as the identifications 
of the alloy Bi$^{\ }_{1-x}$Sb$^{\ }_{x}$
in a certain range of compositions $x$,%
\cite{Fu07,Hsieh08,Hsieh09}
the compounds
Bi$^{\ }_{2}$Te$^{\ }_{3}$,%
~\cite{Zhang09,Chen09}
Sb$^{\ }_{2}$Te$^{\ }_{3}$,%
~\cite{Zhang09}
and
Bi$^{\ }_{2}$Se$^{\ }_{3}$,%
~\cite{Zhang09,Xia09}
and the prediction for another 50 and counting materials 
as three-dimensional $\mathbb{Z}^{\ }_{2}$-topological band insulators
that support surface Dirac fermions,%
~\cite{Lin10,Yan10,Chen10} 
the localization properties of random Dirac fermions 
have become relevant from an experimental point of view.

While all these examples share the massless Dirac spectrum 
as the energy dispersion in the non-interacting and
clean limit,
the effects induced by randomness -- 
weak localization, 
universal conductance fluctuations,
localization, 
metal-insulator transition,
spectral singularities,
etc --
vary with 
(i) the intrinsic symmetries respected by the disorder,
(ii) the dimensionality of the Dirac matrices
representing the Dirac Hamiltonian,
and (iii) the strength and/or correlations in space of the disorder.

When space is effectively zero-dimensional, i.e.,
at the level of RMT, 
ten symmetry classes have originally been identified
and labeled according to the Cartan classification of symmetric spaces
(see Table~\ref{table: Fendley classification}).%
~\cite{Zirnbauer96,Altland97,Heinzner05}

As emphasized in Refs.\ \onlinecite{Fendley00,Fendley01}, the two-dimensional
fermionic replicated NL$\sigma$Ms in eight of the ten
symmetry classes allow for terms of topological origin,
in the form of either $\theta$ terms%
~\cite{theta-term}
or Wess-Zumino-Novikov-Witten (WZNW) terms%
~\cite{Wess71,Novikov82,Witten84}
(see Table~\ref{table: Fendley classification}).
Symmetry classes A, C, and D
support Pruisken
($\theta$) terms.%
\cite{Pruisken84,Senthil98,Senthil00}
Symmetry classes AIII, DIII, and CI 
support WZNW terms.
Finally, symmetry classes AII and CII 
support $\mathbb{Z}^{\ }_2$-topological terms.

WZNW terms in symmetry classes AIII, DIII, and CI appear when
Dirac fermions propagate in the presence of
static vector-gauge-like randomness.%
~\cite{Ludwig94,Nersesyan95,Chamon96a,Chamon96b,Mudry96,Mudry99,Bocquet00,%
Bhaseen01,Ludwig00,Altland02} 
This can only be achieved at the lattice level if
the fermion doubling problem has been overcome, as is the case
with the surface states of three-dimensional 
$\mathbb{Z}$-topological band insulators.

The $\mathbb{Z}^{\ }_2$-topological term in 
symmetry class AII was derived
in the context of disordered graphene
with long-range correlated disorder%
~\cite{Ostrovsky07,Ryu07b}
or two-dimensional surfaces
of three-dimensional $\mathbb{Z}^{\ }_2$ topological band insulators.%
~\cite{Ryu07b}

LeClair and Bernard have extended the RMT classification
by demanding that all perturbations to the two-dimensional Dirac 
Hamiltonian with $\mathrm{N}^{\ }_{\mathrm{f}}$ flavors
preserve the Dirac structure.%
\cite{Bernard02}
In this way, 
the ten-fold classification 
can be refined
by discriminating the parity of $\mathrm{N}^{\ }_{\mathrm{f}}$
for the 3 symmetry classes AIII, DIII, and CI.
These 3 subclasses correspond to the fact that
the replicated principal chiral models (PCMs)
whose target space correspond to symmetry classes
AIII, DIII, and CI, respectively,
can be augmented by WZNW terms. The
realization of any of these additional 3 subclasses in
a lattice model requires overcoming
the fermion doubling problem.

The parity of the flavor number $\mathrm{N}^{\ }_{\mathrm{f}}$ 
of random Dirac fermions also matters for symmetry classes AII and CII.
The fermionic replicated NL$\sigma$Ms derived from the random Dirac
Hamiltonians in symmetry classes AII and CII 
can acquire a $\mathbb{Z}^{\ }_{2}$ topological term on account of the
dimensionality of the Dirac matrices 
(twice the number $\mathrm{N}^{\ }_{\mathrm{f}}$ of flavors) 
that represents the random Dirac Hamiltonian.
Deriving these $\mathbb{Z}^{\ }_{2}$ topological terms from
lattice models is not automatic, for
the fermion doubling problem must be surmounted.

In this paper, by identifying a disordered fermionic model
that gives rise to the 
$\mathbb{Z}^{\ }_2$-topological term in symmetry class CII, 
we complete the derivation for non-interacting 
fermions subject to a weak white-noise correlated
random potential of topological or WZNW terms
in all 10 symmetry classes relevant to 
two-dimensional Anderson localization.
The microscopic fermionic model is realized
by the surface states of a three-dimensional
$\mathbb{Z}^{\ }_{2}$ topological band insulator
in symmetry class CII of Ref.~\onlinecite{Schnyder08}.
(See Ref.\ \onlinecite{Hosur2010} for a particular lattice model 
of a three-dimensional $\mathbb{Z}_2$ topological insulator 
in symmetry class CII.)  

\begin{figure}
\begin{center}
(a)
\includegraphics[height=3cm,clip]{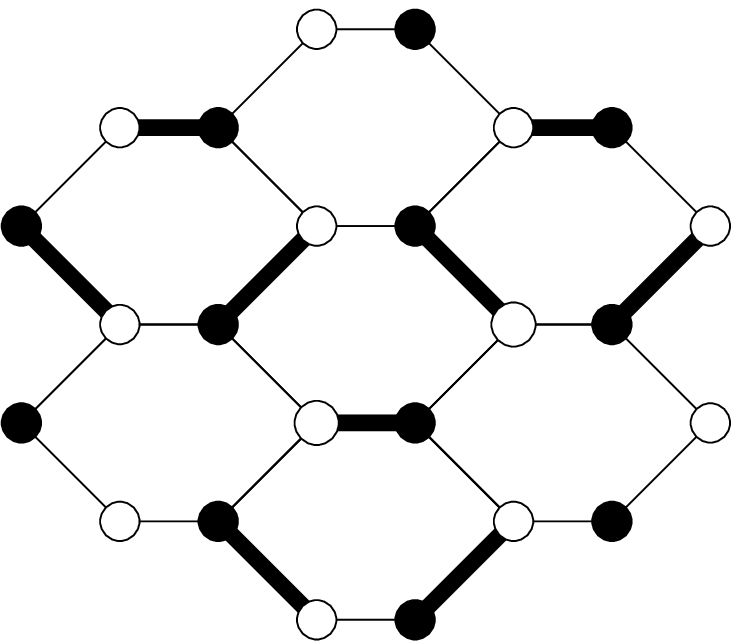}
(b)
\includegraphics[height=3cm,clip]{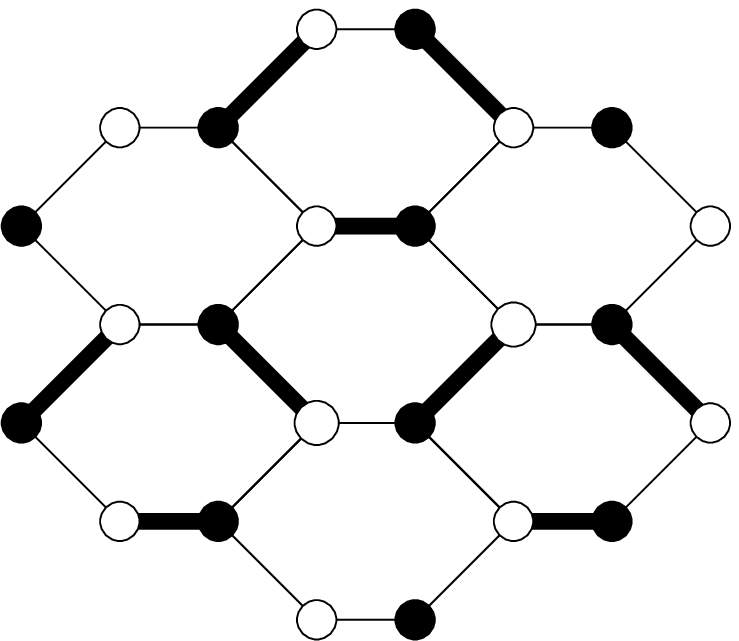}
\\ \vskip 10 true pt
(c)
\includegraphics[height=3cm,clip]{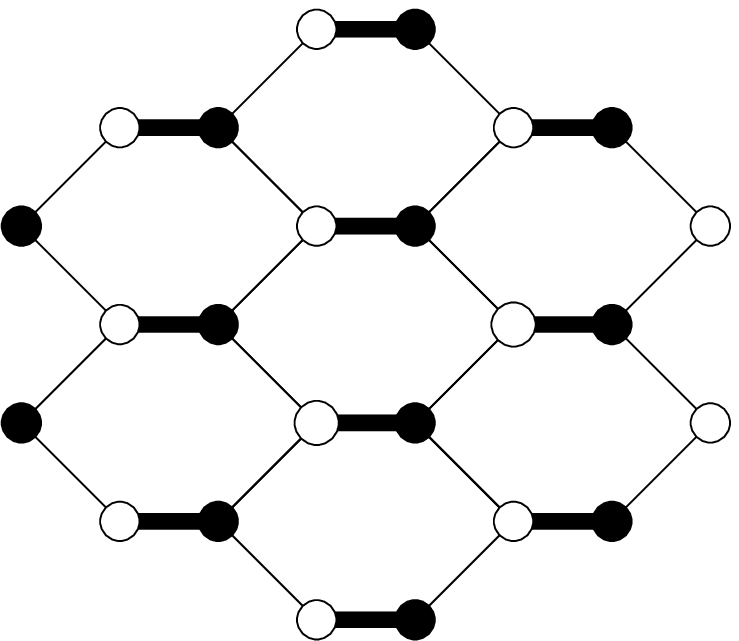}
(d)
\includegraphics[height=3cm,clip]{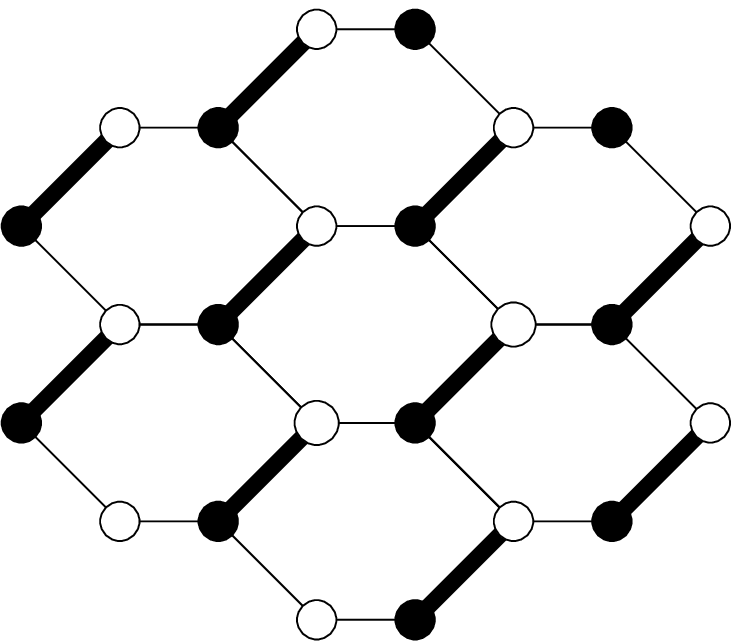}
\caption{
The four independent
dimerization patterns for the real-valued
nearest-neighbor hopping amplitudes
of a spinless electron on the honeycomb lattice
that preserve the sublattice symmetry and the time-reversal symmetry
for a spinless particle.
The two triangular sublattices of the honeycomb lattice
are distinguished by the coloring of their sites
(white or black colored circles).
Strong and weak bonds are depicted by thick and thin lines,
respectively.
The two independent K\'ekule dimerization patterns 
(a) and (b) 
are responsible for the opening of a complex-valued gap $m$
in the continuum approximation by a $4\times4$ Dirac equation.
The two independent columnar dimerization patterns
(c) and (d) 
are responsible for the emergence of an axial vector gauge field
or, equivalently, the complex-valued axial gauge field $a'$
in the continuum approximation by a $4\times4$ Dirac equation.
        }
\label{fig: Graphene+dimerization}
\end{center}
\end{figure}

\subsection{Global phase diagram}

In this paper, we start from the kinetic Hamiltonian
$\mathcal{K}$ for $\mathrm{N}^{\ }_{\mathrm{f}}=2$
flavors of Dirac fermions that make up a (reducible) 
4-dimensional representation of the homogeneous
Lorentz group $\mathrm{SO}(1,2)$. 
We then subject $\mathcal{K}$
to a static and chiral-symmetric random potential
$\mathcal{V}$, 
i.e., the random Dirac Hamiltonian 
$\mathcal{H}=\mathcal{K}+\mathcal{V}$
must anticommute with a unitary matrix $\mathcal{C}$,
$\left\{\mathcal{H}, \mathcal{C}\right\} =0$,
which squares to the identity.
By imposing the condition that $\mathcal{H}$
is invariant under a representation $\mathcal{T}=\mathcal{T}^{T}$
of time reversal for spinless single-particle states,
$\mathcal{H}$
belongs to symmetry class BDI in the 
10-fold classification 
(see Table \ref{table: Fendley classification}).
This corresponds
to an antiunitary time reversal operator 
whose square equals plus the identity.

It is also known that
such a Hamiltonian
 $\mathcal{H}$
describes graphene 
(see Fig.~\ref{fig: Graphene+dimerization})
or the two-dimensional
$\pi$-flux phase, in the presence
of real-valued, nearest-neighbor, 
spin-independent, random hopping amplitudes
when the Fermi energy is at the band center
and once the long-wave-length limit has been taken
with respect to the discrete Fermi points.%
\cite{Hatsugai97,Guruswamy00,Mudry03,Ostrovsky06,Ryu07a,Ryu10a}
For the case of graphene,%
~\cite{CastroNeto08}
static random real-valued nearest-neighbor hopping amplitudes
are induced by neglecting%
~\cite{Lee73}
the dynamics of phonons 
relative to that of the electrons to which they couple.
We emphasize that it is imperative to treat \textit{all channels}
(see Fig.~\ref{fig: Graphene+dimerization})
of disorder compatible with the chiral and time-reversal symmetries.

The {\it first} result of this paper is that
analytical continuation of the 
real-valued random hopping amplitudes to imaginary ones
in the aforementioned bipartite lattice models yields
a random Dirac Hamiltonian that
belongs to symmetry class CII,
as it now turns out to obey the time-reversal symmetry (TRS)
generated by an operator $\mathcal{T}^{\prime}=-\mathcal{T}^{\prime T}$ 
acting on an isospin-1/2 single-particle state.
This corresponds to an antiunitary time reversal operator
whose square equals minus the identity.

{\it Second}, we argue that,
this random Dirac Hamiltonian
captures the (nearly) critical localization properties
of the surface states of a lattice model that,
in the clean limit, realizes a 
three-dimensional $\mathbb{Z}^{\ }_2$-topological
band insulator in symmetry class CII.

\begin{figure*}
\begin{center}
\includegraphics[angle=0,scale=0.65]{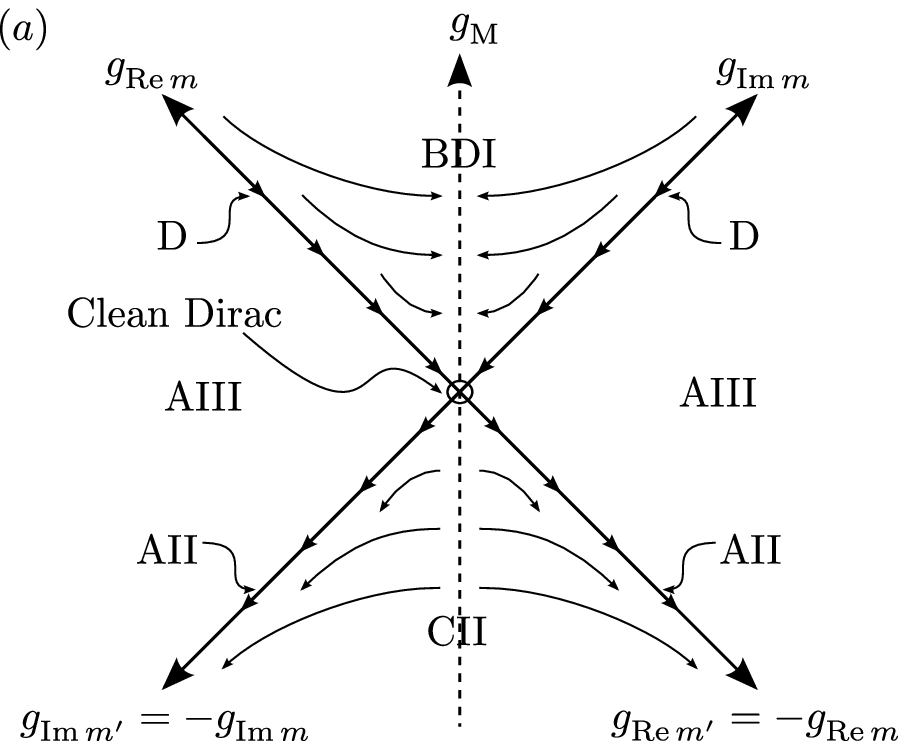}
\hfill
\includegraphics[angle=0,scale=0.6]{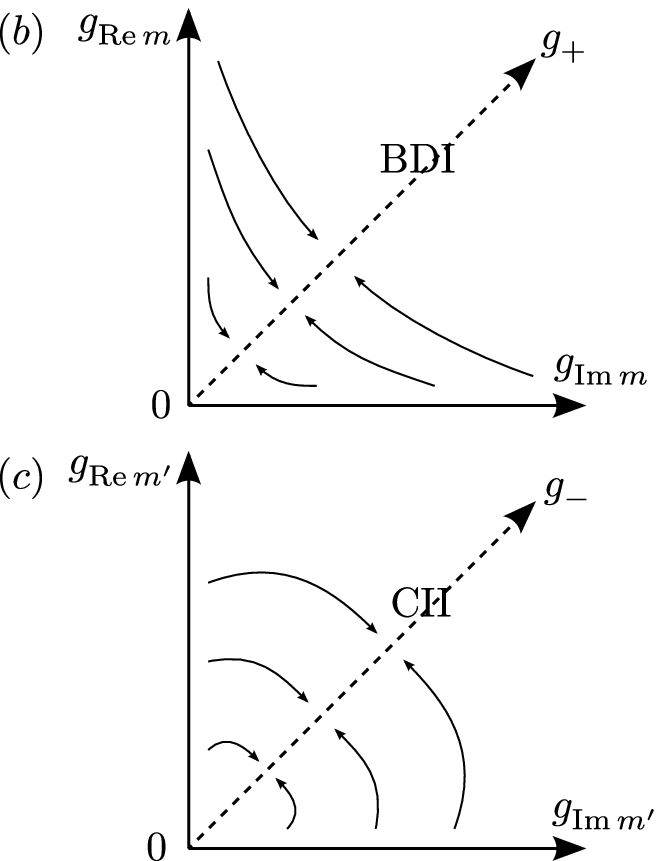}
\hfill
\includegraphics[angle=0,scale=0.6]{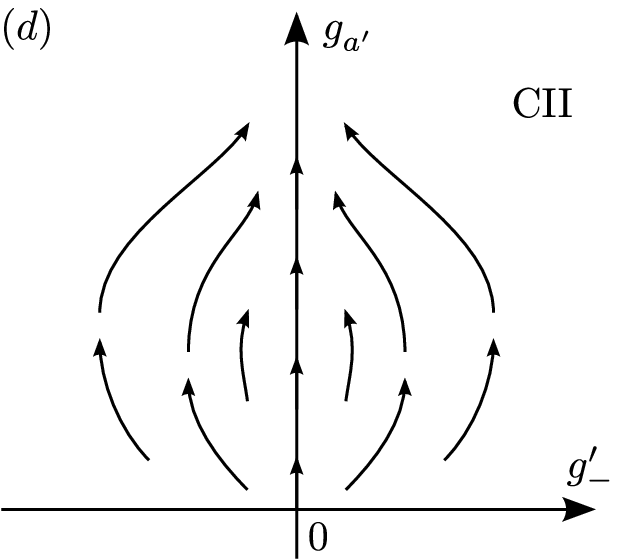}
\caption{
Global phase diagram for random Dirac fermion
defined by Eqs.~(\ref{eq: chiral random dirac hamiltonian}),
(\ref{eq: def HWK BDI}),
(\ref{eq: def HWK CII}),
(\ref{eq: def a a' m m'}),
and
(\ref{eq: def gv gw}). 
(a)
Flows of the coupling constants close to the
clean Dirac point (the origin denoted by an open circle).
Along the boundaries D and AII,
the coupling constant
$g^{\ }_{a'}$ is not generated under the RG, so
$g^{\ }_{a'}=0$ can be imposed in a consistent way.
In fact,  symmetry classes D and AII require\cite{Schnyder08,Bernard02}
$g^{\ }_{a'}=0$.
Away from these boundaries,
$g^{\ }_{a'}$ grows 
under the RG
and we have projected the flows onto the
$g^{\ }_{a'}=0$ plane 
in the regime
where
$g^{\ }_{a'}$ is still small.
In the region denoted BDI of the phase diagram,
there exists a line of (nearly) critical points
denoted by a dashed line as a result of Eq.~(\ref{eq: bet fcts b}).
This line of (nearly) critical points is perturbatively stable under
the RG flow~(\ref{eq: Bernard flows for BDI}).
In the region denoted CII of the phase diagram,
there exists a line of (nearly) critical points
denoted by a dashed line as a result of Eq.~(\ref{eq: bet fcts CII b}).
This line of (nearly) critical points appears to be
perturbatively unstable under
the RG flow~(\ref{eq: Bernard flows for CII})
for small values of
$g^{\ }_{a'}$.
(b)
Infrared flows dictated by
Eq.~(\ref{eq: Bernard flows for BDI})
close to the clean Dirac point when $g^{\ }_{a'}>0$. 
The slopes of the flows on the BDI boundaries
$g^{\ }_{\mathrm{Im}\,m}\,g^{\ }_{\mathrm{Re}\,m}=0$
have changed as compared to the case 
when $g^{\ }_{a'}=0$. 
(c)
Infrared flows dictated by
Eq.~(\ref{eq: Bernard flows for CII})
close to the clean Dirac point 
when $g^{\ }_{a'}>g^{\prime}_{+}$ with
$g^{\prime}_{\pm}:=g^{\ }_{\mathrm{Im}\,m'}\pm g^{\ }_{\mathrm{Re}\,m'}$
and
$g^{\prime}_{+}\geq|g^{\prime}_{-}|$.
The slopes of the flows on the CII boundaries
$g^{\prime}_{\mathrm{Im}\,m}\,g^{\prime}_{\mathrm{Re}\,m}=0$
have changed as compared to the case 
when $g^{\ }_{a'}=0$. 
Moreover, because of the condition 
$g^{\ }_{a'}>g^{\prime}_{+}$,
the RG flows in the quadrant CII are towards
the surface defined by the dashed line of (nearly) critical points
(the $g^{\prime}_{+}$ axis)
and the out-of-plane $g^{\ }_{a'}$ axis.
The plane 
$g^{\ }_{\mathrm{Im}\,m'}$ - $g^{\ }_{a'}$ 
with
$g^{\ }_{\mathrm{Re}\,m'}=0$ and $g^{\ }_{a'}>0$
and the plane
$g^{\ }_{\mathrm{Re}\,m'}$ - $g^{\ }_{a'}$
with
$g^{\ }_{\mathrm{Im}\,m'}=0$ and $g^{\ }_{a'}>0$
are always unstable under the one-loop flow%
~(\ref{eq: Bernard flows for CII}).
(d)
Infrared RG flows of Eq.~(\ref{eq: Bernard flows for CII})
in the surface defined by the $g^{\prime}_{-}$ axis as horizontal axis 
and the $g^{\ }_{a'}$ axis as vertical axis 
of the quadrant CII.
        }
\label{fig: phase diagram}
\end{center}
\end{figure*}

More specifically, we show 
that the phase diagram depicted in 
Fig.~\ref{fig: phase diagram}
encodes the localization properties of 
the random Dirac Hamiltonian 
$\mathcal{H}=\mathcal{K}+\mathcal{V}$
when the chiral-symmetric
random potential $\mathcal{V}$ 
is assigned the three possible independent disorder strengths
$g^{\ }_{\mathrm{Re}\,m}, g^{\ }_{\mathrm{Im}\,m}, g^{\ }_{a'}$
which are not irrelevant under the RG.
Here we discuss a ``global phase diagram'', depicted in
Fig.~\ref{fig: phase diagram}(a),
in the space of these three couplings which is projected onto the
$g^{\ }_{\mathrm{Re}\,m}$\,-\,$g^{\ }_{\mathrm{Im}\,m}$ 
plane (with $g^{\ }_{a'}=0$).
In this phase diagram, disordered Dirac fermion systems
in all three chiral symmetry classes, AIII,
CII, and BDI
occur in 4 quadrants, sharing one corner which
represents the clean Dirac fermion limit.
Also realized in the phase diagram are 
the symmetry classes AII and D
at the boundaries separating the
three  chiral symmetry classes,
whereby the parametrization of class D 
turns out to follow from analytic continuation 
of the relevant disorder strength
that parametrizes class AII in the phase diagram.

The random Dirac Hamiltonian $\mathcal{H}$ 
whose potential $\mathcal{V}$ is restricted to symmetry class AII 
captures the transport properties at long wave lengths 
of the surface states of a disordered
three-dimensional $\mathbb{Z}^{\ }_{2}$-topological
band insulator in symmetry class AII
(say, Bi$^{\ }_{1-x}$Sb$^{\ }_x$).%
\cite{Schnyder08}

The random Dirac Hamiltonian $\mathcal{H}$ 
whose potential $\mathcal{V}$ is restricted to symmetry class D 
captures the transport properties
of the fermionic quasiparticles of
a disordered
two-dimensional chiral $p$-wave superconductor
(say, Sr$^{\ }_2$RuO$^{\ }_{4}$) 
or their counterparts in the random bond Ising model
at long wave lengths.

Located in the center of the phase diagram of 
Fig.~\ref{fig: phase diagram}(a)
is a vertical dashed line.
There exists a sector of the theory
that decouples\cite{Guruswamy00} from the
 random $\mathrm{U}(1)$ gauge potential.
This sector is critical along the dashed line in
Fig.~\ref{fig: phase diagram}(a).
We will call the dashed line in
Fig.~\ref{fig: phase diagram}(a)
a line of nearly-critical points
to account for the non-critical sector
that is not depicted in
Fig.~\ref{fig: phase diagram}(a).

It is argued in Sec.~\ref{sec: topological term in class CII}
that along the dashed line in region CII of 
Fig.~\ref{fig: phase diagram}(a),
the transport properties of $\mathcal{H}$ 
are also encoded by those of a NL$\sigma$M
on the target manifold appropriate for this
symmetry class. 
(Such a possibility was also discussed,
independently and from a different perspective,
in Refs.~\onlinecite{Mitev08}, \onlinecite{Candu09}, \onlinecite{Candu10},
and \onlinecite{Candu11}.) 
Remarkably, the standard kinetic energy of the NL$\sigma$M
must be augmented by a $\mathbb{Z}^{\ }_{2}$-topological term
(see Appendix \ref{appsec: The sign ambiguity of a Pfaffian}).
Here, the necessary requirement for the presence of the 
$\mathbb{Z}^{\ }_{2}$-topological term is 
that the number $\mathrm{N}^{\ }_{\mathrm{f}}$ of flavors 
be two  times an odd integer $n$, i.e.
$\mathrm{N}^{\ }_{\mathrm{f}}=2n$.
However, any purely two-dimensional non-interacting local 
tight-binding Hamiltonian
with Fermi points at the band center
that breaks the spin-rotation symmetry but preserves
the time-reversal and sublattice symmetries 
yields a Dirac equation with 
$\mathrm{N}^{\ }_{\mathrm{f}}=2n$ 
where $n$ is an even integer 
because of the fermion doubling problem.
The fermion doubling problem for fermions
in two dimensions can be circumvented by working
with fermions localized at the two-dimensional
boundary of a three-dimensional crystal, i.e.,
with the boundary states of a topological band insulator
in symmetry class CII. It is the nearly-critical localization
properties of these surface states that are captured by the
dashed line in region CII of
Fig.~\ref{fig: phase diagram}.
Thus, we can view the $\mathbb{Z}^{\ }_{2}$-topological term in the NL$\sigma$M
for symmetry class CII 
as the signature of the physics of (de)localization,
that arises from the existence of boundary states in the clean limit,
the \textit{defining property} of 
three-dimensional $\mathbb{Z}^{\ }_{2}$-topological
band insulators in symmetry class CII.

{\it Third}, we argue that the initial
flow away from the apparently unstable nearly-critical line
in region CII depicted in 
Fig.~\ref{fig: phase diagram}(a) 
is not a crossover flow to the
diffusive metallic fixed point of the NL$\sigma$M
in symmetry class AII 
augmented by a $\mathbb{Z}^{\ }_{2}$ topological term.
Rather, it is the flow depicted in Fig.~\ref{fig: phase diagram}(c) 
that bends back towards the nearly-critical plane
defined by the dashed line and the out-of-plane axis for the coupling
$g^{\ }_{a'}$ as a result of the RG flow of the coupling $g^{\ }_{a'}$
to strong-coupling. This flow on sufficiently large
length scales 
along trajectories in the  three-dimensional coupling space 
is depicted through the two-dimensional cuts
presented in Figs.~\ref{fig: phase diagram}(b), 
\ref{fig: phase diagram}(c),
and~\ref{fig: phase diagram}(d).
The full RG flow along the boundary AII,
a separatrix of the RG flow,
was computed numerically
in Refs.~\onlinecite{Bardarson07} and \onlinecite{Nomura07}
owing to the presence of a
$\mathbb{Z}^{\ }_{2}$-topological term
on the target manifold of the NL$\sigma$M appropriate for  
symmetry class AII.%
\cite{Ostrovsky07,Ryu07b}

Finally,
in the quadrant labeled by BDI, the dashed line also 
represents a line of nearly-critical points.%
\cite{Hatsugai97,Guruswamy00,Mudry03,Ostrovsky06,Ryu10a}
This line of nearly-critical points is stable, without
the reentrant behavior of the kind mentioned in the
preceding paragraph.
The one-loop RG flow along the boundary D,
again a separatrix of the RG flow,
was computed in Refs.%
~\onlinecite{Dotsenko83}, 
\onlinecite{Ludwig87}, and \onlinecite{Shankar87}.

The fact that the quadrant in symmetry class BDI can be analytically 
continued to the quadrant in symmetry class CII 
suggests that  one can compute properties of the latter phase 
from the former one. In particular, 
sets of non-perturbative and exact results 
have been obtained for
e.g., boundary multifractal exponents for the point contact conductance 
on the critical line in symmetry class BDI.%
~\cite{Obuse08,Quella08} 
These results will also apply to the critical line in 
symmetry class CII upon suitable analytical continuation.

\subsection{
Outline
           }

The rest of the paper is organized as follows:
The non-interacting random Dirac fermion model is defined 
in Sec.~\ref{sec; Definitions and phase diagram}.
The main result of this section is captured by 
Fig.~\ref{fig: phase diagram}.
We argue in Secs.~\ref{sec: Projected Thirring model}
and~\ref{sec: topological term in class CII} 
that the generating function for
the moments of the retarded Green's functions
for microscopic parameters corresponding to the
quadrant CII in 
Fig.~\ref{fig: phase diagram}
realizes a replicated fermionic or, alternatively, 
a supersymmetric (SUSY) NL$\sigma$M 
augmented by a $\mathbb{Z}^{\ }_2$-topological term.
We conclude in Sec.~\ref{sec: Discussions}.

\section{ 
Definitions and phase diagram
        }
\label{sec; Definitions and phase diagram}

We begin in Sec.~\ref{subsec: Definitions} 
by defining a non-interacting random Dirac Hamiltonian
and proceed with a symmetry analysis. To identify the axis
of the phase diagram in 
Fig.~\ref{fig: phase diagram}, 
a generating function for the disorder
average over products of $\mathrm{N}$ 
retarded single-particle Green's functions 
is needed. This is done using the supersymmetric 
(SUSY) formalism in Secs.~\ref{subsec: SUSY path integral}
and \ref{subsec: Phase diagram}.
The flows in 
Fig.~\ref{fig: phase diagram}
to or away from the nearly-critical line follow once
it is shown in Sec.~\ref{subsec: nearly-critical line}
that the SUSY generating function defines a 
$\widehat{\mathrm{gl}}(2\mathrm{N}|2\mathrm{N})^{\ }_{ k=1}$ 
SUSY Thirring model studied in 
Refs.~\onlinecite{Guruswamy00} and \onlinecite{Mudry03}.

\subsection{
Definitions
           }
\label{subsec: Definitions}

Common to all the aforementioned microscopic examples
is the existence of 4 Fermi points at the relevant
Fermi energy around which linearization 
in momentum space yields the
continuum Dirac kinetic energy
\begin{equation}
\begin{split}
\mathcal{K}(\bm{p}):=&\,
\begin{pmatrix}
0
&
0
&
0
&
p
\\
0
&
0
&
\bar{p}
&
0
\\
0
&
p
&
0
&
0
\\
\bar{p}
&
0
&
0
&
0
\end{pmatrix}
\\
\equiv&\,
\begin{pmatrix}
0
&
\sigma^{\ }_{x}
p^{\ }_{x}
+
\sigma^{\ }_{y}
p^{\ }_{y}
\\
\sigma^{\ }_{x}
p^{\ }_{x}
+
\sigma^{\ }_{y}
p^{\ }_{y}
&
0
\end{pmatrix}
\\
\equiv&\,
\rho^{\ }_{1}\otimes\sigma^{\ }_{1}\,p^{\ }_{1}
+
\rho^{\ }_{1}\otimes\sigma^{\ }_{2}\,p^{\ }_{2},
\end{split}
\label{eq: def K}
\end{equation}
up to a unitary transformation.
Here, the momentum 
$\bm{p}=
(p^{\ }_{x},p^{\ }_{y})
\equiv
(p^{\ }_{1},p^{\ }_{2})$
is measured relative to the Fermi points
at the band center. The complex
notation 
$p      =p^{\ }_{x}-{i}p^{\ }_{y}$
and
$\bar{p}=p^{\ }_{x}+{i}p^{\ }_{y}$
is occasionally used for conciseness.
The unit $2\times2$ matrix $\sigma^{\ }_{0}$
and the three Pauli matrices
$(\sigma^{\ }_{1} ,\sigma^{\ }_{2},\sigma^{\ }_{3})$
are reserved for the spinor indices of $\mathrm{SO}(1,2)$.
The unit $2\times2$ matrix $\rho^{\ }_{0}$
and the three Pauli matrices
$(\rho^{\ }_{1} ,\rho^{\ }_{2},\rho^{\ }_{3})$
are reserved for the two-dimensional flavor subspace.

This kinetic energy has two interesting properties.
First, it anticommutes with the $4\times4$
unitary and Hermitian matrices
\begin{equation}
\begin{split}
&
\mathcal{C}^{\ }_{1}:=
\rho^{\ }_{3}\otimes\sigma^{\ }_{0},
\qquad
\mathcal{C}^{\ }_{1}\mathcal{C}^{\dag}_{1}=
\mathcal{C}^{\ }_{1}\mathcal{C}^{\   }_{1}=
+
\mathcal{C}^{\ }_{1}\mathcal{C}^{*   }_{1}=1,
\\
&
\mathcal{C}^{\ }_{2}:=
\rho^{\ }_{2}\otimes\sigma^{\ }_{0},
\qquad
\mathcal{C}^{\ }_{2}\mathcal{C}^{\dag}_{2}=
\mathcal{C}^{\ }_{2}\mathcal{C}^{\   }_{2}=
-
\mathcal{C}^{\ }_{2}\mathcal{C}^{*   }_{2}=1,
\\
&
\mathcal{C}^{\ }_{3}:=
\rho^{\ }_{0}\otimes\sigma^{\ }_{3},
\qquad
\mathcal{C}^{\ }_{3}\mathcal{C}^{\dag}_{3}=
\mathcal{C}^{\ }_{3}\mathcal{C}^{\   }_{3}=
+
\mathcal{C}^{\ }_{3}\mathcal{C}^{*   }_{3}=1,
\\
&
\mathcal{C}^{\ }_{4}:=
\rho^{\ }_{1}\otimes\sigma^{\ }_{3},
\qquad
\mathcal{C}^{\ }_{4}\mathcal{C}^{\dag}_{4}=
\mathcal{C}^{\ }_{4}\mathcal{C}^{\   }_{4}=
+
\mathcal{C}^{\ }_{4}\mathcal{C}^{*   }_{4}=1.
\end{split}
\label{eq: 4 C's}
\end{equation}
Second, the operations on $\mathcal{K}$ consisting in
the momentum inversion $\bm{p}\to-\bm{p}$,
complex conjugation,
and matrix multiplication from the left and from the right
by the $4\times4$
unitary and Hermitian matrices
\begin{equation}
\begin{split}
&
\mathcal{T}^{\ }_{1}:=
\rho^{\ }_{3}\otimes\sigma^{\ }_{1},
\qquad
\mathcal{T}^{\ }_{1}\mathcal{T}^{\dag}_{1}=
\mathcal{T}^{\ }_{1}\mathcal{T}^{\   }_{1}=
+
\mathcal{T}^{\ }_{1}\mathcal{T}^{*   }_{1}=1,
\\
&
\mathcal{T}^{\ }_{2}:=
\rho^{\ }_{0}\otimes\sigma^{\ }_{2},
\qquad
\mathcal{T}^{\ }_{2}\mathcal{T}^{\dag}_{2}=
\mathcal{T}^{\ }_{2}\mathcal{T}^{\   }_{2}=
-
\mathcal{T}^{\ }_{2}\mathcal{T}^{*   }_{2}=1,
\\
&
\mathcal{T}^{\ }_{3}:=
\rho^{\ }_{1}\otimes\sigma^{\ }_{2},
\qquad
\mathcal{T}^{\ }_{3}\mathcal{T}^{\dag}_{3}=
\mathcal{T}^{\ }_{3}\mathcal{T}^{\   }_{3}=
-
\mathcal{T}^{\ }_{3}\mathcal{T}^{*   }_{3}=1,
\\
&
\mathcal{T}^{\ }_{4}:=
\rho^{\ }_{2}\otimes\sigma^{\ }_{1},
\qquad
\mathcal{T}^{\ }_{4}\mathcal{T}^{\dag}_{4}=
\mathcal{T}^{\ }_{4}\mathcal{T}^{\   }_{4}=
-
\mathcal{T}^{\ }_{4}\mathcal{T}^{*   }_{4}=1,
\end{split}
\label{eq: 4 T's}
\end{equation}
all yield $\mathcal{K}$ again.
For any $i,j=1,\cdots,4$,
the property 
\begin{equation}
\mathcal{C}^{\ }_{i}\,
\mathcal{K}(\bm{p})\,
\mathcal{C}^{\ }_{i}=
-
\mathcal{K}(\bm{p}),
\end{equation}
that we will call (abusively) chiral symmetry (chS),
is compatible with the property
\begin{equation}
\mathcal{T}^{\ }_{j}\,
\mathcal{K}^{* }(-\bm{p})\,
\mathcal{T}^{\ }_{j}=
\mathcal{K}(\bm{p}),
\end{equation}
that we will call TRS,
if and only if
\begin{equation}
[\mathcal{C}^{\ }_{i},\mathcal{T}^{\ }_{j}]=0.
\end{equation}

In this paper, 
we shall assume that the lattice model from which
$\mathcal{K}(\bm{p})$ 
emerges imposes the chiral symmetry generated by 
\begin{equation}
\mathcal{C}\equiv
\mathcal{C}^{\ }_{1}.
\label{eq: def cal C}
\end{equation}
This chiral symmetry commutes with
\begin{equation}
\mathcal{T}\equiv
\mathcal{T}^{\ }_{1}
\label{eq: def cal T}
\end{equation}
and with
\begin{equation}
\mathcal{T}^{\prime}\equiv
\mathcal{T}^{\ }_{2}.
\end{equation}
(Observe that $\mathcal{T}$ and $\mathcal{T}^{\prime}$ anticommute.
They are not compatible.)
This leads to two possible forms of TRS,
either the one appropriate for particles with integer isospin when
\begin{equation}
\mathcal{T}^{T}=+\mathcal{T}
\end{equation}
is imposed as a symmetry,
or the one for particles with half-integer isospin when
\begin{equation}
\mathcal{T}^{\prime T}=-\mathcal{T}^{\prime}
\end{equation}
is imposed as a symmetry. Again, the choice between $\mathcal{T}$
and $\mathcal{T}^{\prime}$ is dictated by the underlying lattice model.

The most general static random potential
that anticommutes with $\mathcal{C}$ is of the form 
\begin{subequations}
\label{eq: chiral random dirac hamiltonian}
\begin{equation}
\begin{split}
&
\mathcal{V}=
\begin{pmatrix}
0
&
V
\\
V^{\dag}
&
0
\end{pmatrix},
\\
&
V=
\sigma^{\ }_1
A^{\ }_1
+
\sigma^{\ }_2
A^{\ }_2
+
\sigma^{\ }_3
M^{\ }_3 
+
\sigma^{\ }_0
M^{\ }_0,
\label{eq: def mathcal D in chiral random dirac hamiltonian}
\end{split}
\end{equation}
where the complex-valued
\begin{equation}
\begin{split}
&
A^{\ }_{1}=
a^{\ }_{1}
-
ia^{\prime}_{1},
\\
&
A^{\ }_{2}=
a^{\ }_{2}
-
ia^{\prime}_{2},
\\
&
M^{\ }_3=
-
m^{\ }_{3}
-
im^{\prime}_{3}, 
\\
&
M^{\ }_0=
m^{\prime}_0
-
i
m^{\ }_{0},
\end{split}
\label{eq: randomness in HWK}
\end{equation}
represent sources of (static) randomness,
i.e., complex-valued functions of the space
coordinates
$\bm{r}\in\mathbb{R}^{2}$.
(The unusual sign conventions is chosen
to make contact with the notation
of Ref.~\onlinecite{Mudry03}.)
It yields the random Dirac Hamiltonian
\begin{equation}
\begin{split}
\mathcal{H}(\bm{r}):=&\,
\left(
\mathcal{K}
+
\mathcal{V}
\right)(\bm{r})
\\
=&\,
\left(
\begin{array}{cc}
0 
& 
D(\bm{r})
\\
D^{\dag}(\bm{r})
& 
0
\end{array}
\right)
\\
=&\,
-i\rho^{\ }_{1}\otimes\sigma^{\ }_{1}\,\partial^{\ }_{1}
-i\rho^{\ }_{1}\otimes\sigma^{\ }_{2}\,\partial^{\ }_{2}
\\
&\,
+
\rho^{\ }_{1}\otimes\sigma^{\ }_{1}\,a^{\ }_{1}(\bm{r})
+
\rho^{\ }_{1}\otimes\sigma^{\ }_{2}\,a^{\ }_{2}(\bm{r})
\\
&\,
+
\rho^{\ }_{2}\otimes\sigma^{\ }_{1}\,a^{\prime}_{1}(\bm{r})
+
\rho^{\ }_{2}\otimes\sigma^{\ }_{2}\,a^{\prime}_{2}(\bm{r})
\\
&\,
-
\rho^{\ }_{1}\otimes\sigma^{\ }_{3}\,m^{\ }_{3}(\bm{r})
+
\rho^{\ }_{1}\otimes\sigma^{\ }_{0}\,m^{\prime}_{0}(\bm{r})
\\
&\,
+
\rho^{\ }_{2}\otimes\sigma^{\ }_{3}\,m^{\prime}_{3}(\bm{r})
+
\rho^{\ }_{2}\otimes\sigma^{\ }_{0}\,m^{\ }_{0}(\bm{r}).
\end{split}
\label{eq: def mathcal H in chiral random dirac hamiltonian}
\end{equation}
\end{subequations}
By construction,
Hamiltonian~(\ref{eq: def mathcal H in chiral random dirac hamiltonian}) 
is a member of the AIII symmetry class (chiral-unitary symmetry class)
of Anderson localization in two dimensions.

When the disorder~(\ref{eq: randomness in HWK})
is restricted to 
\begin{subequations}
\label{eq: def HWK BDI}
\begin{equation}
A^{\ }_{\mu}= -{i} a^{\prime}_{\mu} \in {i}\mathbb{R}, 
\
M^{\ }_3= -m^{\ }_3 \in \mathbb{R},
\
M^{\ }_0= -{i}m^{\ }_0 \in {i}\mathbb{R},
\label{eq: def randomness if BDI}
\end{equation}
the random Hamiltonian%
~(\ref{eq: def mathcal H in chiral random dirac hamiltonian})
reduces to
\begin{equation}
\begin{split}
\mathcal{H}(\bm{r})=&\,
-i\rho^{\ }_{1}\otimes\sigma^{\ }_{1}\,\partial^{\ }_{1}
-i\rho^{\ }_{1}\otimes\sigma^{\ }_{2}\,\partial^{\ }_{2}
\\
&\,
+
\rho^{\ }_{2}\otimes\sigma^{\ }_{1}\,a^{\prime}_{1}(\bm{r})
+
\rho^{\ }_{2}\otimes\sigma^{\ }_{2}\,a^{\prime}_{2}(\bm{r})
\\
&\,
-
\rho^{\ }_{1}\otimes\sigma^{\ }_{3}\,m^{\ }_{3}(\bm{r})
+
\rho^{\ }_{2}\otimes\sigma^{\ }_{0}\,m^{\ }_{0}(\bm{r})
\end{split}
\end{equation}
and hence is invariant under the time-reversal
\begin{eqnarray}
  \mathcal{T}
  \mathcal{H}^*(\bm{r})
  \mathcal{T}=
  \mathcal{H}(\bm{r}),
  \qquad
  \mathcal{T}:=
  \rho^{\ }_3\otimes\sigma^{\ }_1,
\end{eqnarray}
\end{subequations}
for any realization of the disorder~(\ref{eq: def randomness if BDI}).
Accordingly,
this Hamiltonian is a
member of the BDI symmetry class 
(chiral-orthogonal symmetry class)
in Anderson localization.

On the other hand, when the disorder~(\ref{eq: randomness in HWK}) 
is restricted to  
\begin{subequations}
\label{eq: def HWK CII}
\begin{equation}
A^{\ }_{\mu}= -{i} a^{\prime}_{\mu} \in {i}\mathbb{R},
\ 
M^{\ }_3= -{i}m^{\prime}_3 \in {i}\mathbb{R},
\
M^{\ }_0= m^{\prime}_0 \in \mathbb{R},
\label{eq: def randomness if CII}
\end{equation}
the random Hamiltonian%
~(\ref{eq: def mathcal H in chiral random dirac hamiltonian})
reduces to
\begin{equation}
\begin{split}
\mathcal{H}(\bm{r})=&\,
-i\rho^{\ }_{1}\otimes\sigma^{\ }_{1}\,\partial^{\ }_{1}
-i\rho^{\ }_{1}\otimes\sigma^{\ }_{2}\,\partial^{\ }_{2}
\\
&\,
+
\rho^{\ }_{2}\otimes\sigma^{\ }_{1}\,a^{\prime}_{1}(\bm{r})
+
\rho^{\ }_{2}\otimes\sigma^{\ }_{2}\,a^{\prime}_{2}(\bm{r})
\\
&\,
+
\rho^{\ }_{2}\otimes\sigma^{\ }_{3}\,m^{\prime}_{3}(\bm{r})
+
\rho^{\ }_{1}\otimes\sigma^{\ }_{0}\,m^{\prime}_{0}(\bm{r})
\end{split}
\end{equation}
and hence is invariant under the time-reversal
\begin{eqnarray}
  \mathcal{T}'
  \mathcal{H}^*(\bm{r})
  \mathcal{T}'=
  \mathcal{H}(\bm{r}),
  \qquad
  \mathcal{T}':=
  \rho^{\ }_0\otimes\sigma^{\ }_2,
\end{eqnarray}
\end{subequations}
for any realization of the disorder~(\ref{eq: def randomness if CII}).
Accordingly, this Hamiltonian is a
member of the CII symmetry class 
(chiral-symplectic symmetry class)
in Anderson localization.

The BDI case~(\ref{eq: def HWK BDI})
can be derived as the continuum limit
of a real-valued, nearest-neighbor, spin-independent,
and random hopping model 
on a bipartite lattice; the honeycomb lattice
of graphene or the square lattice with a $\pi$-flux phase say.%
\cite{Hatsugai97} 
The four-dimensional subspace associated to the 
$\rho$'s and $\sigma$'s originates from the 2-sublattices structure
and the 2 non-equivalent Fermi points at the band center.
The electronic spin plays here no role besides an overall degeneracy
factor as spin-orbit coupling is neglected.
In the context of graphene, 
the random fields 
$a^{\prime}_{1}$
and
$a^{\prime}_{2}$
are called ripples
[see Fig.\ \ref{fig: Graphene+dimerization}(c-d)],%
\cite{ripples}
while the random masses
$m^{\ }_{3}$ and $m^{\ }_{0}$
are smooth bond fluctuations about the Kekul\'e 
dimerization pattern of the nearest-neighbor hopping
amplitude
[see Fig.\ \ref{fig: Graphene+dimerization}(a-b)].~\cite{Hou07}
In the context of the $\pi$-flux phase, the random fields 
$a^{\prime}_{1}$
and
$a^{\prime}_{2}$
are smooth fluctuations of the nearest-neighbor
hopping amplitudes about the two wave vectors for
the two independent staggered dimerization patterns,
while the random masses
$m^{\ }_{3}$ and $m^{\ }_{0}$
are smooth bond fluctuations about the 
two independent columnar dimerization pattern.%
~\cite{Fradkin91}
The CII case~(\ref{eq: def HWK CII})
can be derived as the 
restriction to a two-dimensional boundary
of a disordered, three-dimensional
$\mathbb{Z}^{\ }_{2}$-topological band insulator
in the chiral-symplectic class of Anderson localization.%
~\cite{Schnyder08}

When the disorder is restricted to
\begin{subequations}
\label{eq: class D}
\begin{equation}
A^{\ }_{\mu}=0,
\qquad
M^{\ }_{3}=-m^{\ }_{3}\in\mathbb{R},
\qquad
M^{\ }_{0}=
0,
\label{eq: strong conditions class D}
\end{equation}
we observe that
Hamiltonian~(\ref{eq: def mathcal H in chiral random dirac hamiltonian})
reduces to
\begin{equation}
\mathcal{H}=
\rho^{\ }_{1}\otimes D,
\end{equation}
with
\begin{equation}
D=D^\dagger,
\qquad
\sigma_1 D^* \sigma_1=-D,
\label{eq: def D}
\end{equation}
and can be thought of as a random Hamiltonian
belonging to the symmetry class D 
(BdG Hamiltonians with both 
time-reversal symmetry and spin-1/2 rotation symmetry broken)
in Anderson localization,
for $\mathcal{H}$
is then unitarily equivalent to
\begin{equation}
\left(
\begin{array}{cc}
D
&
0
\\
0
&
-D
\end{array}
\right)
\label{(D,-D)}
\end{equation}
\end{subequations}
with the unitary transformation
$(\rho^{\ }_{0}+i\rho^{\ }_{2})\otimes\sigma^{\ }_{0}/\sqrt2$.

Finally, when the disorder is restricted to
\begin{subequations}
\label{eq: AII limit random Dirac Hamiltonian}
\begin{equation}
A^{\ }_{\mu}=
0,
\qquad
M^{\ }_{3}=
0,
\qquad
M^{\ }_{0}=
m^{\prime}_{0}
\in\mathbb{R},
\label{eq: strong conditions class AII}
\end{equation}
we observe that
Hamiltonian~(\ref{eq: def mathcal H in chiral random dirac hamiltonian})
reduces to
\begin{equation}
\mathcal{H}=
\rho^{\ }_{1}\otimes D
\end{equation}
with
\begin{equation}
D=D^\dagger,
\qquad
\sigma^{\ }_{2}D^{*}\sigma^{\ }_{2}=D,
\end{equation}
\end{subequations}
and can be thought of as a random Hamiltonian belonging
to symmetry class AII
(a spin-$1/2$ electron with
time-reversal symmetry but without spin-rotation symmetry) 
in Anderson localization,
for $\mathcal{H}$ can then be brought to
the block diagonal form
(\ref{(D,-D)})
by the same unitary transformation used to reach
(\ref{(D,-D)}).

All four symmetry conditions are summarized in
Table~\ref{table I}. The defining conditions 
on classes D and AII can be made
slightly more general than in Eqs.%
~(\ref{eq: strong conditions class D})
and%
~(\ref{eq: strong conditions class AII}) 
as will become clear at the end of 
Sec.~\ref{subsec: Phase diagram}.

\begin{table}
\caption{
\label{table I}
Symmetry conditions on the static random fields 
in the Hamiltonian~(\ref{eq: chiral random dirac hamiltonian}).
For the symmetry classes D and AII, $M^{\ }_{3}M^{\ }_{0}=0$ must hold.
        }
\begin{ruledtabular}
\begin{tabular}{cccccc}
&
AIII
&
BDI
&
CII
&
D
&
AII
\\
\hline
$
A^{\ }_{1}
$
&
$
a^{\ }_{1}
-
{i}a^{\prime}_{1}
$
&
$
-
{i}a^{\prime}_{1}
$
&
$
-
{i}a^{\prime}_{1}
$
&
$
0
$
&
$
0
$
\\
$
A^{\ }_{2}
$
&
$
a^{\ }_{2}
-
{i}a^{\prime}_{2}
$
&
$
-
{i}a^{\prime}_{2}
$
&
$
-
{i}a^{\prime}_{2}
$
&
$
0
$
&
$
0
$
\\
$
M^{\ }_{3}
$
&
$
-
m^{\ }_{3}
-
{i}
m^{\prime}_{3}
$
&
$
-
m^{\ }_{3}
$
&
$
-
{i}
m^{\prime}_{3}
$
&
$
-
m^{\ }_{3}
$
&
$
-{i}m^{\prime}_{3}
$
\\
$
M^{\ }_{0}
$
&
$
m^{\prime}_{0}
-
{i}
m^{\     }_{0}
$
&
$
-
{i}
m^{\     }_{0}
$
&
$
m^{\prime}_{0}
$
&
$
-{i}m^{\ }_{0}
$
&
$
m^{\prime}_{0}
$
\end{tabular}
\end{ruledtabular}
\end{table}

\subsection{ 
Path integral representation of the single-particle 
Green's function
           }
\label{subsec: SUSY path integral}

In Anderson localization,
physical quantities are expressed 
by (products of) the retarded 
($+{i}\eta$, $\eta>0$)
and advanced 
($-{i}\eta$)
Green's functions 
\begin{equation}
  \mathcal{G}^{\mathrm{R}/\mathrm{A}}(E):=
  (E\pm{i}\eta -\mathcal{H})^{-1}.
\end{equation}
At the band center $E=0$,
the retarded and advanced Green's functions are
related by the chiral symmetry through
\begin{eqnarray}
  \mathcal{C}\,
\mathcal{G}^{\mathrm{R}}(E=0)\,\mathcal{C}=
  -\mathcal{G}^{\mathrm{A}}(E=0).
\label{eq: R to A G}
\end{eqnarray}
Hence, any arbitrary product of retarded or advanced 
Green's function
at the band center equates, up to a sign, 
a product of retarded Green's functions at the band center.
{}From now on we will omit the energy argument 
of the Green's function,
bearing in mind that it is always fixed to the band center $E=0$.

Because of Eq.~(\ref{eq: R to A G}),
it suffices to introduce functional
integrals for the retarded Green's function
defined with the help of 
the SUSY partition function 
\begin{subequations}
\label{eq: def Z}
\begin{equation}
\begin{split}
&
Z:= 
Z^{\ }_{\mathrm{F}}\times Z^{\ }_{\mathrm{B}},
\\
&
  Z^{\ }_{\mathrm{F}}:=
  \int\mathcal{D}[\bar{\chi},\chi]
  \exp   
  \left(
  {i}
  \int \mathrm{d}^2\,r\,
  \bar{\chi}
  \left(
    {i}\eta
    -\mathcal{H}
  \right)
  \chi
  \right),
\\
&
  Z^{\ }_{\mathrm{B}}:=
  \int\mathcal{D}[\bar{\xi},\xi]
  \exp   
  \left(
  {i}
  \int \mathrm{d}^2\,r\,
  \bar{\xi}
  \left(
    {i}\eta
    -\mathcal{H}
  \right)
  \xi
  \right).
\end{split}
\label{eq: def Z a}
\end{equation}
Here,
$(\bar{\chi},\chi)$ is a pair of two independent 
four-component fermionic fields,
and
$(\bar{\xi},\xi)$ 
is a pair of four-component bosonic fields related by complex conjugation.
For any $\eta>0$, 
\begin{equation}
Z=1
\end{equation}
\end{subequations}
holds. The matrix elements of the retarded Green's 
function can be represented as
\begin{equation}
{i}
\mathcal{G}^{\mathrm{R}} (\boldsymbol{r},\boldsymbol{r}')=
  \langle
  \chi(\boldsymbol{r})\bar{\chi}(\boldsymbol{r}')
  \rangle
  =
  \langle
  \xi(\boldsymbol{r})\bar{\xi}(\boldsymbol{r}')
  \rangle
\label{eq: first rep G(r,r')}
\end{equation}
with $\langle\cdots\rangle$ denoting the expectation value taken with
the partition function $Z$. 

We now perform the change of integration variables 
from
$\bar{\chi},\chi$ 
to 
$\bar{\psi}^{\mathrm{a}\dag}_{\ },\bar{\psi}^{\ }_{\mathrm{a}},
 \psi^{\mathrm{a}\dag}_{\ },\psi^{\ }_{\mathrm{a}}$
in the fermionic sector and
from
$\bar{\xi},\xi$
to
$\bar{\beta}^{\mathrm{a}\dag}_{\ },\bar{\beta}^{\ }_{\mathrm{a}},
 \beta^{\mathrm{a}\dag}_{\ },\beta^{\ }_{\mathrm{a}}$
in the bosonic sector
where $\mathrm{a}=1,2$ and,
\begin{equation}
\begin{split}
&
\bar{\chi}=:
\frac{1}{\sqrt{2\pi}}
\begin{pmatrix}
\bar{\psi}^{1\dag}_{\ }
&
\psi^{1\dag}_{\ }
&
-i\bar{\psi}^{\ }_2
&
-i\psi^{\ }_2
\end{pmatrix},
\\
&
\chi=:
\frac{1}{\sqrt{2\pi}}
\begin{pmatrix}
+i
\psi^{2\dag}_{\ }
&
+i
\bar{\psi}^{2\dag}_{\ }
&
\psi^{\ }_1
&
\bar{\psi}^{\ }_{1}
\end{pmatrix}^T,
\\
&
\bar{\xi}=:
\frac{1}{\sqrt{2\pi}}
\begin{pmatrix}
\bar{\beta}^{1\dag}_{\ }
&
\beta^{1\dag}_{\ }
&
-i
\bar{\beta}^{\ }_{2}
&
-i
\beta^{\ }_2
\end{pmatrix},
\\
&
\xi=:
\frac{1}{\sqrt{2\pi}}
\begin{pmatrix}
-i
\beta^{2\dag}_{\ }
&
-i
\bar{\beta}^{2\dag}_{\ }
&
\beta^{\ }_1
&
\bar{\beta}^{\ }_1
\end{pmatrix}^T.
\end{split}
\label{eq: from chi and xi to psi and beta}
\end{equation}
Any correlation function
such as the retarded Green's function~(\ref{eq: first rep G(r,r')})
is, under this or any similar change of integration variable,
to be computed with
the SUSY partition function
\begin{equation}
\begin{split}
Z=&\,
\int
\mathcal{D}[\bar\psi,\psi,\bar\beta,\beta]
\frac{\mathcal{D}\bar\psi}{\mathcal{D}\bar\chi}
\frac{\mathcal{D}    \psi}{\mathcal{D}    \chi}
\frac{\mathcal{D}\bar\xi}{\mathcal{D}\bar\beta}
\frac{\mathcal{D}    \xi}{\mathcal{D}    \beta}
\\
&\,
\times
\exp
\left(
{i}
\int \mathrm{d}^2\,r\,
\bar\chi(\bar\psi,\psi)
\left({i}\eta-\mathcal{H}\right)
    \chi(\bar\psi,\psi)
\right)
\\
&\,
\times
\exp
\left(
{i}
\int \mathrm{d}^2\,r\,
\bar\xi(\bar\beta,\beta)
\left({i}\eta-\mathcal{H}\right)
    \xi(\bar\beta,\beta)
\right).
\end{split}
\label{eq: relabeling integration variables}
\end{equation}
The message conveyed by Eq.%
~(\ref{eq: relabeling integration variables}) 
is that we are free to relabel
\textit{all} integration variables in Eq.~(\ref{eq: def Z a})
independently from each other, 
provided the correct book keeping with  
the integration variables in the convergent path integral%
~(\ref{eq: def Z a})
is kept.
In this context the symbols $\bar{\ }$ and
$\dag$ on the right-hand side of
Eq.~(\ref{eq: from chi and xi to psi and beta}) 
are only distinctive labels, i.e., 
here they are not to be confused
with complex conjugation.
The change of integration variable%
~(\ref{eq: from chi and xi to psi and beta}) is
made to bring the effective action to a form identical 
to that found in Ref.~\onlinecite{Mudry03}
in which important symmetries\cite{Guruswamy00} of 
the partition function in the limit $\eta=0$ become
manifest.

We also introduce 
\begin{equation}
\begin{split}
&
a\equiv
a^{\ }_1 - {i} a^{\ }_2
\equiv
\mathrm{Re}\,A^{\ }_1 - {i}\,\mathrm{Re}\,A^{\ }_2
,
\\
&
a'\equiv
a^{\prime}_1 - {i} a^{\prime}_2
\equiv
-\mathrm{Im}\,A^{\ }_1 + {i}\,\mathrm{Im}\,A^{\ }_2 
,
\\
&
m\equiv
m^{\ }_0 - {i}m^{\ }_3
\equiv
-\mathrm{Im}\,M^{\ }_0 + {i}\,\mathrm{Re}\,M^{\ }_3
,
\\
&
m'\equiv
m^{\prime}_0 - {i}m^{\prime}_3
\equiv
\mathrm{Re}\,M^{\ }_0 + {i}\,\mathrm{Im}\,M^{\ }_3,
\end{split}
\label{eq: def a a' m m'}
\end{equation}
and their complex conjugate 
$\bar{a}$,
$\bar{a}'$,
$\bar{m}$,
and
$\bar{m}'$,
in terms of which symmetry class BDI is defined by the conditions
\begin{equation}
a=0,\qquad
a'\in\mathbb{C},\qquad
m\in\mathbb{C},\qquad
m'=0,
\end{equation}
while symmetry class CII is defined by the conditions
\begin{equation}
a=0,\qquad
a'\in\mathbb{C},\qquad
m=0,\qquad
m'\in\mathbb{C}.
\end{equation}
The boundary 
\begin{equation}
a=a'=0,
\qquad
\mathrm{Re}\,m=0,
\qquad
m'=0
\label{eq: strong conditions class D bis}
\end{equation}
between the symmetry classes
BDI and AIII belongs to symmetry class D.
The boundary 
\begin{equation}
a=a'=0,
\qquad
m=0,
\qquad
\mathrm{Im}\,m'=0
\label{eq: strong conditions class AII bis}
\end{equation}
between the symmetry classes
CII and AIII belongs to the symmetry class AII.
All four symmetry conditions are summarized in
Table~\ref{table II}. The defining conditions 
on the symmetry classes D and AII can be made
slightly more general than in Eqs.%
~(\ref{eq: strong conditions class D bis})
and%
~(\ref{eq: strong conditions class AII bis}) 
as will become clear at the end of 
Sec.~\ref{subsec: Phase diagram}.

\begin{table}
\caption{
\label{table II}
Symmetry conditions on the static random fields 
in the generating function~(\ref{eq: final partition fct HWK}).
        }
\begin{ruledtabular}
\begin{tabular}{cccccc}
&
AIII
&
BDI
&
CII
&
D
&
AII
\\
\hline
$
a
$
&
$
   a^{\ }_{1}
-
{i}a^{\ }_{2}
$
&
$
0
$
&
$
0
$
&
$
0
$
&
$
0
$
\\
$
a^{\prime}
$
&
$
a^{\prime}_{1}
-
{i}
a^{\prime}_{2}
$
&
$
a^{\prime}_{1}
-
{i}
a^{\prime}_{2}
$
&
$
a^{\prime}_{1}
-
{i}
a^{\prime}_{2}
$
&
$
0
$
&
$
0
$
\\
$
m
$
&
$
m^{\ }_{0}
-
{i}
m^{\ }_{3}
$
&
$
m^{\ }_{0}
-
{i}
m^{\ }_{3}
$
&
$
0
$
&
$
m^{\ }_{0}
m^{\ }_{3}=0
$
&
$
0
$
\\
$
m^{\prime}
$
&
$
m^{\prime}_{0}
-
{i}
m^{\prime}_{3}
$
&
$
0
$
&
$
m^{\prime}_{0}
-
{i}
m^{\prime}_{3}
$
&
$
0
$
&
$
m^{\prime}_{0}
m^{\prime}_{3}=0
$
\end{tabular}
\end{ruledtabular}
\end{table}

With these changes of variables, 
the partition function 
$Z=Z^{\ }_{\mathrm{F}}\times Z^{\ }_{\mathrm{B}}$
at $E=0$ can be written as
\begin{subequations}
\label{eq: final partition fct HWK}
\begin{equation}
\begin{split}
&
Z^{\ }_{\mathrm{F}}= 
\int
\mathcal{D}\!
\left[
\bar{\psi}^{\mathrm{a}\dag}_{\ },
\psi^{\mathrm{a}\dag}_{\ },
\bar{\psi}^{\ }_{\mathrm{a}},
\psi^{\ }_{\mathrm{a}}
\right]
\exp\!
\left(
-
\int \mathrm{d}^2\,r
\left(
\mathcal{L}^{\ }_{\mathrm{F}}
+
\mathcal{L}^{{i}\eta}_{\mathrm{F}}
\right)
\right),
\\
&
Z^{\ }_{\mathrm{B}} =
\int
\mathcal{D}\!
\left[
\bar{\beta}^{\mathrm{a}\dag}_{\ },
\beta^{\mathrm{a}\dag}_{\ },
\bar{\beta}^{\   }_{\mathrm{a}},
\beta^{\   }_{\mathrm{a}}
\right]
\exp\!
\left(
-
\int \mathrm{d}^2\,r
\left(
\mathcal{L}^{\      }_{\mathrm{B}}
+
\mathcal{L}^{{i}\eta}_{\mathrm{B}}
\right)
\right),
\end{split}
\end{equation}
with the effective action for the fermionic part given by
\begin{equation}
\begin{split}
&
\mathcal{L}^{\ }_{\mathrm{F}}=
\frac{1}{2\pi}
\sum_{\mathrm{a}=1}^2 
\Big\{
\hphantom{+}\!
\bar{\psi}^{\mathrm{a}\dag}
\left[
2\partial
- 
{i}
(-1)^{\mathrm{a}}\,
a  
+ 
a^{\prime}
\right]
\bar{\psi}^{\ }_{\mathrm{a}}
\\
&
\hphantom{\mathcal{L}^{\ }_F=\frac{1}{2\pi}\sum_{a=1}^2}
+
\psi^{\mathrm{a}\dag}
\left[
2\bar{\partial}
- 
{i} 
(-1)^{\mathrm{a}} 
\bar{a}
+ 
\bar{a}^{\prime}
\right]
\psi^{\ }_{\mathrm{a}}
\\
&
\hphantom{\mathcal{L}^{\ }_F=\frac{1}{2\pi}\sum_{a=1}^2}
+
\left[
m
+
(-1)^{\mathrm{a}+1}
{i}
m'
\right]
\bar{\psi}^{\mathrm{a}\dag} 
\psi^{\ }_{\mathrm{a}}
\\
&
\hphantom{\mathcal{L}^{\ }_F=\frac{1}{2\pi}\sum_{a=1}^2}
+
\left[
\bar{m}
+
(-1)^{\mathrm{a}+1}
{i}
\bar{m}^{\prime}
\right]
\psi^{\mathrm{a}\dag}
\bar{\psi}^{\ }_{\mathrm{a}}
\Big\}
\end{split}
\label{eq: mathcal L f no eta}
\end{equation}
and
\begin{equation}
\mathcal{L}^{{i}\eta}_{\mathrm{F}}=
\frac{{i}\eta}{2\pi}
\left(
\bar\psi^{1\dag}
\psi^{2\dag}
+
\psi^{1\dag}
\bar\psi^{2\dag}
-
\bar\psi^{\ }_{2}
\psi^{\ }_{1}
-
\psi^{\ }_{2}
\bar\psi^{\ }_{1}
\right),
\label{eq: mathcal L f with eta}
\end{equation}
and the bosonic part of the effective action given by
\begin{equation}
\begin{split}
&
\mathcal{L}^{\ }_{\mathrm{B}}=
\frac{1}{2\pi}
\sum_{\mathrm{a}=1}^2 
\Big\{
\hphantom{+}\!
\bar{\beta}^{\mathrm{a}\dag}
\left[
2\partial
- 
{i}
(-1)^{\mathrm{a}}\,
a  
+ 
a^{\prime}
\right]
\bar{\beta}^{\ }_{\mathrm{a}}
\\
&
\hphantom{\mathcal{L}^{\ }_F=\frac{1}{2\pi}\sum_{a=1}^2}
+
\beta^{\mathrm{a}\dag}
\left[
2\bar{\partial}
-
{i}
(-1)^{\mathrm{a}} 
\bar{a}
+ 
\bar{a}^{\prime}
\right]
\beta^{\ }_{\mathrm{a}}
\\
&
\hphantom{\mathcal{L}^{\ }_F=\frac{1}{2\pi}\sum_{a=1}^2}
+
\left[
m
+
(-1)^{\mathrm{a}+1}
{i}
m'
\right]
\bar{\beta}^{\mathrm{a}\dag} 
\beta^{\ }_{\mathrm{a}}
\\
&
\hphantom{\mathcal{L}^{\ }_F=\frac{1}{2\pi}\sum_{a=1}^2}
+
\left[
\bar{m}
+
(-1)^{\mathrm{a}+1}
{i}
\bar{m}^{\prime}
\right]
\beta^{\mathrm{a}\dag}
\bar{\beta}^{\ }_{\mathrm{a}}
\Big\}
\end{split}
\end{equation}
and
\begin{equation}
\mathcal{L}^{{i}\eta}_{\mathrm{B}}=
\frac{  
{i}\eta 
     }
     {
2\pi
     }
\left(
-
\bar\beta^{1\dag}
\beta^{2\dag}
-
\beta^{1\dag}
\bar\beta^{2\dag}
-
\bar\beta^{\ }_{2}
\beta^{\ }_{1}
-
\beta^{\ }_{2}
\bar\beta^{\ }_{1}
\right),
\end{equation}
\end{subequations}
where $2\partial=\partial_1-i\partial_2$
and $2\bar\partial=\partial_1+i\partial_2$.
The asymmetry between fermions and bosons in 
$\mathcal{L}^{{i}\eta}_{\mathrm{F}}$
and
$\mathcal{L}^{{i}\eta}_{\mathrm{B}}$,
a consequence of the asymmetry between
the $\psi$'s and $\beta$'s
on the right-hand side of
Eq.~(\ref{eq: from chi and xi to psi and beta}),
is the price to be paid in order to make a 
$\mathrm{GL}(2|2)$ 
supersymmetry of
$\mathcal{L}^{\ }_{\mathrm{F}}
+\mathcal{L}^{\ }_{\mathrm{B}}$
explicit,
as is shown in 
Refs.~\onlinecite{Guruswamy00}
and \onlinecite{Mudry03}.%
~\cite{Schnyder09}

The $\mathrm{N}$-th moment of the retarded single-particle Green's 
function evaluated at the band center is obtained by allowing the index 
$\mathrm{a}$ to run from 1 to $2\mathrm{N}$ 
in Eq.~(\ref{eq: final partition fct HWK}).

\subsection{
Phase diagram
           }
\label{subsec: Phase diagram}

We now assume that the disorder potentials 
are white-noise correlated following 
the Gaussian laws
with vanishing mean and nonvanishing variances
\begin{subequations}
\label{eq: def gv gw}
\begin{equation}
\begin{split}
&
\overline{
w(\boldsymbol{r})
         }=0,
\qquad
\overline{
w(\boldsymbol{r})
w(\boldsymbol{r}')}=
g^{\ }_{w} 
\delta^{(2)}(\boldsymbol{r}-\boldsymbol{r}').
\end{split}
\end{equation}
Here,
$\delta^{(2)}(\boldsymbol{r}-\boldsymbol{r}')$ 
is the two-dimensional delta function,
$\overline{(\cdots)}$ represents disorder averaging, 
\begin{equation}
\begin{split}
w\in W:=& \{
\mathrm{Re}\,a, \,
\mathrm{Im}\,a, \,
\mathrm{Re}\,a', \,
\mathrm{Im}\,a',
\\
&\;
\mathrm{Re}\,m, \,
\mathrm{Im}\,m, \,
\mathrm{Re}\,m',\,
\mathrm{Im}\,m'\},
\end{split}
\label{eq: def w labels}
\end{equation}
\end{subequations}
and 
the disorder strengths $g^{\ }_w$ \textit{are all positive}.
We shall treat symmetry class BDI defined by
\begin{equation}
g^{\ }_{\mathrm{Re}\, a}=g^{\ }_{\mathrm{Im}\, a}=
g^{\ }_{\mathrm{Re}\, m'}=g^{\ }_{\mathrm{Im}\, m'}=0
\end{equation}
and symmetry class CII defined by
\begin{equation}
g^{\ }_{\mathrm{Re}\, a}=g^{\ }_{\mathrm{Im}\, a}=
g^{\ }_{\mathrm{Re}\, m}=g^{\ }_{\mathrm{Im}\, m}=0.
\end{equation}
Their boundaries 
\begin{equation}
\begin{split}
0=&\,
g^{\ }_{\mathrm{Re}\, a}=g^{\ }_{\mathrm{Im}\, a}=
g^{\ }_{\mathrm{Re}\, a'}=g^{\ }_{\mathrm{Im}\, a'}
\\
=&\,
g^{\ }_{\mathrm{Re}\, m'}=g^{\ }_{\mathrm{Im}\, m'}=
g^{\ }_{\mathrm{Re}\, m}
\end{split}
\label{eq: strong conditions class D bis bis}
\end{equation}
and
\begin{equation}
\begin{split}
0=&\,
g^{\ }_{\mathrm{Re}\, a}=g^{\ }_{\mathrm{Im}\, a}=
g^{\ }_{\mathrm{Re}\, a'}=g^{\ }_{\mathrm{Im}\, a'}
\\
=&\,
g^{\ }_{\mathrm{Re}\, m}=g^{\ }_{\mathrm{Im}\, m}=
g^{\ }_{\mathrm{Im}\, m'}
\end{split}
\label{eq: strong conditions class AII bis bis}
\end{equation}
to symmetry class AIII
are in symmetry class D
and in symmetry class AII,
respectively.
All four symmetry conditions are summarized in
Table~\ref{table III}. The defining conditions 
on classes D and AII can be made
slightly more general than in Eqs.%
~(\ref{eq: strong conditions class D bis bis})
and%
~(\ref{eq: strong conditions class AII bis bis}) 
as will become clear shortly.

\begin{table}
\caption{
\label{table III}
Symmetry conditions on the (positive) variances of
the static random fields from Table~\ref{table II}.
For symmetry class D, 
$g^{\ }_{\mathrm{Re}\, m}g^{\ }_{\mathrm{Im}\, m}=0$.
For symmetry class AII, 
$g^{\ }_{\mathrm{Re}\, m'}g^{\ }_{\mathrm{Im}\, m'}=0$.
        }
\begin{ruledtabular}
\begin{tabular}{ccccc}
AIII
&
BDI
&
CII
&
D
&
AII
\\
\hline
$
g^{\ }_{\mathrm{Re}\,a}
$
&
$
0
$
&
$
0
$
&
$
0
$
&
$
0
$
\\
$
g^{\ }_{\mathrm{Im}\,a}
$
&
$
0
$
&
$
0
$
&
$
0
$
&
$
0
$
\\
$
g^{\ }_{\mathrm{Re}\,a^{\prime}}
$
&
$
g^{\ }_{\mathrm{Re}\,a^{\prime}}
$
&
$
g^{\ }_{\mathrm{Re}\,a^{\prime}}
$
&
$
0
$
&
$
0
$
\\
$
g^{\ }_{\mathrm{Im}\,a^{\prime}}
$
&
$
g^{\ }_{\mathrm{Im}\,a^{\prime}}
$
&
$
g^{\ }_{\mathrm{Im}\,a^{\prime}}
$
&
$
0
$
&
$
0
$
\\
$
g^{\ }_{\mathrm{Re}\,m}
$
&
$
g^{\ }_{\mathrm{Re}\,m}
$
&
$
0
$
&
$
g^{\ }_{\mathrm{Re}\,m}
$
&
$
0
$
\\
$
g^{\ }_{\mathrm{Im}\,m}
$
&
$
g^{\ }_{\mathrm{Im}\,m}
$
&
$
0
$
&
$
g^{\ }_{\mathrm{Im}\,m}
$
&
$
0
$
\\
$
g^{\ }_{\mathrm{Re}\,m'}
$
&
$
0
$
&
$
g^{\ }_{\mathrm{Re}\,m'}
$
&
$
0
$
&
$
g^{\ }_{\mathrm{Re}\,m'}
$
\\
$
g^{\ }_{\mathrm{Im}\,m'}
$
&
$
0
$
&
$
g^{\ }_{\mathrm{Im}\,m'}
$
&
$
0
$
&
$
g^{\ }_{\mathrm{Im}\,m'}
$
\end{tabular}
\end{ruledtabular}
\end{table}

The phase diagram for the random Dirac fermions
defined by Eqs.~(\ref{eq: chiral random dirac hamiltonian}),
(\ref{eq: def HWK BDI}),
(\ref{eq: def HWK CII}),
(\ref{eq: def a a' m m'}),
and
(\ref{eq: def gv gw})
belongs to the 8-dimensional parameter space
\begin{equation}
\Omega^{\ }_{\mathrm{AIII}}:=
\left\{
g^{\ }_{w}\in\mathbb{R}
\, | \,
0\leq
g^{\ }_{w}<
\infty,
\;
w\in W
\right\}
\end{equation}
with the origin representing the clean limit.
Imposing on $\Omega^{\ }_{\mathrm{AIII}}$
the constraints summarized in 
Table~\ref{table III}
yields the 4-dimensional subspaces
\begin{equation}
\Omega^{\ }_{\mathrm{BDI}}
\subset\Omega^{\ }_{\mathrm{AIII}},
\qquad
\Omega^{\ }_{\mathrm{CII}}
\subset\Omega^{\ }_{\mathrm{AIII}},
\end{equation}
and the one-dimensional subspaces
\begin{equation}
\Omega^{\ }_{\mathrm{D}}
\subset\Omega^{\ }_{\mathrm{AIII}},
\qquad
\Omega^{\ }_{\mathrm{AII}}
\subset\Omega^{\ }_{\mathrm{AIII}}.
\end{equation}

We are going to analyze the phase diagram and the
projected RG flows of its
couplings through two-dimensional cuts in
$\Omega^{\ }_{\mathrm{AIII}}$
which we will depict with Fig.%
~\ref{fig: phase diagram}.
All those cuts belong to the 6-dimensional subspace
\begin{equation}
\Omega^{\perp}:=
\{
g^{\ }_{w}
\in
\Omega^{\ }_{\mathrm{AIII}}
|\,
0=
g^{\ }_{\mathrm{Re}\,a}=
g^{\ }_{\mathrm{Im}\,a}
\}.
\end{equation}
The cuts will involve a plane with the variance of
the gauge potential $a'$ set to either 
zero in Fig.%
~\ref{fig: phase diagram}(a)
or a nonvanishing value in Figs.%
~\ref{fig: phase diagram}(b)
and~\ref{fig: phase diagram}(c).
We shall also represent the effect of the RG flow to strong
coupling of the variance of $a'$ 
on the coupling 
$g^{\prime}_{-}:=
g^{\ }_{\mathrm{Im}\,m'}
-
g^{\ }_{\mathrm{Re}\,m'}
$
in Fig.%
~\ref{fig: phase diagram}(d).

To this end, we observe that
the quadrant
\begin{equation}
g^{\ }_{\mathrm{Re}\,m}>0,
\qquad
g^{\ }_{\mathrm{Im}\,m}>0,
\end{equation}
belongs to symmetry class BDI
in Fig.~\ref{fig: phase diagram}(a).
The  quadrant 
\begin{equation}
g^{\ }_{\mathrm{Re}\,m}<0,
\qquad
g^{\ }_{\mathrm{Im}\,m}<0,
\label{eq: def BDI quadrant bis}
\end{equation}
in Fig.~\ref{fig: phase diagram}(a)
belongs to symmetry class CII as we now demonstrate.
This is expected from the fact that 
$m'_{0,3}$ present in the CII model is the imaginary counterpart
of $m_{0,3}$ present in the BDI model.

We begin with the Lagrangian~(\ref{eq: mathcal L f no eta})
on which we perform the transformation
\begin{equation}
\begin{split}
&
\bar{\psi}^{2\dag} \to -\bar{\psi}^{2\dag},
\qquad
\bar{\psi}^{\ }_{2} \to -\bar{\psi}^{\ }_{2}.
\end{split}
\label{eq: trsf 1}
\end{equation}
Under this transformation
\begin{equation}
\begin{split}
&
\sum_{\mathrm{a}=1}^{2}
(-1)^{\mathrm{a}+1}
\bar{\psi}^{\mathrm{a}\dag}
\psi^{\ }_{\mathrm{a}}
\to
\sum_{\mathrm{a}=1}^{2}
\bar{\psi}^{\mathrm{a}\dag}
\psi^{\ }_{\mathrm{a}},
\\
&
\sum_{\mathrm{a}=1}^{2}
(-1)^{\mathrm{a}+1}
\psi^{\mathrm{a}\dag}
\bar{\psi}^{\ }_{\mathrm{a}}
\to
\sum_{\mathrm{a}=1}^{2}
\psi^{\mathrm{a}\dag}
\bar{\psi}^{\ }_{\mathrm{a}},
\\
&
\sum_{\mathrm{a}=1}^{2}
\psi^{\mathrm{a}\dag}
\bar{\psi}^{\ }_{\mathrm{a}}
\to
\sum_{\mathrm{a}=1}^{2}
(-1)^{\mathrm{a}+1}
\psi^{\mathrm{a}\dag}
\bar{\psi}^{\ }_{\mathrm{a}},
\\
&
\sum_{\mathrm{a}=1}^{2}
\bar{\psi}^{\mathrm{a}\dag}
\psi^{\ }_{\mathrm{a}}
\to
\sum_{\mathrm{a}=1}^{2}
(-1)^{\mathrm{a}+1}
\bar{\psi}^{\mathrm{a}\dag}
\psi^{\ }_{\mathrm{a}},
\end{split}
\end{equation}
while all other terms 
in Lagrangian~(\ref{eq: mathcal L f no eta})
remain unchanged.
We conclude that Lagrangian~(\ref{eq: mathcal L f no eta})
remains unchanged by combining
transformation~(\ref{eq: trsf 1})
with the transformation
\begin{equation}
\mathrm{Re}\,m\,
\longleftrightarrow
{i}\,\mathrm{Re}\,m' ,
\qquad
\mathrm{Im}\,m\,
\longleftrightarrow
{i}\,\mathrm{Im}\,m'.
\label{eq: trsf 2}
\end{equation}
As the same argument carries through in the bosonic sector
by combining transformation~(\ref{eq: trsf 2}) with
\begin{equation}
\begin{split}
&
\bar{\beta}^{2\dag} \to -\bar{\beta}^{2\dag},
\qquad
\bar{\beta}^{\ }_{2} \to -\bar{\beta}^{\ }_{2},
\end{split}
\label{eq: trsf 3}
\end{equation}
we conclude that a disorder realization in 
symmetry class CII is obtained from the analytical continuation%
~(\ref{eq: trsf 2})
of a disorder realization in symmetry class BDI
when $\eta=0$
[$\mathcal{L}^{i\eta}_\mathrm{F}$ and
$\mathcal{L}^{i\eta}_\mathrm{B}$
are not invariant under the transformations%
~(\ref{eq: trsf 1}), (\ref{eq: trsf 2}), and (\ref{eq: trsf 3})].
Upon disorder averaging,
the analytical continuation%
~(\ref{eq: trsf 2})
amounts to mapping the CII quadrant
\begin{equation}
g^{\ }_{\mathrm{Re}\,m'}>0,
\qquad
g^{\ }_{\mathrm{Im}\,m'}>0
\label{eq: def CII quadrant}
\end{equation}
one-to-one into the quadrant%
~(\ref{eq: def BDI quadrant bis}) 
through the mapping
\begin{eqnarray}
g^{\ }_{\mathrm{Re}\,m'}
\to
-g^{\ }_{\mathrm{Re}\,m},
\quad
g^{\ }_{\mathrm{Im}\,m'}
\to
-g^{\ }_{\mathrm{Im}\,m},
\end{eqnarray}
that relates the positive variances 
$g^{\ }_{\mathrm{Re}\,m'}$
and
$g^{\ }_{\mathrm{Im}\,m'}$
in symmetry class CII
to the negative variances 
$g^{\ }_{\mathrm{Re}\,m}$
and
$g^{\ }_{\mathrm{Im}\,m}$.
The remaining quadrants in 
Fig.~\ref{fig: phase diagram}(a) 
\begin{equation}
0<g^{\ }_{\mathrm{Re}\,m},
\qquad
0>g^{\ }_{\mathrm{Im}\,m}=-g^{\ }_{\mathrm{Im}\,m'}
\label{eq: def AIII quadrant 1}
\end{equation}
and
\begin{equation}
0>g^{\ }_{\mathrm{Re}\,m}=-g^{\ }_{\mathrm{Re}\,m'},
\qquad
0<g^{\ }_{\mathrm{Im}\,m}
\label{eq: def AIII quadrant 2}
\end{equation}
belong to symmetry class AIII
as their corresponding disorder potential
$\rho^{\ }_{2}\otimes
(\sigma^{\ }_{0}m^{\ }_{0}+\sigma^{\ }_{3}m^{\prime}_{3})$
and
$\rho^{\ }_{1}\otimes
(\sigma^{\ }_{0}m^{\prime}_{0}-\sigma^{\ }_{3}m^{\ }_{3})$
are not invariant under
neither the time-reversal operation
$\mathcal{T}$ nor the time-reversal operation $\mathcal{T}'$.

The one-dimensional boundary
\begin{equation}
0=g^{\ }_{\mathrm{Re}\,m},
\qquad
0<g^{\ }_{\mathrm{Im}\,m}
\label{eq: def D boundary}
\end{equation}
of the BDI quadrant, 
\begin{equation}
0<g^{\ }_{\mathrm{Re}\,m},
\qquad
0<g^{\ }_{\mathrm{Im}\,m},
\label{eq: def BDI quadrant}
\end{equation}
belongs to symmetry class D according to 
Eq.~(\ref{eq: strong conditions class D bis bis}).
The one-dimensional boundary
\begin{equation}
0<g^{\ }_{\mathrm{Re}\,m'},
\qquad
0=g^{\ }_{\mathrm{Im}\,m'}
\label{eq: def AII boundary}
\end{equation}
of the CII quadrant~(\ref{eq: def CII quadrant})
belongs to symmetry class AII according to 
Eq.~(\ref{eq: strong conditions class AII bis bis}).
The one-dimensional boundaries
\begin{equation}
0<g^{\ }_{\mathrm{Re}\,m},
\qquad
0=g^{\ }_{\mathrm{Im}\,m}
\label{eq: def D boundary bis}
\end{equation}
and
\begin{equation}
0=g^{\ }_{\mathrm{Re}\,m'},
\qquad
0<g^{\ }_{\mathrm{Im}\,m'}
\label{eq: def AII boundary bis}
\end{equation}
also belong to symmetry classes
D and AII, respectively, as follows
from the mirror symmetry about the line
\begin{equation}
\mathbb{R}\ni g^{\ }_{\mathrm{M}}\equiv
g^{\ }_{\mathrm{Re}\,m}=
g^{\ }_{\mathrm{Im}\,m}.
\label{eq: def mirror symmetry line}
\end{equation}
To derive this mirror symmetry, one
observes, when $\eta=0$, 
the invariance of the generating function%
~(\ref{eq: final partition fct HWK})
under the combined
transformations
($\mathrm{a}=1,2$)
\begin{equation}
\begin{split}
&
\bar\psi^{\mathrm{a}\dag} 
\to 
\bar\psi^{\mathrm{a}\dag},
\qquad
\bar\psi^{\ }_{\mathrm{a}} 
\to 
\bar\psi^{\ }_{\mathrm{a}},
\\
&
\psi^{\mathrm{a}\dag} 
\to 
-{i}\psi^{\mathrm{a}\dag},
\qquad
\psi^{\ }_{\mathrm{a}} 
\to 
+{i}\psi^{\ }_{\mathrm{a}},
\\
&
\bar\beta^{\mathrm{a}\dag} 
\to 
\bar\beta^{\mathrm{a}\dag},
\qquad
\bar\beta^{\ }_{\mathrm{a}}
\to 
\bar\beta^{\ }_{\mathrm{a}},
\\
&
\beta^{\mathrm{a}\dag} 
\to 
-{i}\beta^{\mathrm{a}\dag},
\qquad
\beta^{\ }_{\mathrm{a}} 
\to 
+{i}\beta^{\ }_{\mathrm{a}},
\\
&
\mathrm{Re}\,m\to
\mathrm{Im}\,m,
\qquad
\mathrm{Im}\,m\to
-
\mathrm{Re}\,m,
\\
&
\mathrm{Re}\,m'\to
\mathrm{Im}\,m',
\qquad
\mathrm{Im}\,m'\to
-
\mathrm{Re}\,m'.
\end{split}
\label{eq: exchange Re and Im}
\end{equation}
However, the signs of the random fields
$\mathrm{Re}\,m$,
$\mathrm{Im}\,m$,
$\mathrm{Re}\,m'$,
and
$\mathrm{Im}\,m'$
are innocuous after disorder averaging, 
for these fields are Gaussian
distributed with a vanishing mean
according to Eq.~(\ref{eq: def gv gw}). Hence,
a mirror symmetry along the vertical axis in
Fig.~\ref{fig: phase diagram}(a)
must hold.

The RG flows along the boundaries
D and AII are known and shown in 
Fig.~\ref{fig: phase diagram}(a).
In symmetry class D, the RG flow is to the clean Dirac limit
(see Refs.~\onlinecite{Dotsenko83}, 
\onlinecite{Ludwig87}, \onlinecite{Shankar87},
\onlinecite{Senthil00} and \onlinecite{Ryu06a}),
while the RG flow is to the metallic fixed point in 
symmetry class AII
(see Refs.~\onlinecite{Ryu07b}, \onlinecite{Bardarson07}, 
and \onlinecite{Nomura07}).%
\cite{footnote on AII, footnote on AII B}
The random vector potentials
$a^{\ }_{1}-{i}a^{\prime}_{1}$ and $a^{\ }_{2}-{i}a^{\prime}_{2}$.
are not generated under the RG on the boundaries D and AII.

The RG flows away from the boundaries D shown in 
Fig.~\ref{fig: phase diagram}(a) 
are consistent with the fact that
the line~(\ref{eq: def mirror symmetry line}) 
is a stable line of 
nearly-critical points
in the BDI quadrant.
As we show below, they also follow from a one-loop stability analysis
summarized in Fig.~\ref{fig: phase diagram}(b).
The RG flows away from the boundaries AII shown in 
Fig.~\ref{fig: phase diagram}(a)
are a more subtle matter. They are drawn to be consistent with
the fact that
the nearly-critical line~(\ref{eq: def mirror symmetry line}) 
appears to be unstable in the CII quadrant of 
Fig.~\ref{fig: phase diagram}(a)
when the approximation $g^{\ }_{a'}\approx0$ is used.
However, as we show below, relaxing this approximation
and allowing the RG flow to reach 
length scales such that $g^{\ }_{a'}$ becomes sufficiently large
changes the flow depicted in
Fig.~\ref{fig: phase diagram}(a)
to that depicted in
Fig.~\ref{fig: phase diagram}(c).
This change is a consequence of the flow
depicted in
Fig.~\ref{fig: phase diagram}(d).

\subsection{
The plane
$
\mathbb{R}\ni
g^{\ }_{\mathrm{M}}\equiv
g^{\ }_{\mathrm{Re}\,m}=g^{\ }_{\mathrm{Im}\,m}
$
and
$g^{\ }_{a'}\geq0$
           }
\label{subsec: nearly-critical line}

Consider the line~(\ref{eq: def mirror symmetry line}) 
in Fig.~\ref{fig: phase diagram}(a).
By combining the results of 
Refs.~\onlinecite{Guruswamy00}
and 
\onlinecite{Mudry03} 
with the results of Sec.~\ref{subsec: Phase diagram}, 
we are going to show that this line is a line of 
nearly-critical points.
To this end, we shall assume that rotation symmetry is
preserved at the statistical level. This means that
we can assume 
\begin{equation}
g^{\ }_{\mathrm{Re}\,a}=
g^{\ }_{\mathrm{Im}\,a}\equiv
g^{\ }_{a},
\qquad
g^{\ }_{\mathrm{Re}\,a'}=
g^{\ }_{\mathrm{Im}\,a'}\equiv
g^{\ }_{a'}.
\end{equation}

\subsubsection{
The plane
$
g^{\ }_{\mathrm{M}}\equiv
g^{\ }_{\mathrm{Re}\,m}=
g^{\ }_{\mathrm{Im}\,m}
\geq0
$
and
$g^{\ }_{a'}\geq0$
              }
\label{subsec: nearly-critical line case BDI}

We begin with the plane
\begin{equation}
0<
g^{\ }_{\mathrm{M}}\equiv
g^{\ }_{\mathrm{Re}\,m}=g^{\ }_{\mathrm{Im}\,m},
\qquad
0\leq g^{\ }_{a'},
\label{eq: half-line BDI}
\end{equation}
in Fig.~\ref{fig: phase diagram}
along which the generating function for the average
retarded Green's function, which is nothing but the
$\widehat{\mathrm{gl}}(2|2)^{\ }_{k=1}$ Thirring model
studied in Refs.~\onlinecite{Guruswamy00} and \onlinecite{Mudry03}.
Indeed, by setting $\eta=0$ in Eq.~(\ref{eq: final partition fct HWK})
and integrating over the random potentials,
one finds the partition function
\begin{subequations}
\label{eq: gl22 perturbed by two current-current int BDI}
\begin{equation}
\begin{split}
&
Z^{\ }_{\widehat{\mathrm{gl}}(2|2)^{\ }_{1}}=
\int\mathcal{D}[\psi^{\dag},\psi,
\bar{\psi}^\dag,\bar{\psi}]
\exp
\left(
-S^{\ }_{\widehat{\mathrm{gl}}(2|2)^{\ }_{1}}
\right),
\\
&
S^{\ }_{\widehat{\mathrm{gl}}(2|2)^{\ }_{1}}=
S^{\ }_{0}
+
\int \frac{\mathrm{d}\,\bar{z}\,\mathrm{d}\,z}{2\pi{i}}
\left(
\frac{g^{\ }_{a'}}{2\pi} 
\mathcal{O}^{\ }_{a'}
+
\frac{g^{\ }_{\mathrm{M}}}{2\pi}
\mathcal{O}^{\ }_{\mathrm{M}}
\right),
\\
&
\mathcal{O}^{\ }_{a'}=
- 
     J^{\,\mathrm{A}}_{\mathrm{A}}\,
(-1)^\mathrm{A}\, 
\bar J^{\,\mathrm{B}}_{\mathrm{B}}\,
(-1)^\mathrm{B},
\\
& 
\mathcal{O}^{\ }_{\mathrm{M}}=
-
     J^{\,\mathrm{B}}_{\mathrm{A}}
\bar J^{\,\mathrm{A}}_{\mathrm{B}}
(-1)^\mathrm{A}.
\end{split}
\label{eq: gl22 perturbed by two current-current int BDI a}
\end{equation}
The action
\begin{equation}
S^{\ }_{0}:=
\int \frac{\mathrm{d}\,\bar{z}\,\mathrm{d}\,z}{4\pi{i}}
\left(
\bar\psi^{\mathrm{A}\dag}\,
2\partial\,
\bar\psi^{\ }_{\mathrm{A}}
+
\psi^{\mathrm{A}\dag}\,
2\bar\partial\,
\psi^{\ }_{\mathrm{A}}
\right)
\label{eq: free action}
\end{equation}
($\bar{z}\equiv r^{\ }_{1}-{i}r^{\ }_{2}$, 
    $z\equiv r^{\ }_{1}+{i}r^{\ }_{2}$)
is the action 
in Eq.~(\ref{eq: final partition fct HWK}) 
without disorder when $\eta=0$. 
The capitalized index $\mathrm{A}=1,\cdots,4$
carries a grade which is either 0 for $\mathrm{A}=1,2$
or 1 for $\mathrm{A}=3,4$.
It is the grade of the indices $\mathrm{A}$ and $\mathrm{B}$ 
that enters  expressions such as 
$(-)^{\mathrm{A}}$ or $(-)^{\mathrm{A}\mathrm{B}}$.
The grade $0$ ($1$) thus corresponds to the bosons (fermions).%
\cite{Ryu06b}
We are using the summation convention over repeated
indices $\mathrm{A},\mathrm{B}=1,\cdots,4$.
We also have defined the supercurrents
\begin{equation}
J^{\,\mathrm{B}}_{\mathrm{A}}:=
\psi^{\ }_{\mathrm{A}}
\psi^{\mathrm{B}\dag},
\qquad
\bar J^{\,\mathrm{B}}_{\mathrm{A}}:=
\bar \psi^{\ }_{\mathrm{A}}
\bar \psi^{\mathrm{B}\dag}
\label{eq: def GL(2|2) currents as free spinors}
\end{equation}
\end{subequations}
where $\mathrm{A},\mathrm{B}=1,\cdots,4$
and 
$\psi^{\ }_{\mathrm{A}}$,
$\bar{\psi}^{\ }_{\mathrm{A}}$,
$\psi^{\mathrm{A}\dag}$,
and
$\bar{\psi}^{\mathrm{A}\dag}$
now denote
bosons for $\mathrm{A}=1,2$ and fermions for $\mathrm{A}=3,4$.
(By allowing the graded indices 
$\mathrm{A}$ and $\mathrm{B}$ to run from 1 to 
$4\mathrm{N}$, we can compute the $\mathrm{N}$-th
moment of the retarded single-particle Green's function.)

Observe that the integration measure in Eq.%
~(\ref{eq: gl22 perturbed by two current-current int BDI a})
and the free action~(\ref{eq: free action}) 
are both invariant under the \textit{local}
chiral $\mathrm{GL}(2|2)\times\mathrm{GL}(2|2)$
transformation
\begin{subequations}
\label{eq: def GL(2|2) times GL(2|2) trsf}
\begin{equation}
\begin{split}
\bar\psi^{\mathrm{A}\dag}\to
\bar\psi^{\mathrm{B}\dag} 
L^{-1\,\mathrm{A}}_{\ \ \mathrm{B}},
\qquad
\bar\psi^{\ }_{\mathrm{A}}\to
L^{\ \mathrm{B}}_{\mathrm{A}}
\bar\psi^{\ }_{\mathrm{B}}
\end{split}
\label{eq: def GL(2|2) times GL(2|2) trsf a}
\end{equation}
and
\begin{equation}
\begin{split}
\psi^{\mathrm{A}\dag}\to
\psi^{\mathrm{B}\dag} 
R^{-1\mathrm{A}}_{\ \ \mathrm{B}},
\qquad
\psi^{\ }_{\mathrm{A}}\to
R^{\ \mathrm{B}}_{\mathrm{A}}
\psi^{\ }_{\mathrm{B}}
\end{split}
\label{eq: def GL(2|2) times GL(2|2) trsf b}
\end{equation}
for any anti-holomorphic  $L(\bar{z})$ and holomorphic $R(z)$ 
in the fundamental representation of
$\mathrm{GL}(2|2)$.
The transformation law of the currents
under~(\ref{eq: def GL(2|2) times GL(2|2) trsf a})
and (\ref{eq: def GL(2|2) times GL(2|2) trsf b})
is
\begin{equation}
\begin{split}
J^{\ \mathrm{B}}_{\mathrm{A}}\to
R^{\ \mathrm{C}}_{\mathrm{A}}
J^{\ \mathrm{D}}_{\mathrm{C}}
R^{-1\mathrm{B}}_{\ \ \mathrm{D}},
\qquad
\bar J^{\ \mathrm{B}}_{\mathrm{A}}\to
L^{\ \mathrm{C}}_{\mathrm{A}}
\bar J^{\ \mathrm{D}}_{\mathrm{C}}
L^{-1\mathrm{B}}_{\ \ \mathrm{D}}.
\end{split}
\label{eq: def GL(2|2) times GL(2|2) trsf c}
\end{equation}
Hence, the Thirring model%
~(\ref{eq: gl22 perturbed by two current-current int BDI})
is invariant under the global diagonal subgroup 
of the global transformation%
~(\ref{eq: def GL(2|2) times GL(2|2) trsf a})
and~(\ref{eq: def GL(2|2) times GL(2|2) trsf b})
defined by choosing 
\begin{equation}
R=L
\label{eq: def diagonal GL(2|2) trsf}
\end{equation}
\end{subequations}
in Eqs.~(\ref{eq: def GL(2|2) times GL(2|2) trsf a})
and~(\ref{eq: def GL(2|2) times GL(2|2) trsf b})
to be independent of space.
It can be shown that the $\eta$ term
responsible for the convergence of the integrals in the bosonic sector 
that has been neglected so far
breaks this symmetry down to the subsupergroup $\mathrm{OSp}(2|2)$.
In fact, the symmetry-breaking pattern
$\mathrm{GL}(2|2)\to\mathrm{OSp}(2|2)$
occurs due to superfield bilinears acquiring
an expectation value with the consequence of
a diverging density of states (DOS) at the band center.%
~\cite{Guruswamy00,Mudry03}

The (infrared) beta functions for the
couplings $g^{\ }_{a'}$ and $g^{\ }_{\mathrm{M}}$
have been computed non-perturbatively in
Ref.~\onlinecite{Guruswamy00}. They are%
~\cite{footnote-erratum}
\begin{subequations}
\label{eq: bet fcts}
\begin{equation}
\beta^{\ }_{g^{\ }_{ a' }}:=
\frac{\mathrm{d}g^{\ }_{ a'}}{\mathrm{d}l} = 
\frac{1}{\pi}
\left(
\frac{
g^{\ }_{\mathrm{M}}
     }
     {
1
+
g^{\ }_{\mathrm{M}}/\pi
     }
\right)^{2}
\label{eq: bet fcts a}
\end{equation}
and
\begin{equation}
\beta^{\ }_{g^{\ }_{\mathrm{M}}}:= 
\frac{\mathrm{d}g^{\ }_{\mathrm{M}}}{\mathrm{d}l}= 
0.
\label{eq: bet fcts b}
\end{equation}
\end{subequations}
Observe that the coupling constant
$0<g^{\ }_{\mathrm{M}}$ does not flow 
(we emphasize that this is a non-perturbative result)
while the coupling constant $g^{\ }_{a'}$
flows to strong coupling even when it is initially zero.
This is what is meant with the statement
that the plane 
defined by Eq.~(\ref{eq: half-line BDI})
(and its projection onto a half-line)
is nearly-critical:
it is critical (in spite of the flow of the coupling
$g^{\ }_{a'}$) 
for all correlation functions of fields 
that are unaffected by the flow of $g^{\ }_{a'}$.
The half-line~(\ref{eq: half-line BDI})
in Fig.~\ref{fig: phase diagram}(a)
belongs to the 
2-dimensional symmetry class BDI
in the ten-fold   classification of Anderson localization
(see Refs.\ \onlinecite{Zirnbauer96,Altland97,Heinzner05}
and Appendix%
~\ref{subsec: Patterns of symmetry breaking and supermanifolds}).

\subsubsection{
The plane
$
g^{\ }_{\mathrm{M}'}\equiv
g^{\ }_{\mathrm{Re}\,m'}=
g^{\ }_{\mathrm{Im}\,m'}
\geq0
$
and
$g^{\ }_{a'}\geq0$
              }
\label{subsec: nearly-critical line case CII}

We continue with the plane
\begin{equation}
\begin{split}
g^{\ }_{\mathrm{M}'}\equiv
g^{\ }_{\mathrm{Re}\,m'}=
g^{\ }_{\mathrm{Im}\,m'}
\geq0,
\qquad
g^{\ }_{a'}\geq0,
\end{split}
\end{equation}
in Fig.~\ref{fig: phase diagram}.
The half line 
obtained from the projection to $g^{\ }_{a'}=0$
of this plane
is also a line of nearly-critical points
that now belongs to the two-dimensional symmetry class CII in 
the ten-fold classification of Anderson localization
(see Refs.\ \onlinecite{Zirnbauer96,Altland97,Heinzner05}). 
Indeed, the counterpart
to Eq.~(\ref{eq: gl22 perturbed by two current-current int BDI})
is
\begin{equation}
\begin{split}
&
Z^{\ }_{\widehat{\mathrm{gl}}(2|2)^{\ }_{1}}=
\int\mathcal{D}[\psi^{\dag},\psi,
\bar{\psi}^\dag,\bar{\psi}]
\exp
\left(
-S^{\ }_{\widehat{\mathrm{gl}}(2|2)^{\ }_{1}}
\right),
\\
&
S^{\ }_{\widehat{\mathrm{gl}}(2|2)^{\ }_{1}}=
S^{\ }_{0}
+
\int \frac{\mathrm{d}\,\bar{z}\,\mathrm{d}\,z}{2\pi{i}}
\left(
\frac{g^{\ }_{a'}}{2\pi} 
\mathcal{O}^{\ }_{a'}
-
\frac{g^{\ }_{\mathrm{M}'}}{2\pi}
\mathcal{O}^{\ }_{\mathrm{M}}
\right),
\\
&
\mathcal{O}^{\ }_{a'}=
- 
     J^{\,\mathrm{A}}_{\mathrm{A}}\,
(-1)^\mathrm{A}\, 
\bar J^{\,\mathrm{B}}_{\mathrm{B}}\,
(-1)^\mathrm{B},
\\
& 
\mathcal{O}^{\ }_{\mathrm{M}}=
-
     J^{\,\mathrm{B}}_{\mathrm{A}}
\bar J^{\,\mathrm{A}}_{\mathrm{B}}
(-1)^\mathrm{A},
\end{split}
\label{eq: gl22 perturbed by two current-current int CII}
\end{equation}
as follows from the analytical continuation
$g^{\ }_{\mathrm{M}}\to-g^{\ }_{\mathrm{M}'}$
of Eq.~(\ref{eq: gl22 perturbed by two current-current int BDI}) 
or by explicit integration over the random potentials
in Eq.~(\ref{eq: final partition fct HWK}) with $\eta=0$,
whereby one must account for the extra imaginary
number multiplying the random mass $m'$
for symmetry class CII relative to the random mass $m$
for symmetry class BDI
in Eq.~(\ref{eq: final partition fct HWK}).
Accordingly, the counterparts of Eq.~(\ref{eq: bet fcts}) 
are
\begin{subequations}
\label{eq: bet fcts CII}
\begin{equation}
\beta^{\ }_{g^{\ }_{ a' }}=
\frac{1}{\pi}
\left(
\frac{
g^{\ }_{\mathrm{M}'}
     }
     {
1
-
g^{\ }_{\mathrm{M}'}/\pi
     }
\right)^{2}
\label{eq: bet fcts CII a}
\end{equation}
and
\begin{equation}
\beta^{\ }_{g^{\ }_{\mathrm{M}'}}= 
0
\label{eq: bet fcts CII b}
\end{equation}
where one must impose the condition
\begin{equation}
0\leq g^{\ }_{\mathrm{M}'}<\pi
\end{equation}
\end{subequations}
to avoid the pole in the beta function for $g^{\ }_{a'}$.

\subsection{
Conjectured RG flows in Fig.~\ref{fig: phase diagram}
           }

We are now going to justify why we have conjectured the
RG flows depicted in Fig.~\ref{fig: phase diagram}.
More precisely, we make the following claims.
\begin{itemize}
\item
The boundaries D and AII in the plane $g^{\ }_{a'}=0$
are RG separatrices.

\item
The plane defined by the dashed line in
Fig.~\ref{fig: phase diagram}(a)
and the out-of-plane $g^{\ }_{a'}$ axis
is a stable nearly-critical plane in that all RG trajectories
from region BDI or CII, 
except the fine-tuned RG flows along the separatrix D and AII,
reach this plane asymptotically in the infrared limit.
\item
The rationale that allows us to deduce 
from one-loop flows
nonperturbative  statements
is that
the anomalous scaling dimension of the operator
that couples to the `asymmetry coupling'
$
g^{\ }_{-}\equiv
g^{\ }_{\mathrm{Re}\,m}-g^{\ }_{\mathrm{Im}\,m}
$
in the quadrant BDI, or  to
$ g^{\prime}_{-}\equiv g^{\ }_{\mathrm{Re}\,m'}-g^{\ }_{\mathrm{Im}\,m'} $
in the quadrant CII,
is known to all orders in $g^{\ }_{a'}$.
(By definition, 
$g^{\ }_{-}=0$ on the dashed line in Fig.~\ref{fig: phase diagram}.)

\end{itemize}

To substantiate these three claims, we treat first
the BDI case and then the CII case.

The stability analysis in region BDI of Fig.~\ref{fig: phase diagram}
is determined by the one-loop RG equations%
~\cite{footnote-stability-line-gM} 
\begin{subequations}
\label{eq: Bernard flows for BDI}
\begin{equation}
\begin{split}
&
\beta^{\ }_{g^{\ }_{a'}}=
\frac{
g^{2 }_{+}
-
g^{2 }_{-}
     }
{4\pi},
\\
&
\beta^{\ }_{g^{\ }_{+}}=
-
\frac{
g^{2 }_{-}
     }
{4\pi},
\\
&
\beta^{\ }_{g^{\ }_{-}}=
-
\frac{
\left(
g^{\ }_{+}
+
g^{\ }_{a'}
\right)
g^{\ }_{-}
      }
{4\pi},
\end{split}
\label{eq: Bernard flows for BDI a}
\end{equation}
where
\begin{equation}
g^{\ }_{\pm}:=
g^{\ }_{\mathrm{Im}\, m}
\pm
g^{\ }_{\mathrm{Re}\, m},
\qquad
g^{\ }_{+}\geq
|g^{\ }_{-}|.
\label{eq: Bernard flows for BDI b}
\end{equation}
\end{subequations}
These one-loop flows
must respect the conditions
$g^{\ }_{a'}\geq0$
and
$g^{\ }_{+}\geq|g^{\ }_{-}|$
and
i$g^{\ }_{\mathrm{Re}\, m'}$-$g^{\ }_{\mathrm{Im}\, m'}$ order to represent the effects of disorder
on the underlying microscopic Dirac Hamiltonian
and are valid in the close vicinity to the clean Dirac point
$g^{\ }_{a'}=g^{\ }_{+}=g^{\ }_{-}=0$
denoted by an empty circle in 
Fig.~\ref{fig: phase diagram}(a).
In the regime $g^{\ }_{a'}=0$ and $g^{\ }_{+},g^{\ }_{-}\ll1$,
the line defined by any one of the two boundaries D 
from Fig.~\ref{fig: phase diagram}(a) 
becomes the separatrix of a Kosterlitz-Thouless flow
\begin{equation}
\begin{split}
&
\beta^{\ }_{g^{\ }_{+}}=
-
\frac{
g^{2}_{-}
     }
{4\pi},
\\
&
\beta^{\ }_{g^{\ }_{-}}=
-
\frac{
g^{\ }_{+}
g^{\ }_{-}
      }
{4\pi}.
\end{split}
\label{eq: BDI RG flows for 2 regimes a}
\end{equation}
In Fig.~\ref{fig: phase diagram}(a),
we plotted the Kosterlitz-Thouless flows%
~(\ref{eq: BDI RG flows for 2 regimes a})
which accurately capture the flows%
~(\ref{eq: Bernard flows for BDI})
when $g^{\ }_{a'}\approx0$. However,
in the region BDI defined by the condition
$g^{\ }_{+}>|g^{\ }_{-}|$,
the variance $g^{\ }_{a'}$ flows to strong coupling
and the RG flows follow three-dimensional trajectories.
We depict them by using a two-dimensional projection in Fig.~\ref{fig: phase diagram}(b).
The perturbative flows in the region BDI
from Eq.~(\ref{eq: Bernard flows for BDI a})
after projection to the
$g^{\ }_{\mathrm{Re}\, m}$-$g^{\ }_{\mathrm{Im}\, m}$ plane
are depicted in
Fig.~\ref{fig: phase diagram}(b)
for
$g^{\ }_{a'} \not = 0$.
These flows depict the instability of the BDI boundaries
$g^{\ }_{\mathrm{Re}\, m}\geq0$, $g^{\ }_{\mathrm{Im}\, m}=0$
and
$g^{\ }_{\mathrm{Re}\, m}=0$, $g^{\ }_{\mathrm{Im}\, m}\geq0$
to an infrared flow towards the nearly-critical plane
$g^{\ }_{-}=0$.%
~\cite{footnote-slope flow away boundaries}
It can be shown 
by adapting nonperturbative results 
from Ref.~\onlinecite{Guruswamy00}
that the beta function for the
 coupling 
$g^{\ }_{-}$
in Eq. (\ref{eq: Bernard flows for CII a})
holds to all orders in $g^{\ }_{a'}$ and to linear order
in $g^{\ }_{-}$.%
~\cite{footnote-nonperturbative relations line gM}
Hence, we conjecture that the infrared flows are from
the BDI boundaries to the nearly-critical plane spanned
by the dashed line and the out-of-plane axis $g^{\ }_{a'}$
in Fig.~\ref{fig: phase diagram}(b) for the entire quadrant BDI.

The stability analysis of the region
CII of Fig.~\ref{fig: phase diagram}
is determined by the one-loop RG equations%
~\cite{footnote-stability-line-gM} 
\begin{subequations}
\label{eq: Bernard flows for CII}
\begin{equation}
\begin{split}
&
\beta^{\ }_{g^{\ }_{a'}}=
\frac{
g^{\prime2}_{+}
-
g^{\prime2}_{-}
     }
{4\pi},
\\
&
\beta^{\ }_{g^{\prime}_{+}}=
+
\frac{
g^{\prime2}_{-}
     }
{4\pi},
\\
&
\beta^{\ }_{g^{\prime}_{-}}=
+
\frac{
\left(
g^{\prime}_{+}
-
g^{\ }_{a'}
\right)
g^{\prime}_{-}
      }
{4\pi},
\end{split}
\label{eq: Bernard flows for CII a}
\end{equation}
where
\begin{equation}
g^{\prime}_{\pm}:=
g^{\ }_{\mathrm{Im}\, m'}
\pm
g^{\ }_{\mathrm{Re}\, m'},
\qquad
g^{\prime}_{+}\geq
|g^{\prime}_{-}|.
\label{eq: Bernard flows for CII b}
\end{equation}
\end{subequations}
These one-loop flows
must respect the conditions
$g^{\ }_{a'}\geq0$
and
$g^{\prime}_{+}\geq|g^{\prime}_{-}|$
in order to represent the effects of disorder
on the underlying microscopic Dirac Hamiltonian
and are valid in the close vicinity to the clean Dirac point
$g^{\ }_{a'}=g^{\prime}_{+}=g^{\prime}_{-}=0$
denoted by an empty circle in 
Fig.~\ref{fig: phase diagram}(a).
In the regime $g^{\ }_{a'}=0$ and $g^{\prime}_{+},g^{\prime}_{-}\ll1$,
the line defined by any one of the two boundaries AII
from Fig.~\ref{fig: phase diagram}(a) 
becomes the separatrix of the
Kosterlitz-Thouless flow
\begin{equation}
\begin{split}
&
\beta^{\ }_{g^{\prime}_{+}}=
+
\frac{
g^{\prime2}_{-}
     }
{4\pi},
\\
&
\beta^{\ }_{g^{\prime}_{-}}=
+
\frac{
g^{\prime}_{+}
g^{\prime}_{-}
      }
{4\pi}.
\end{split}
\label{eq: CII RG flows for 2 regimes a}
\end{equation}
In Fig.~\ref{fig: phase diagram}(a), 
we plotted the Kosterlitz-Thouless flows%
~(\ref{eq: CII RG flows for 2 regimes a})
which accurately capture the flows%
~(\ref{eq: Bernard flows for CII})
when $g^{\ }_{a'}\approx0$. However,
in the region CII defined by the condition
$g^{\prime}_{+}>|g^{\prime}_{-}|$,
the variance $g^{\ }_{a'}$ flows to strong coupling
and the RG flows follow three-dimensional trajectories.
We depict them
by using two-dimensional projections in Figs.~\ref{fig: phase diagram}(c) 
and~\ref{fig: phase diagram}(d).
The perturbative flows in the region CII
from Eq.~(\ref{eq: Bernard flows for CII a})
after projection to the  
$g^{\ }_{\mathrm{Re}\, m'}$-$g^{\ }_{\mathrm{Im}\, m'}$
plane
are depicted in
Fig.~\ref{fig: phase diagram}(c)
when $g^{\ }_{a'}$ is large. 
These flows show the instability
 of the CII boundaries
$g^{\ }_{\mathrm{Re}\, m'}\geq0$, $g^{\ }_{\mathrm{Im}\, m'}=0$
and
$g^{\ }_{\mathrm{Re}\, m'}=0$, $g^{\ }_{\mathrm{Im}\, m'}\geq0$
to any $g^{\ }_{a'}>0$.%
~\cite{footnote-slope flow away boundaries}
Moreover, these flows 
also show the infrared flow 
towards the nearly-critical plane
$g^{\prime}_{-}=0$
due to a reversal in the direction along the $g^{\prime}_{-}$ axis
of the infrared flows caused by the growth
of $g^{\ }_{a'}$ as is depicted in
Fig.~\ref{fig: phase diagram}(d).
It can be shown 
by adapting nonperturbative results 
from Ref.~\onlinecite{Guruswamy00}
that the change in the sign of the beta function for the coupling 
$g^{\prime}_{-}$
holds to all orders in $g^{\ }_{a'}$ and to linear order
in $g^{\prime}_{-}$.%
~\cite{footnote-nonperturbative relations line gM}
Hence, we conjecture that the infrared flows
emerging  from
the CII boundaries continue to the nearly-critical plane 
$g^{\prime}_{-}=0$
in Fig.~\ref{fig: phase diagram}(c) for the entire quadrant CII.

\section{
Projected Thirring model
        }
\label{sec: Projected Thirring model}

We now proceed by discussing the dashed line in
Fig.~\ref{fig: phase diagram},
$g^{\ }_{\mathrm{Re}\, m}$=$g^{\ }_{\mathrm{Im}\, m}$
and
$g^{\ }_{\mathrm{Re}\, m'}$=$g^{\ }_{\mathrm{Im}\, m'}$.

If we are only interested in correlation functions that are 
not affected by the flow (to strong coupling) of 
$g^{\ }_{\mathrm{a}'}$,
we can set $g^{\ }_{\mathrm{a}'}=0$
in Sec.~\ref{sec; Definitions and phase diagram}.
This is because, along the dashed line, the coupling
$g^{\ }_{\mathrm{a}'}$
turns out\cite{Guruswamy00}
to never feed into the RG equations
for the remaing two couplings, 
$g^{\ }_{\mathrm{Re}\, m}$ and $g^{\ }_{\mathrm{Im}\, m}$,
or
$g^{\ }_{\mathrm{Re}\, m'}$ and $g^{\ }_{\mathrm{Im}\, m'}$.

A mathematically consistent way to achieve this is to
replace the affine Lie superalgebra
$\widehat{\mathrm{gl}}(2|2)^{\ }_{1}$
by its affine Lie subsuperalgebra
$\widehat{\mathrm{psl}}(2|2)^{\ }_{1}$,%
~\cite{Bershadsky99,Berkovits99}
i.e., the $\widehat{\mathrm{gl}}(2|2)^{\ }_{1}$
Thirring models%
~(\ref{eq: gl22 perturbed by two current-current int BDI})
and%
~(\ref{eq: gl22 perturbed by two current-current int CII})
are combined into the $\widehat{\mathrm{psl}}(2|2)^{\ }_{1}$
Thirring model defined by
\begin{subequations}
\label{eq: psl22 perturbed by one current-current int}
\begin{equation}
\begin{split}
&
Z^{\ }_{\widehat{\mathrm{gl}}(2|2)^{\ }_{1}}=
\int\mathcal{D}[\psi^{\dag},\psi,
\bar{\psi}^\dag,\bar{\psi}]
\exp
\left(
-S^{\ }_{\widehat{\mathrm{gl}}(2|2)^{\ }_{1}}
\right),
\\
&
S^{\ }_{\widehat{\mathrm{gl}}(2|2)^{\ }_{1}}=
S^{\ }_{0}
+
\int \frac{\mathrm{d}\,\bar{z}\,\mathrm{d}\,z}{2\pi{i}}\,
\frac{g^{\ }_{\mathrm{M}}}{2\pi}
\mathcal{O}^{\ }_{\mathrm{M}},
\\
&
\mathcal{O}^{\ }_{\mathrm{M}}=
-
     J^{\,\mathrm{B}}_{\mathrm{A}}
\bar J^{\,\mathrm{A}}_{\mathrm{B}}
(-1)^\mathrm{A},
\end{split}
\end{equation}
subject to the 
$\widehat{\mathrm{psl}}(2|2)^{\ }_{1}$
constraints
\begin{equation}
0=
J^{\ \mathrm{A}}_{\mathrm{A}}(-)^{\mathrm{A}}=
\bar J^{\ \mathrm{A}}_{\mathrm{A}}(-)^{\mathrm{A}}
\label{eq: sl constraint} 
\end{equation}
and
\begin{equation}
0=
J^{\ \mathrm{A}}_{\mathrm{A}}=
\bar J^{\ \mathrm{A}}_{\mathrm{A}}
\label{eq: psl constraint} 
\end{equation}
\end{subequations}
along the now critical line $g^{\ }_{\mathrm{M}}\in\mathbb{R}$.
The constraint~(\ref{eq: sl constraint}) justifies setting
$g^{\ }_{a'}=0$.
The sign of the variance 
$g^{\ }_{\mathrm{M}}$
distinguishes symmetry class BDI 
($g^{\ }_{\mathrm{M}}>0$)
from symmetry class CII
($g^{\ }_{\mathrm{M}}<0$).
The graded index $\mathrm{A}$ runs
from 1 to $4\mathrm{N}$ when dealing with 
the $\mathrm{N}$-th moment
of the single-particle Green's function.

\section{
Relationship to a NL$\sigma$M
        }
\label{sec: topological term in class CII}

So far, we have relied on a description of the
global phase diagram and, in particular, of the vertical 
dashed line of nearly-critical points
in region CII of Fig.~\ref{fig: phase diagram}
that makes explicit the Dirac structure underlying the
clean limit of the theory. In this section, 
we seek an alternative description of this line,
in particular far away from the clean Dirac limit.

To this end, we first observe that
we can derive a replicated NL$\sigma$M by integrating out
replicated Dirac fermions in favor of Goldstone modes as is done in Appendix%
~\ref{appsec: The sign ambiguity of a Pfaffian}.
We find a replicated NL$\sigma$M augmented by a term of topological origin,
the $\theta$ term at $\theta=\pi$.
The same calculation also applies to the 
supersymmetric formulation of the disordered system, 
yielding a $\theta$ term at $\theta=\pi$ for the NL$\sigma$M 
defined on the SUSY target manifold given in 
Eq.~(\ref{eq: def CII target space})
below.

Without the $\theta$ term at $\theta=\pi$, 
this NL$\sigma$M was already derived starting from a different microscopic
model within the chiral symplectic symmetry class CII by 
Gade in Ref.~\onlinecite{Gade91-93}. This NL$\sigma$M has
two coupling constants $t^{\ }_{\mathrm{M}'}$ and $t^{\ }_{a'}$
that are positive numbers, in addition to the 
topological coupling $\theta=\pi$.
The labels of these couplings are chosen to convey
the fact that
$t^{\ }_{\mathrm{M}'}$
does not flow (Ref.~\onlinecite{Gade91-93})
whereas
$t^{\ }_{a'}$ does flow away from its value 0 at the Gaussian fixed point 
(Ref.~\onlinecite{Gade91-93}),
by analogy to the flow of the couplings
$g^{\ }_{\mathrm{M}'}$ 
and
$g^{\ }_{a'}$
in Eq.~(\ref{eq: gl22 perturbed by two current-current int CII}),
respectively. The topological coupling does not flow,
for it can only take discrete values.

The question we want to address in this section
is what is the relationship between
this NL$\sigma$M 
with a $\theta$ term at $\theta=\pi$
and the Thirring model defined in 
Eq.~(\ref{eq: gl22 perturbed by two current-current int CII}).
We are going to argue that they are dual in a sense that will become 
more precise as we proceed. To this end, 
we shall rely on the SUSY description used to represent the
Thirring model defined in 
Eq.~(\ref{eq: gl22 perturbed by two current-current int CII}).

We begin by establishing the relevant pattern of symmetry breaking.
The field theory%
~(\ref{eq: gl22 perturbed by two current-current int CII})
is a $\mathrm{GL}(2|2)$ principal chiral model
augmented by a WZNW term at level $k=1$
when the couplings $g^{\ }_{\mathrm{M}'}=g^{\ }_{a'}=0$.%
~\cite{Ryu10a}
This means that the theory at
$g^{\ }_{\mathrm{M}'}=g^{\ }_{a'}=0$
is invariant under the symmetry supergroup
\begin{equation}
\mathrm{GL}(2|2)\times\mathrm{GL}(2|2).
\end{equation}
The current-current perturbations 
for any $g^{\ }_{\mathrm{M}'}>0$
in Eq.%
~(\ref{eq: gl22 perturbed by two current-current int CII})
lower this symmetry down to the diagonal supergroup
\begin{equation}
\mathrm{GL}(2|2).
\end{equation} 
In turn, the symmetry 
$\mathrm{GL}(2|2)$
can be further reduced 
if fermion bilinears acquire an expectation value,
as must be the case if the global DOS is nonvanishing at the band center
due to the disorder. This is in fact what happens 
if the analysis of Refs.%
~\onlinecite{Guruswamy00} and \onlinecite{Mudry03} 
along the nearly-critical line in the BDI quadrant of
Fig.~\ref{fig: phase diagram} 
is repeated for the case at hand,
with the remaining residual symmetry being
\begin{equation}
\mathrm{OSp}(2|2).
\end{equation} 
The Goldstone modes associated with this pattern of
symmetry breaking generate the supermanifold
\begin{equation}
\mathrm{GL}(2|2)/\mathrm{OSp}(2|2),
\label{eq: def CII target space}
\end{equation} 
which is nothing but the SUSY target space for a NL$\sigma$M model
in symmetry class CII 
(see Ref.\ \onlinecite{Zirnbauer96}
and Appendix%
~\ref{subsec: Patterns of symmetry breaking and supermanifolds} 
of this paper).
The critical vertical dashed line 
in quadrant CII of Fig.~\ref{fig: phase diagram}
arises from removing the sector
$\mathrm{GL}(1;\mathbb{R})\times \mathrm{U}(1)$
from the field theory%
~(\ref{eq: gl22 perturbed by two current-current int CII}).
The ensuing projected field theory is given by Eq.%
~(\ref{eq: psl22 perturbed by one current-current int}).
The corresponding operation on the target space%
~(\ref{eq: def CII target space}) 
of the NL$\sigma$M for symmetry class CII yields the manifold%
~\cite{Read01}
\begin{eqnarray}
\mathrm{PSL}(2|2)/
\mathrm{OSp}(2|2)
&\sim& 
\mathrm{PSL}(2|2)/\mathrm{SU}(2|1) 
\nonumber\\
&\sim&
\mathrm{U}(2|2)/[\mathrm{U}(1)\times \mathrm{U}(2|1)] 
\nonumber\\
&\sim&
\mathbb{C}P^{2|1}.
\label{eq: def projected CII target space}
\end{eqnarray}
We have used here the isomorphism between $\mathrm{OSp}(2|2)$ and 
$\mathrm{SU}(2|1)$.
By setting all fermionic coordinates to zero on this SUSY manifold,
one obtains the bosonic submanifold given by
\begin{eqnarray}
\hbox{Boson-Boson     (BB)}
&&
\hbox{Fermion-Fermion (FF)}
\nonumber\\
\hbox{(non-compact)}
&&
\hbox{(compact)}
\nonumber\\
\mathrm{SU}^*(2)/\mathrm{Sp}(2) 
& \ \ \times \ \ &
\mathrm{SU}(2)/\mathrm{SO}(2).
\label{eq: def projected CII target space fermions gone}
\end{eqnarray}
(The definition of the group $\mathrm{U}^{*}(2)$ is given
in Appendix~\ref{app sec: The Lie group U{*}(2)}.)
We close this symmetry analysis 
by recalling~\cite{Lundell92}
that the second homotopy group 
of the compact part of the submanifold%
~(\ref{eq: def projected CII target space fermions gone})
is not trivial and given by
\begin{equation} 
\pi^{\ }_{2}[\mathrm{SU}(2)/\mathrm{SO}(2)]= 
\mathbb{Z}^{\ }_{2}.
\end{equation}

\begin{figure}
\begin{center}
(a)
\includegraphics[height=4cm,clip]{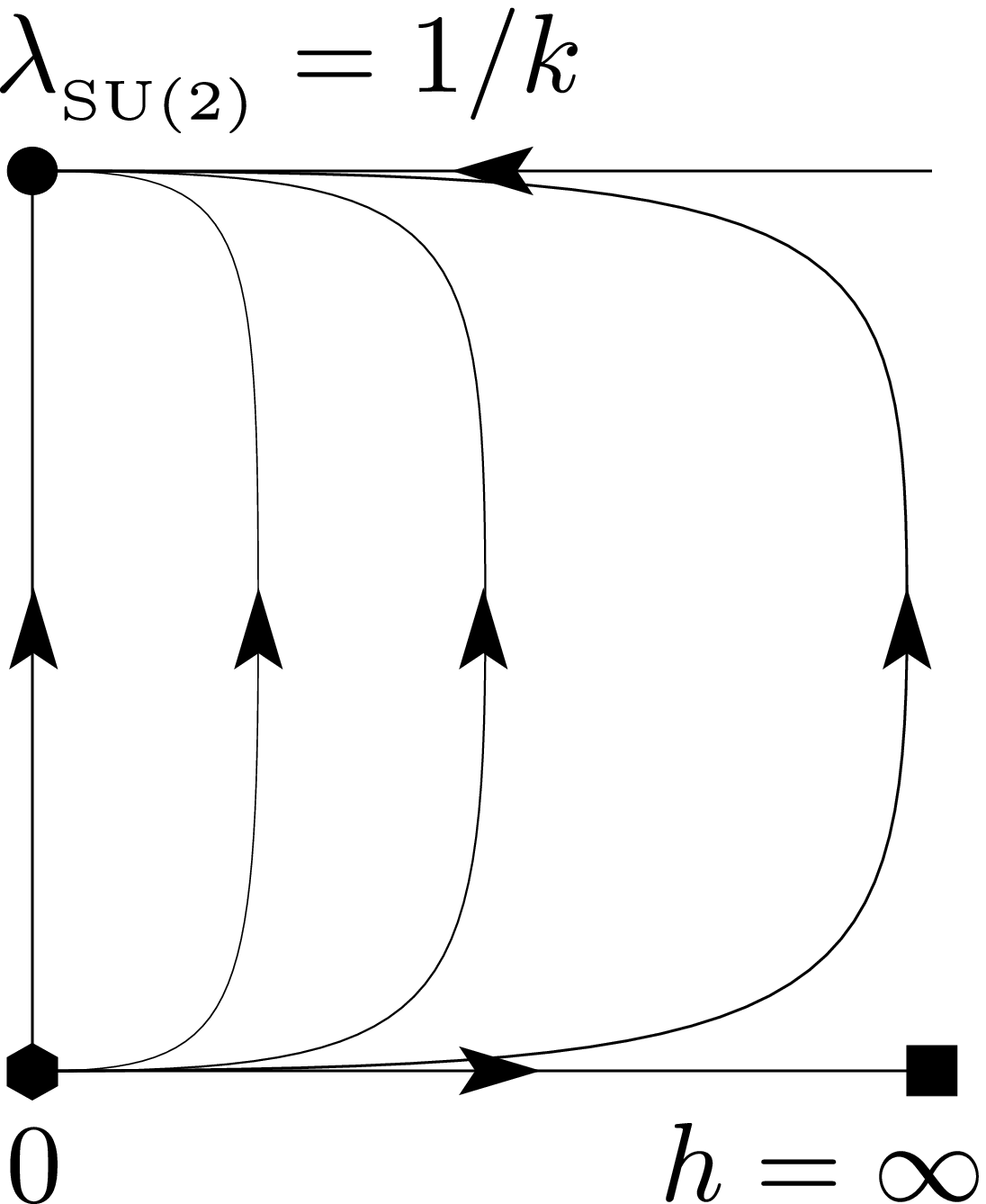}
(b)
\includegraphics[height=4cm,clip]{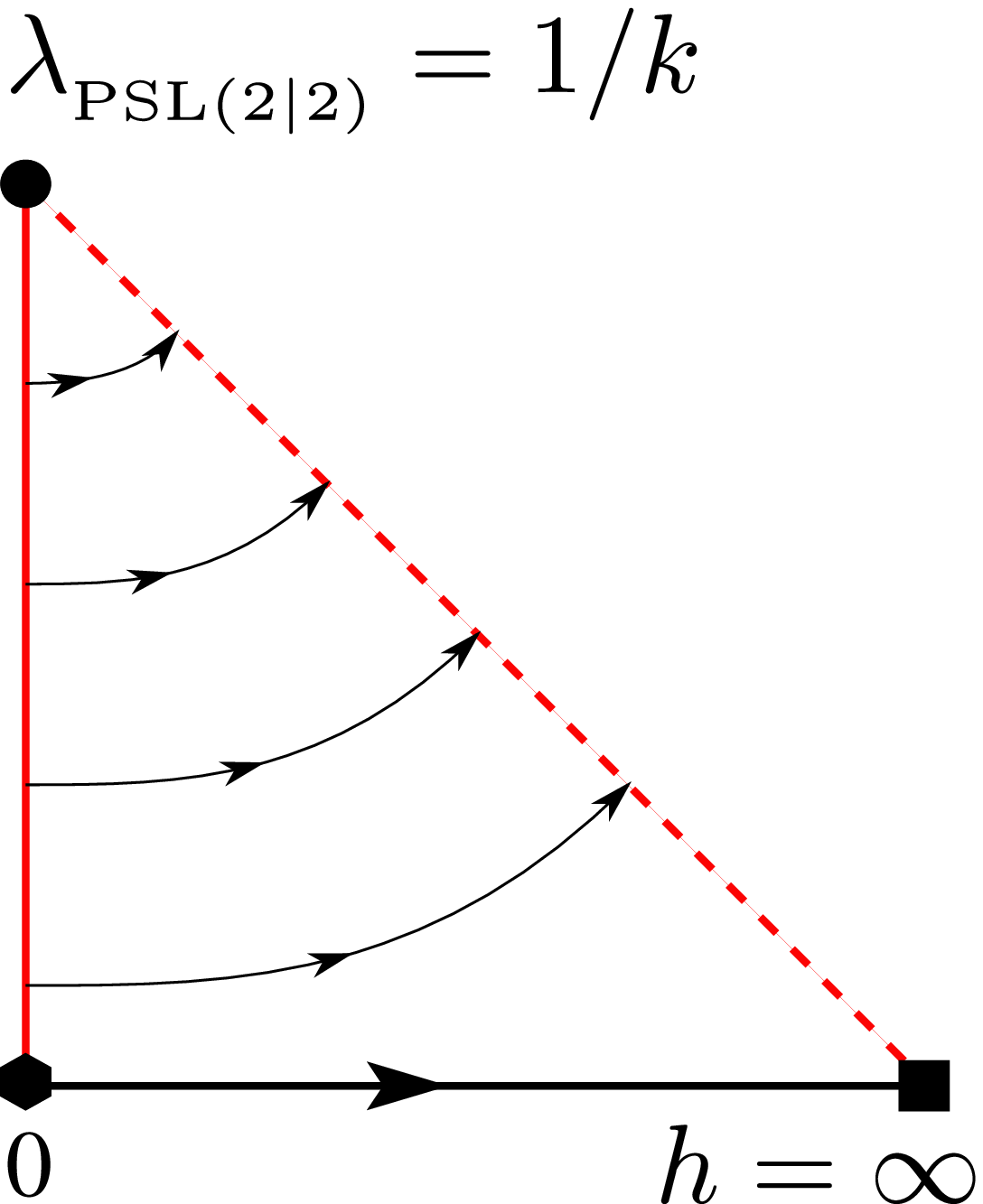}
\caption{(Color online)
(a)
Phase diagram for the SU(2) principal chiral model
with the coupling constant $\lambda^{\ }_{\mathrm{SU(2)}}$
that is (i) augmented by the WZNW term at level $k=1$ 
and (ii) perturbed
by a symmetry-breaking potential with coupling 
$h^{\ }$.
The critical WZNW theory is at 
$\lambda^{\ }_{\mathrm{SU(2)}}=1/k$
and is represented on the upper-left corner of 
the phase diagram by a filled circle. The flow along
the upper boundary of the phase diagram is that of
the marginally irrelevant current-current perturbation.
The lower-left corner of the phase diagram
is the Gaussian fixed point
of the SU(2) principal chiral model
augmented by a WZNW term at level $k=1$,
it is depicted by a filled hexagon.
The lower-right corner of the phase diagram is the
Gaussian fixed point of the
O(3) NL$\sigma$M 
with $\theta$ term at $\theta=\pi$, 
it is depicted by a filled square.
(b) Same as in panel (a) except for the replacement
of SU(2) by PSL(2$|$2) and of SU(2)${/}$U(1) by $\mathbb{C}P^{2|1}$
under the assumption that there is no more relevant perturbations
than the exactly marginal current-current perturbation at the
upper-left corner of the phase diagram. The left vertical boundary%
~\cite{Bershadsky99,Berkovits99}
and the diagonal boundary%
~\cite{Guruswamy00} 
are now lines of critical points (colored in red).
The diagonal boundary that
connects the upper-left to the lower-right
corner is a line of critical points that
is argued to have a dual representation 
in terms of a Thirring model on the one hand or a
NL$\sigma$M 
with a $\theta$ term at $\theta=\pi$
on the other hand.
        }
\label{fig: phase diagram for WZWN versus NLSM}
\end{center}
\end{figure}

We are now going to argue that, under
certain natural assumptions detailed below,
the vertical dashed line of nearly-critical points
in region CII of Fig.~\ref{fig: phase diagram}
is described by a NL$\sigma$M with a 
$\theta$ term at $\theta=\pi$ on the
$\mathbb{C}P^{2|1}$
target space [Eq.\ (\ref{eq: def projected CII target space})].

To understand what could prevent the identification of
the vertical dashed line of nearly-critical points
as realizing the $\mathbb{C}P^{2|1}$ 
NL$\sigma$M with $\theta$ term at $\theta=\pi$,
we are first going to review 
the connection between the O(3) NL$\sigma$M with the
$\theta$ term at $\theta=\pi$ and the $\mathrm{SU(2)}^{\ }_{1}$
WZNW field theory perturbed by the current-current interaction.%
~\cite{Affleck87}

The O(3) NL$\sigma$M with $\theta$-term at $\theta=\pi$
captures the low-energy and long-wave-length excitations
of antiferromagnetic spin-1/2 Heisenberg spin chains.
This field theory is related to the 
$\mathrm{SU(2)}^{\ }_{1}$ WZNW field theory by  
perturbing the latter with a symmetry-breaking 
potential (coupling constant $h$), 
which has the effect of changing the target
manifold of the principal chiral model, at $h=0$, 
to that of the NL$\sigma$M, at $h=\infty$. 
(See Fig.\ \ref{fig: phase diagram for WZWN versus NLSM}.)
When the WZNW model is near its 
weakly-coupled ultra-violet (UV) Gaussian fixed point,
the flow of the coupling $h$ away from this Gaussian fixed point is 
the strongest and brings the theory 
into the vicinity of the weakly coupled
(UV, Gaussian) fixed point of the O(3) NL$\sigma$M
augmented by a $\theta$ term at $\theta=\pi$. 
In the vicinity of
the $\mathrm{SU(2)}^{\ }_{1}$ WZNW critical point,
the symmetry-breaking potential (coupling $h$) reduces to the marginally irrelevant
current-current interaction up to more irrelevant interactions
(some discrete symmetries must here be invoked).
When the coupling constant $\lambda$ of the 
SU(2) principal chiral model 
augmented by the level $k=1$ WZNW term
is close to its critical value $\lambda = 1/k=1$, 
the symmetry-breaking potential
generates RG flows 
that are close to those of 
O(3) NL$\sigma$M with a $\theta$ term at $\theta=\pi$.
When the coupling constant of the 
SU(2) principal chiral model 
augmented by the level $k=1$ WZNW term
is small,
the symmetry-breaking potential generates RG flows
that drive the theory very close to the
weakly coupled 
(Gaussian) fixed point 
of the O(3) NL$\sigma$M 
with a $\theta$ term at $\theta=\pi$.
The envelope of all these RG flows can be thought of as the RG
flow from the Gaussian fixed point of the
O(3) NL$\sigma$M
with a $\theta$ term at $\theta=\pi$
to the $\mathrm{SU(2)}^{\ }_{1}$ WZNW critical point.

The same argument can also be used to relate
the principal chiral models
defined on the groups 
$\mathrm{SU}(N)$
and
$\mathrm{SO}(M+N)$,
augmented by a $k=1$ WZNW term,
to the NL$\sigma$M 
with the target manifold $\mathrm{SU}(N)/\mathrm{SO}(N)$
and $\mathrm{SO}(M+N)/\mathrm{SO}(M)\times\mathrm{SO}(N)$,
respectively, when
augmented by  a $\theta$ term at $\theta=\pi$.
This argument is confirmed
by exact results obtained from Bethe-Ansatz
integrability for these 
NL$\sigma$Ms.%
~\cite{Zamolodchikov92,Fendley01}

\begin{figure}
\begin{center}
(a)
\includegraphics[height=4cm,clip]{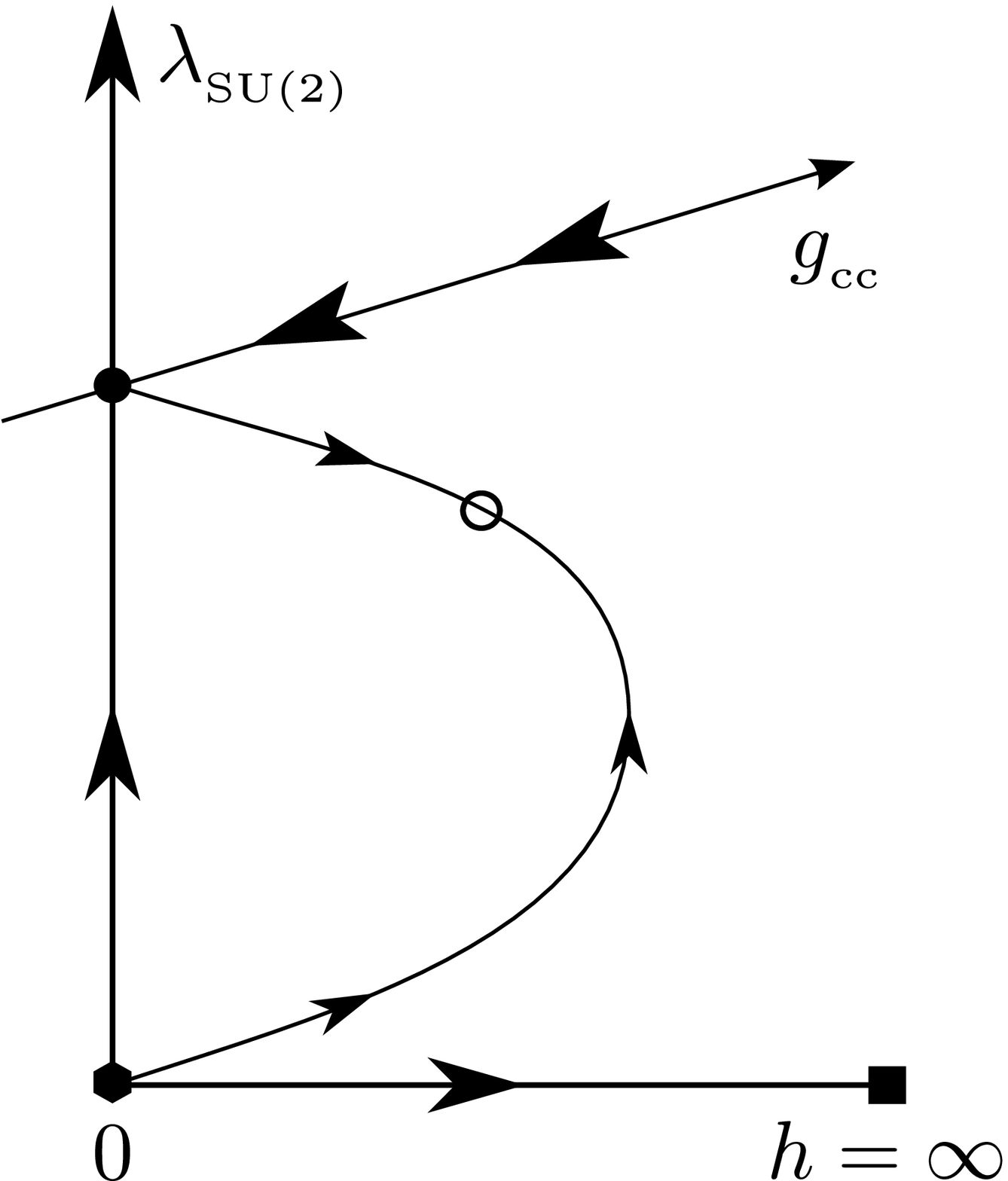}
(b)
\includegraphics[height=4cm,clip]{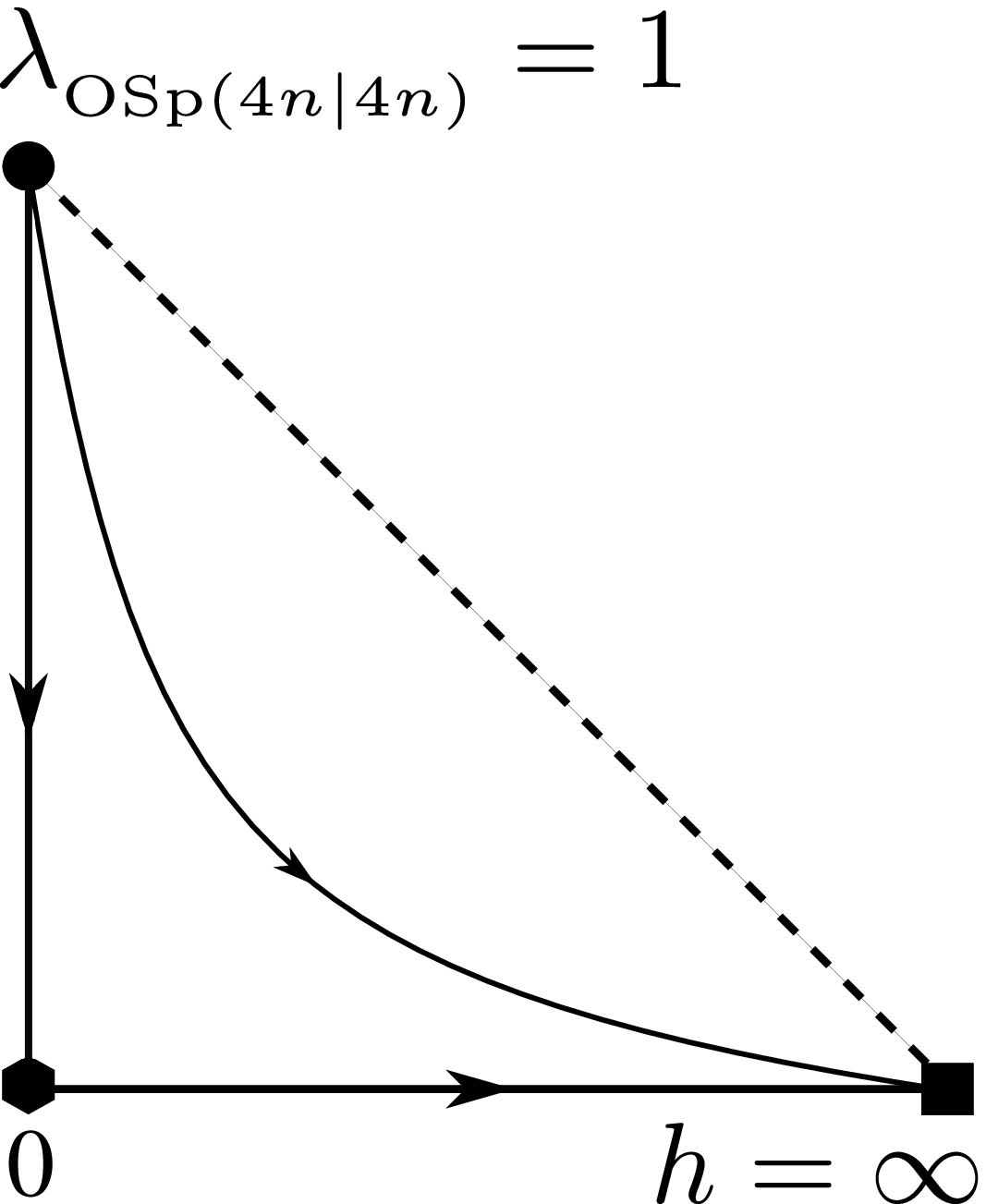}
\caption{
(a)
Phase diagram for the SU(2) principal chiral model
with the coupling constant $\lambda^{\ }_{\mathrm{SU(2)}}$
that is (i) augmented by the WZNW term at level $k>1$ 
and (ii) perturbed
by a symmetry-breaking potential with coupling 
$h^{\ }$.
The critical WZNW theory is at 
$\lambda^{\ }_{\mathrm{SU(2)}}=1/k$
and is represented in the phase diagram by a filled circle. 
The Gaussian fixed point
of the SU(2) principal chiral model
augmented by a WZNW term at level $k>1$
is depicted by a filled hexagon.
The Gaussian fixed point of the
O(3) NL$\sigma$M 
with $\theta$ term at $\theta=\pi$
is depicted by a filled square. The fact that there are
operators more relevant than the current-current interaction
induced by the symmetry-breaking potential 
is indicated by the presence of a third axis 
in coupling space. This third axis
quantifies the running of the current-current coupling constant
$g^{\ }_{\mathrm{cc}}$
that is marginally irrelevant. The critical point
of the O(3) NL$\sigma$M 
augmented by a $\theta$ term
at $\theta=\pi$ is depicted by an open circle.
(b) 
Counterpart to Fig.~\ref{fig: phase diagram for WZWN versus NLSM}(b)
for the case of the WZNW model on $\mathrm{OSp}(4n|4n)$ at level $k=1$
perturbed by a symmetry-breaking potential that projects this WZNW model
to the NL$\sigma$M model in the symmetry class AII.
        }
\label{fig: phase diagram for WZWN versus NLSM if k>1}
\end{center}
\end{figure}

On the other hand, when the level is larger than one 
(for example, $k>1$ arises from a fine-tuned half-integer spin chain
with spin larger than 1/2), the symmetry-breaking potential
permits (on symmetry grounds) the appearance of terms more
relevant than the current-current interactions in the vicinity of the
$\mathrm{SU(2)}^{\ }_{k>1}$ WZNW fixed point.%
~\cite{Affleck87}
Correspondingly, the flow 
of the O(3) NL$\sigma$M 
with a $\theta$ term at $\theta=\pi$
will not reach 
the $\mathrm{SU(2)}^{\ }_{k>1}$
WZNW critical point, 
but reaches  a different, intermediate fixed point
describing the critical behavior of the O(3) NL$\sigma$M 
with the $\theta$ term at $\theta=\pi$
[see Fig.~\ref{fig: phase diagram for WZWN versus NLSM if k>1}(a)].
This also happens in
the $\mathrm{OSp}(2|2)/\mathrm{GL}(1|1)$ 
NL$\sigma$M with theta term at $\theta=\pi$
describing the critical behavior of the spin-quantum-Hall transition.%
~\cite{Gruzberg99,Ludwig00}

After these preliminary comments, we proceed
to the case of interest with
$\mathrm{PSL}(2|2)$
symmetry.
If we assume that no relevant or marginal interactions
other than the marginal current-current interaction
are allowed 
at the $\mathrm{PSL(2|2)}$ WZNW critical point
at level $k=1$
when the
$\mathrm{PSL}(2|2) \times \mathrm{PSL}(2|2)$
symmetry 
of the WZNW model is lowered to its
diagonal $\mathrm{PSL}(2|2)$ symmetry
upon introduction of the symmetry breaking potential
(a natural assumption for the level $k=1$ case under consideration), 
then we obtain with 
Fig.~\ref{fig: phase diagram for WZWN versus NLSM}(b)
the desired relation between the $\mathrm{PSL(2|2)}$ WZNW theory perturbed by
the current-current interaction and the 
$\mathbb{C}P^{2|1}$ principal chiral model 
with $\theta$ term at $\theta=\pi$.%
~\cite{appendix_potential}
By analogy with the 
$\mathrm{SU(2)}^{\ }_{k>1}$ WZNW critical point,
we do not expect this assumption to be fulfilled when the level $|k|>1$.

A similar projection from the WZNW model onto the NL$\sigma$M 
with theta term at $\theta=\pi$
can also be implemented for symmetry class AII,
in complete analogy with the case of the projection 
discussed above from the WZNW model to the 
NL$\sigma$M in symmetry class CII.
For the case of symmetry class AII, consider
the WZNW model on $\mathrm{OSp}(4n|4n)$ at level $k=1$. 
In this case, the coupling constant of the principal chiral model
flows away from the WZNW fixed point down 
towards the weakly coupled WZNW model 
[see Fig.~\ref{fig: phase diagram for WZWN versus NLSM if k>1}(b)]. 
Now, we project again to the NL$\sigma$M model in the symmetry class AII
with the help of the 
corresponding symmetry-breaking potential (with coupling constant $h$). 
When this is done for the weakly coupled WZNW model,
this yields the weakly coupled 
NL$\sigma$M in class AII, the Wess-Zumino term turning
into a theta term at $\theta=\pi$ on the AII target space
[see Fig.~\ref{fig: phase diagram for WZWN versus NLSM if k>1}(b)]. 
On the other hand, the most relevant operator
in the vicinity of the
fixed point of the WZNW model on $\mathrm{OSp}(4n|4n)$ at level $k=1$
which has the symmetries of the symmetry breaking potential
is the current-current interaction between the Noether currents. 
This operator is marginally relevant.
Thus, the RG flow emerging from the unstable 
fixed point of the WZNW model on 
$\mathrm{OSp}(4n|4n)$ at level $k=1$ 
ends up in the infrared at the weakly coupled
NL$\sigma$M in symmetry class AII
[see Fig.~\ref{fig: phase diagram for WZWN versus NLSM if k>1}(b)].
This is one way of understanding that the 
NL$\sigma$M in class AII with the $\mathbb{Z}^{\ }_{2}$ 
term always flows to weak coupling
(as discussed in Refs.\ \onlinecite{Bardarson07, Nomura07}),
for it simply inherits this feature from the RG flow of 
the underlying WZNW model.

In summary, based on this reasoning 
we argue that the line of nearly-critical points 
in region CII of Fig.~\ref{fig: phase diagram}
(the vertical dashed line
in region CII of Fig.~\ref{fig: phase diagram})
has two descriptions;
one in terms of the $\mathrm{PSL}(2|2)$ WZNW model
perturbed by current-current interactions,
and one in terms of the 
$\mathbb{C}P^{2|1}$
NL$\sigma$M at $\theta=\pi$ 
($\mathbb{Z}^{\ }_2$ topological term).
These descriptions are dual to each other in the sense 
that, in the vicinity of the origin of our global phase diagram 
in Fig.\ \ref{fig: phase diagram}
the $\mathrm{PSL}(2|2)$ WZNW model is weakly perturbed,
whereas the 
$\mathbb{C}P^{2|1}$
NL$\sigma$M is strongly interacting.
On the other hand, for large values of the coupling constant
$g^{\ }_{\mathrm{M}'}$
of the current-current interaction about the Dirac point, a measure of the
distance downwards along the dotted line away from the clean Dirac point
at the center of Fig.\ \ref{fig: phase diagram},
the resulting Thirring
model is strongly interacting whereas the 
$\mathbb{C}P^{2|1}$
NL$\sigma$M is weakly interacting.
We recall that, 
because the coupling constant $g^{\ }_{\mathrm{M}'}$ is exactly marginal,
and so is the coupling constant of the corresponding
NL$\sigma$M,\cite{Gade91-93}
it is possible to \textit{continuously} interpolate between these 
two limits by tuning $g^{\ }_{\mathrm{M}'}$. 
(The possibility of such a duality was also discussed,
independently and from a different perspective,
in Ref.~\onlinecite{Mitev08}, \onlinecite{Candu09},
and \onlinecite{Candu10}.) 

\section{
Discussion
        }
\label{sec: Discussions}

\subsection{
$\mathbb{Z}^{\ }_{2}$ topological term
in the symmetry class CII
of two-dimensional Anderson localization
           }

A systematic study of random Dirac fermions 
in $d$-dimensional space provides a road-map
to uncovering universal properties of
Anderson localization. This is so because
random Dirac fermions build a bridge between
models for Anderson localization that
are defined on lattices -- and thus are non-universal --
and effective field theories (NL$\sigma$Ms)
that solely depend on the underlying symmetries 
and dimensionality of space -- and as such are universal.

In one-dimensional space, Dirac fermions generically emerge after
linearization of the energy dispersion around the Fermi energy 
in the clean limit. The effects of weak static disorder are
then elegantly encoded by a description of quasi-one-dimensional 
quantum transport in terms of diffusive processes
on non-compact symmetric spaces.%
\cite{Huffmann90,Brouwer00,Mudry00,Titov01,Brouwer03,Brouwer05}
This long-wave length description
is sufficiently fine to account for non-perturbative effects
such as parity effects in the numbers of propagating channels
in the chiral symmetry classes AIII, CII, and BDI.%
~\cite{Brouwer98}
A parity effect can also be derived for the symplectic 
symmetry class AII in quasi-one dimension.%
~\cite{Zirnbauer92,Takane04}
Although the latter 
parity effect is not generic in quasi-one-dimensional
space because of the fermion-doubling obstruction, it is generic on
one-dimensional boundaries of 
two-dimensional $\mathbb{Z}^{\ }_{2}$-topological
band insulators.%
\cite{Obuse07}

Dirac fermions are the exception
rather than the rule in band theory when the 
dimensionality of space $d$ is larger than one. 
Fine-tuning between the lattice and the hopping amplitudes 
is needed to select a linear energy dispersion. 
There is a parallel to this fact in the context
of Anderson localization.

For example, in two-dimensional space,
the symmetries respected by the static disorder 
do not enforce, on their own, the presence of
WZNW or $\mathbb{Z}^{\ }_{2}$-topological terms 
in the NL$\sigma$M effective long-wave length description of the
physics of localization. 

Ludwig \textit{et al}.\cite{Ludwig94}
(Nersesyan \textit{et al}.\cite{Nersesyan95}) 
have shown that non-perturbative effects
can modify  the localization properties encoded by the 
two-dimensional NL$\sigma$M with a WZNW term
in symmetry class AIII
when studying the random Dirac Hamiltonian with 
$\mathrm{N}^{\ }_{\mathrm{f}}=1$
($\mathrm{N}^{\ }_{\mathrm{f}}>1$)
flavors. Analogous physics can appear in symmetry classes
DIII and CI in two spatial dimensions.%
~\cite{Schnyder08,Schnyder09b,Schnyder09a}
However, because of the fermion-doubling
obstruction, these conditions cannot be met
in purely two-dimensional lattice models for Anderson localization.

On the other hand, 
they can always be fulfilled on the two-dimensional boundaries
of three-dimensional topological band insulators
(that are characterized by an integer topological index).%
~\cite{Schnyder08,Ryu10b}

A similar situation holds
for the $\mathbb{Z}^{\ }_{2}$-topological terms.
The number of Dirac flavors $\mathrm{N}^{\ }_{\mathrm{f}}$
matters crucially to obtain a
$\mathbb{Z}^{\ }_{2}$-topological term in symmetry class AII
as shown by Ryu \textit{et al.}\ in 
Ref.~\onlinecite{Ryu07b}.
In the present paper, we have completed 
the derivation of topological terms of
two-dimensional NL$\sigma$M by constructing the
$\mathbb{Z}^{\ }_{2}$-topological term for a NL$\sigma$M in 
symmetry class CII 
as a sign ambiguity in the Pfaffian of disordered Majorana spinors.
Our derivation suggests that this
$\mathbb{Z}^{\ }_{2}$-topological term
cannot arise from two-dimensional local lattice models
of Anderson localization because of the 
fermion-doubling obstruction, but requires a three-dimensional
topological band insulator with two-dimensional boundaries.

\subsection{
Global phase diagram at the band center
           }

The main results of this paper are summarized in
Fig.~\ref{fig: phase diagram}. They apply to
$\mathrm{N}^{\ }_{\mathrm{f}}=2$ flavors
of random Dirac fermions. 

Figure~\ref{fig: phase diagram} should be compared
with Fig.\ 9 from Ref.~\onlinecite{Ludwig94}
that captures the phase diagram
for $\mathrm{N}^{\ }_{\mathrm{f}}=1$ flavors
of random Dirac fermions or, more precisely,
with its projection onto the plane 
$g^{\ }_{\mathrm{A}}=0$
in Ref.~\onlinecite{Ludwig94}
($\Delta^{\ }_{\mathrm{A}}=0$ in the notation of 
Ref.~\onlinecite{Ludwig94}).
The phase diagram in Fig.~\ref{fig: phase diagram}
is also obtained after projecting a three-dimensional flow
to a two-dimensional subspace of in coupling constant space.

The three chiral phases AIII, BDI, and CII 
in Fig.~\ref{fig: phase diagram}
meet at the origin of the phase diagram.
This meeting point realizes the clean Dirac limit. 
We showed that analytical continuation of the disorder 
at the level of the Dirac fermions allows
one to move between the BDI and CII phases.
However, at the microscopic scale of the 
two-dimensional lattice model that realizes the BDI phase,
this analytical continuation is meaningless.
This is yet another manifestation of the fermion-doubling
obstruction. A realization as a local lattice model of the 
CII phase in Fig.~\ref{fig: phase diagram} must go through
the two-dimensional boundary of a three-dimensional 
topological band insulator. 

The quadrant BDI in Fig.~\ref{fig: phase diagram}
is fairly well understood
if we assume that the perturbative
flows to the nearly-critical dashed line
extend all the way to the boundary D.
Bulk%
~\cite{Mudry03} 
and boundary%
~\cite{Obuse08} multifractality
and an analytic dependence of the conductance on the disorder
strength $g^{\ }_{\mathrm{M}}$ 
at the band center $E=0$%
~\cite{Ryu07a}
are governed in the thermodynamic limit
by their dependence on $g^{\ }_{\mathrm{M}}$ 
along the nearly-critical dashed line.

The dashed line in region CII 
of Fig.~\ref{fig: phase diagram}
is a line of nearly-critical points, 
each of which is captured by a projected Thirring model. 
We have argued that the strong coupling regime of this theory
is ``dual'' to a weakly-coupled NL$\sigma$M 
augmented by a  $\mathbb{Z}^{\ }_{2}$ topological term 
with the target space of symmetry class CII.

We would like to emphasize that the RG
flows in the CII quadrant of Fig.~\ref{fig: phase diagram}, 
\textit{first away from and then to} 
the nearly-critical plane 
defined by the dashed line and the out-of-plane $g^{\ }_{a'}$ axis,
are perturbative in $g^{\prime}_{\pm}$
(nonperturbative in $g^{\ }_{a'}$).
They are derived under the assumption
that no relevant or marginal interactions
other than the current-current interactions are allowed.
The continuation of these flows to strong coupling 
is a conjecture.
At strong coupling, it is tempting to ask if
these two-parameter flows might be captured by a NL$\sigma$M.
Evidently, a NL$\sigma$M 
whose target space is a symmetric target space 
will not do since it would only be characterized by one
running coupling constant. A NL$\sigma$M 
on a \textit{homogeneous} but \textit{not symmetric}
target space with two independent coupling constants in addition
to the Gade term would do. (We refer the reader to Ref.%
~\onlinecite{Friedan85} 
for a systematic study of NL$\sigma$M on \textit{Riemannnian
manifolds}, of which homogeneous and symmetric spaces are 
special examples, as is explained in the context of disordered
systems in Ref.~\onlinecite{footnote: homogeneous space}.)
We propose that this scenario is captured by
a NL$\sigma$M with the following
homogeneous, but not symmetric target space 
($n$ is an integer,
see Appendix \ref{appsec: homogeneous vs symmetric})
\begin{equation}
\label{HomeogeneousSpace}
\mathrm{GL}(2n|2n)/[\mathrm{OSp}(n|n)\times \mathrm{OSp}(n|n)].
\end{equation}
The situation here is analogous to the NL$\sigma$M
discussed in Ref.~\onlinecite{footnote: homogeneous space}
in the context of the random-bond Ising model in two dimensions.
The NL$\sigma$M on the homogeneous target space in 
Eq.~(\ref{HomeogeneousSpace})
has two coupling constants [in addition to the coupling
constant of the ``Gade'' term 
(``projected out'' in our global phase diagram in 
Fig. \ref{fig: phase diagram}), of the kind that we previously
denoted by $g^{\ }_{a'}$ in the present paper].
The NL$\sigma$M on this \textit{homogeneous} space
interpolates between the two NL$\sigma$Ms on the \textit{symmetric}
target spaces corresponding to symmetry classes CII and
AII, which are specific limits within the 2-parameter
coupling constant space of the
NL$\sigma$M on this \textit{homogeneous} space (in a manner
analogous to the situation discussed in 
Ref.~\onlinecite{footnote: homogeneous space}).
See Appendix E for more details.

\begin{figure}
  \begin{center}
(a)\includegraphics[width=3.5cm,clip]{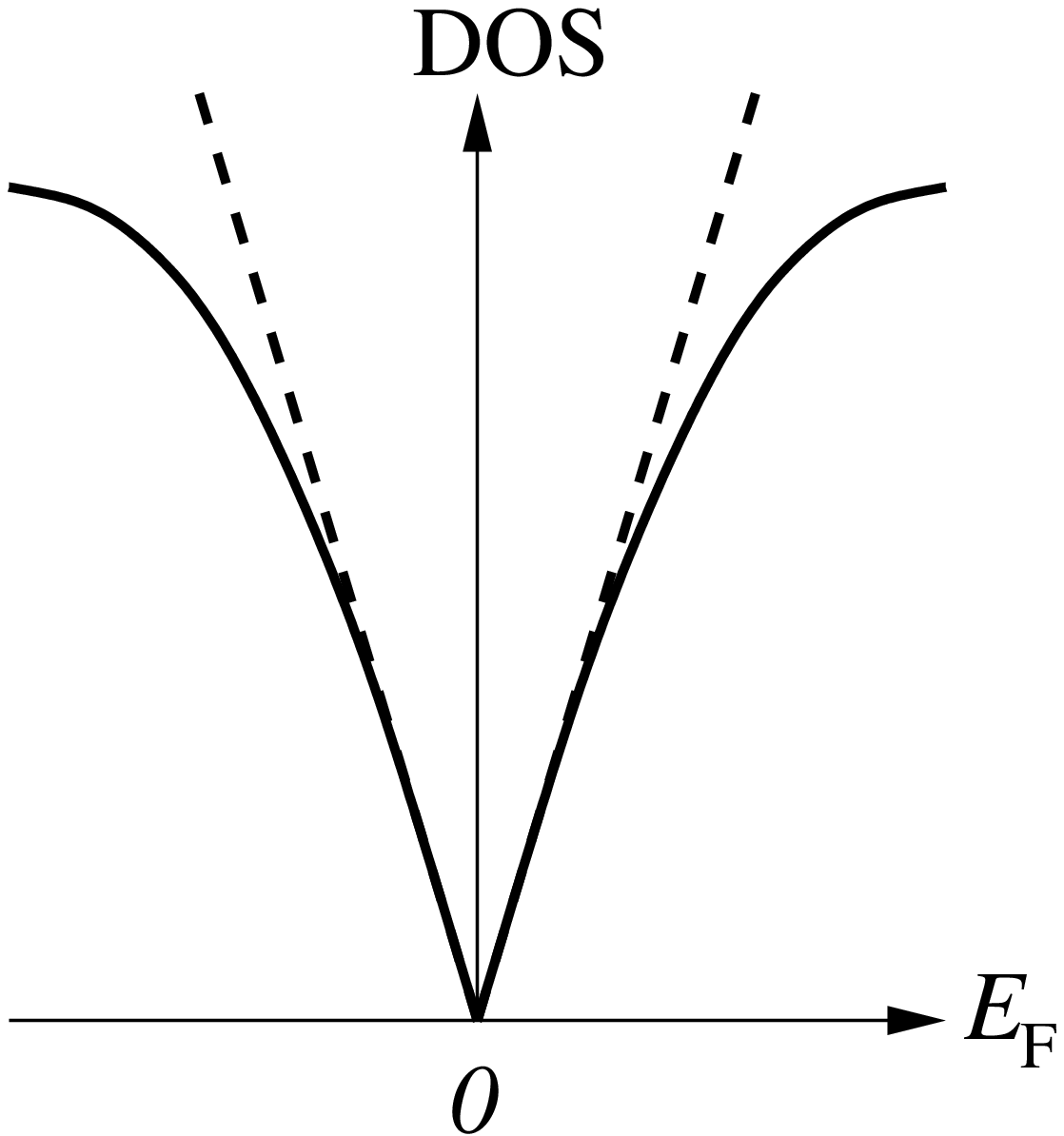} 
(b)\includegraphics[width=3.5cm,clip]{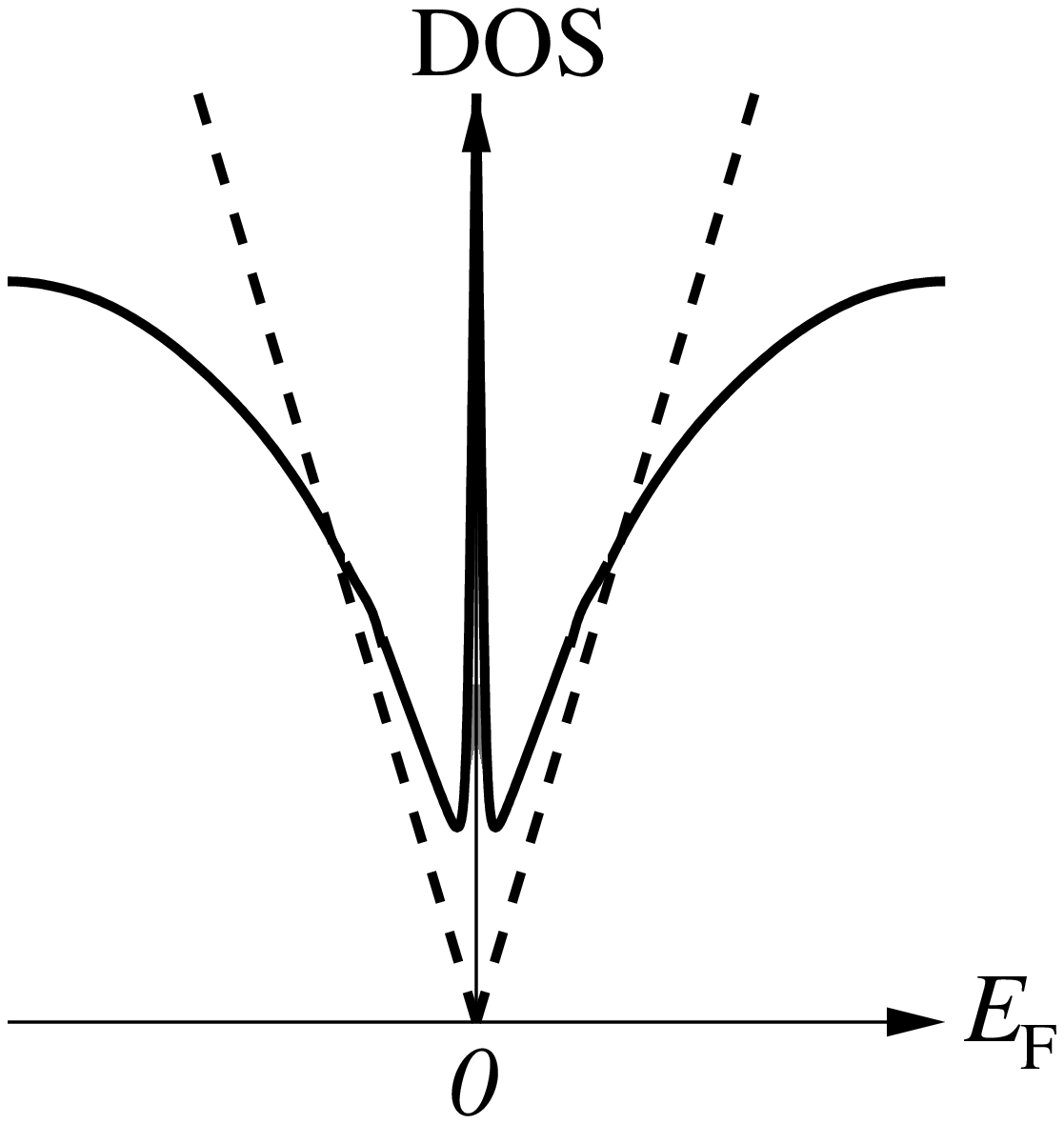}
\caption{
\label{fig: DOS CII broken by E}
(a) Global density of states (DOS) of a two-dimensional  
disorder-free tight-binding model
with sublattice symmetry and an even number equal to
or larger than
2 of non-equivalent discrete Fermi points
at the band center. The dashed line is the 
global DOS of the corresponding
disorder-free Dirac fermions in two dimensions.
(b) Effect of weak disorder for symmetry class AII
without the $\mathbb{Z}^{\ }_{2}$ topological term
(the surface states of a three-dimensional time-reversal-symmetric 
weak topological insulator),
the symmetry class relevant to the quadrant CII when
perturbed by a finite chemical potential.
The band center is a critical energy at which,
according to Eq.~(\ref{eq: diverging DOS AII}),
the global DOS diverges.
This critical energy separates two metallic phases.
       }
\end{center}
\end{figure}

\subsection{
Weak breaking of the chiral symmetry in the vicinity of the band center
           }

The effects of a finite Fermi energy $E^{\ }_{\mathrm{F}}$ 
on the physics of localization 
for the quadrants BDI and CII in Fig.~\ref{fig: phase diagram} 
are dramatic in that, in both cases, 
a finite $E^{\ }_{\mathrm{F}}$ 
breaks the chiral symmetry chS.

Turning on a  finite Fermi energy $E^{\ }_{\mathrm{F}}$ 
in the quadrant BDI in Fig.~\ref{fig: phase diagram}
reduces the symmetry class to AI.
All states at finite $E^{\ }_{\mathrm{F}}$ are then localized.%
~\cite{Evers08}
The band center is a quantum critical point separating two
insulating phases, one defined by 
$E^{\ }_{\mathrm{F}}<0$ 
and another one defined by 
$E^{\ }_{\mathrm{F}}>0$,
very much as is the case in the integer quantum Hall effect (IQHE)
(see Ref.~\onlinecite{Evers08} for a review on
plateau transitions in the IQHE).
The global density of states $\nu(E^{\ }_{\mathrm{F}})$
diverges as
\begin{equation}
\nu(E^{\ }_{\mathrm{F}})\sim
\frac{1}{|E^{\ }_{\mathrm{F}}|}
\exp\left(-c|\ln|E^{\ }_{\mathrm{F}}||^{2/3}\right)
\end{equation}
with $c$ a non-universal number when 
$E^{\ }_{\mathrm{F}}$ approaches the band center.%
~\cite{Mudry03}

Turning on a finite Fermi energy $E^{\ }_{\mathrm{F}}$ 
in the quadrant CII in Fig.~\ref{fig: phase diagram}
reduces the symmetry class to AII but
\textit{without} the $\mathbb{Z}^{\ }_{2}$ topological term.
Indeed, the random Dirac Hamiltonian at a finite
chemical potential has now \textit{two flavors}
that are coupled by the disorder.
This corresponds to two Dirac cones
in any underlying microscopic model that are 
generically coupled by the disorder.
The global density of states $\nu(E^{\ }_{\mathrm{F}})$
is again diverging according to the law
\begin{equation}
\nu(E^{\ }_{\mathrm{F}})\sim
\frac{1}{|E^{\ }_{\mathrm{F}}|}
\exp\left(-c'|\ln|E^{\ }_{\mathrm{F}}||^{2/3}\right)
\label{eq: diverging DOS AII}
\end{equation}
with $c'$ a non-universal number when 
$E^{\ }_{\mathrm{F}}$ approaches the band center.%
~\cite{Mudry03}
The state at the band center is critical.
(The robustness to strong disorder
of the critical behavior of the band center
in the chiral classes is well documented in 
quasi-one and two dimensions.%
~\cite{Motrunich02})
However, contrary to the quadrant BDI in Fig~\ref{fig: phase diagram},
the band center is not any more a quantum critical point separating
two insulating phases. Indeed, the localized nature as a function
of the chemical potential of these two-dimensional states 
is that of the surface states of 
a three-dimensional time-reversal-symmetric weak topological insulator.
The issue of Anderson localization as a function of the chemical
potential for such surface states
was recently discussed in Refs.~\onlinecite{Ringel11} and 
\onlinecite{Mong11}.
According to the numerical study in Ref.~\onlinecite{Mong11}
(corresponding to the case of a mean value $\bar{m}=0$ of the random mass 
$m$ in Ref.~\onlinecite{Mong11}), 
these surface states remain extended (metallic)
in the presence of disorder even though the characteristic energy 
at which the upturn of the diverging global DOS becomes sizable relative 
to the clean DOS shown in Fig.~\ref{fig: DOS CII broken by E}(a)
is exponentially small for weak disorder.%
~\cite{Mudry03}

\section*{Acknowledgments}

This work has been supported by the National Science Foundation (NSF) under
Grant No.\ PHY05-51164 
and in part by the NSF under DMR-0706140 (AWWL)
and by a Grant-in-Aid for Scientific Research from the 
Japan Society for the Promotion of Science
(Grant No.\ 21540332).
AWWL thanks the organizers of the workshop
"Workshop on Applied 2D Sigma Models",
held at DESY (Hamburg/GERMANY), November 10-14, 2008,
for the opportunity to present
the results of the work reported in the present paper to
an interdisciplinary audience.
SR, CM, and AF are grateful to the Kavli Institute for Theoretical Physics
for its hospitality, where this paper was completed.
We thank P.\ M.\ Ostrovsky and A.\ D.\ Mirlin for 
discussions on Anderson localization in the ``Chiral''
symmetry classes.

\appendix

\section{
The sign ambiguity of a Pfaffian
        }
\label{appsec: The sign ambiguity of a Pfaffian}

In this section, we are going to argue
that the dashed line in the
phase diagram of Fig.~\ref{fig: phase diagram}
that belongs to the chiral symplectic class CII
has the particularity that, 
within the fermionic replica NL$\sigma$M representation,
there appears a $\mathbb{Z}^{\ }_2$-topological term
in addition to the standard kinetic energy.

To this end, it will be useful to enlarge the dimensionality
of the representation of the Dirac Hamiltonian
by a factor of 2 in order to treat the isospin-1/2 TRS.
To avoid ambiguities, we will use the Greek letters for the Pauli
matrices acting on the three relevant two-dimensional subspaces
-- $\rho$ in flavor subspace, 
$\sigma$ in Lorentz subspace,
and $\tau$ in the time-reversal subspace to be introduced below -- 
as subindices to specify the chosen representations.
In this section we use indices $x,y,z$, instead of $1,2,3$,
in the $\sigma$ and $\tau$ subspaces.
For example, we shall denote the
Dirac Hamiltonian~(\ref{eq: chiral random dirac hamiltonian}) 
when Eq.~(\ref{eq: def HWK CII}) holds by
\begin{subequations}
\label{eq: conventions sec CII}
\begin{equation}
\begin{split}
&
\mathcal{H}^{\ }_{\rho,\sigma}:=
\begin{pmatrix}
0 
& 
D^{\ }_{\sigma} 
\\
D^{\dag}_{\sigma} 
& 
0
\end{pmatrix}^{\ }_{\rho},
\\
&
D^{\ }_{\sigma}:=
\sigma^{\ }_{x}
\left(
-{i} \partial^{\ }_x + A^{\ }_x
\right)
+
\sigma^{\ }_{y}
\left(
-{i} \partial^{\ }_y + A^{\ }_y
\right)
\\
&\hphantom{D^{\ }_{\sigma}:=}
+ 
\sigma^{\ }_{z}\,M^{\ }_z 
+ 
\sigma^{\ }_0\,M^{\ }_{0},
\label{eq: random Dirac Hamiltonian}
\end{split}
\end{equation}
where
$A^{\ }_{\mu}=-{i}a'_{\mu} \in {i}\mathbb{R}$,
$M^{\ }_z = -{i}m'_z \in {i}\mathbb{R}$,
$M^{\ }_0 = m'_0 \in \mathbb{R}$,
and with the simultaneous chS
\begin{equation}
(\rho^{\ }_{z}\otimes\sigma^{\ }_{0})
\mathcal{H}^{\ }_{\rho,\sigma}
(\rho^{\ }_{z}\otimes\sigma^{\ }_{0})=
-\mathcal{H}^{\ }_{\rho,\sigma}
\label{eq: SLS}
\end{equation}
and isospin-1/2 TRS
\begin{equation}
({i}\rho^{\ }_{0}\otimes\sigma^{\ }_{y}) 
\mathcal{H}^{T}_{\rho,\sigma} 
(-{i}\rho^{\ }_{0}\otimes\sigma^{\ }_{y})=
\mathcal{H}^{\ }_{\rho,\sigma}.
\label{eq: TRS}
\end{equation}
\end{subequations}

\subsection{
Fermionic functional integral representation of 
the retarded Green's function
           }

The generating function for 
the retarded Green's function
is the partition function
\begin{subequations}
\begin{equation}
\begin{split}
Z:=&\,
\int\mathcal{D}\left[\bar{\chi},\chi\right]
\exp
\left(
-
\int \mathrm{d}^2\,r\,
\mathcal{L}
\right),
\\ 
\mathcal{L}:=&
-
{i} 
\bar{\chi}
({i}\eta -\mathcal{H})^{\ }_{\rho,\sigma}
\chi.
\end{split}
\end{equation}
Here, we have chosen 
\begin{equation}
\bar{\chi}\equiv
\begin{pmatrix}
\bar{\chi}^{\ }_{1}
&
\bar{\chi}^{\ }_{2}
\end{pmatrix}^{\ }_{\rho}
\equiv
\begin{pmatrix}
\bar{\chi}^{\ }_{1\uparrow}
&
\bar{\chi}^{\ }_{1\downarrow}
&
\bar{\chi}^{\ }_{2\uparrow}
&
\bar{\chi}^{\ }_{2\downarrow}
\end{pmatrix}^{\ }_{\rho,\sigma}
\end{equation}
to be a 4-component row spinor with Grassmann-valued entries.
Similarly,
\begin{equation}
\chi\equiv
\begin{pmatrix}
\chi^{\ }_{1}
\\
\chi^{\ }_{2}
\end{pmatrix}^{\ }_{\rho}
\equiv
\begin{pmatrix}
\chi^{\ }_{1\uparrow}
\\
\chi^{\ }_{1\downarrow}
\\
\chi^{\ }_{2\uparrow}
\\
\chi^{\ }_{2\downarrow}
\end{pmatrix}^{\ }_{\rho,\sigma}
\end{equation}
\end{subequations}
is a 4-component column spinor with 
Grassmann-valued entries.
All 8 Grassmann-valued entries labeled by the 
flavor indices $1$ and $2$ 
on which the matrices 
$(\rho^{\ }_{0},\rho^{\ }_{x},\rho^{\ }_{y},\rho^{\ }_{z})$
act and by the Lorentz
indices $\uparrow$ and $\downarrow$
on which the matrices 
$(\sigma^{\ }_{0},\sigma^{\ }_{x},\sigma^{\ }_{y},\sigma^{\ }_{z})$
act are independent. 
For the retarded Green's function, $\eta>0$.

It is useful to make the TRS (\ref{eq: TRS}) explicit. 
To this end, following Ref.~\onlinecite{Efetov80},
we make the manipulation
\begin{subequations}
\label{eq: tau sigma rep}
\begin{equation}
\begin{split}
\mathcal{L}=&\,
-{i}\,
\bar{\chi}
({i}\eta -\mathcal{H})^{\ }_{\rho,\sigma}
\chi
\\
=&\,
+{i}\,
\chi^{T}
({i}\eta -\mathcal{H})^{T}_{\rho,\sigma}
\bar\chi^{T}
\\
=&\,
-{i}\,
\chi^T 
(-{i}\rho^{\ }_{0}\otimes\sigma^{\ }_y)
({i}\eta -\mathcal{H})^{\ }_{\rho,\sigma}
(-{i}\rho^{\ }_{0}\otimes\sigma^{\ }_y \bar{\chi}^T)
\\
=&\,
-{i}\,
\bar{\Psi}
({i}\eta -\mathcal{H})^{\ }_{\rho,\tau,\sigma}
\Psi
\end{split}
\end{equation}
by which we have doubled the number of Grassmann-valued entries
in $\bar\Psi$ and $\Psi$ through the definitions
\begin{equation}
\begin{split}
({i}\eta -\mathcal{H})^{\ }_{\rho,\tau,\sigma}
:=&\,
({i}\eta -\mathcal{H})^{\ }_{\rho,\sigma}
\otimes\tau^{\ }_{0},
\end{split}
\label{eq: def kernel in rho, tau, sigma rep}
\end{equation}
and
\begin{equation}
\begin{split}
\bar{\Psi}:=&\,
\frac{1}{\sqrt{2}}
\left(
\begin{array}{cccc}
\bar{\chi}^{\ }_{\uparrow}
&
\bar{\chi}^{\ }_{\downarrow}
&
\chi^{T}_{\downarrow}
&
-
\chi^{T}_{\uparrow}
\end{array}
\right)^{\ }_{\tau,\sigma},
\\
\Psi:=&\,
\frac{1}{\sqrt{2}}
\left(
\begin{array}{c}
\chi^{\ }_{\uparrow} 
\\
\chi^{\ }_{\downarrow} 
\\
-\bar{\chi}^{T}_{\downarrow} 
\\
\bar{\chi}^{T}_{\uparrow} 
\end{array}
\right)^{\ }_{\tau,\sigma}.
\end{split}
\end{equation}
\label{eq: def TRS grading}
\end{subequations}

\noindent
Here, the subindex $\tau$ denotes the, by now, 
explicit time-reversal subspace
that is spanned by
the unit $2\times2$ matrix $\tau^{\ }_{0}$
and the three Pauli matrices
$(\tau^{\ }_{1} ,\tau^{\ }_{2},\tau^{\ }_{3})$.  
Of course, the number of independent Grassmann-valued
entries remains unchanged in the
representation~(\ref{eq: def TRS grading})
as the TRS~(\ref{eq: TRS}) is now represented by the constraint
\begin{equation}
\bar{\Psi}=
\Psi^{T}
(-{i}\rho^{\ }_{0}\otimes\tau^{\ }_{x}\otimes\sigma^{\ }_{y}).
\label{eq: TRS bis}
\end{equation}
On the other hand, the representation of the chS~(\ref{eq: SLS}) 
is
\begin{equation}
\rho^{\ }_z\otimes\tau^{\ }_{0}\otimes\sigma^{\ }_{0}  
\mathcal{H}^{\ }_{\rho,\tau,\sigma}
\rho^{\ }_z\otimes\tau^{\ }_{0}\otimes\sigma^{\ }_{0}   
=
-\mathcal{H}^{\ }_{\rho,\tau,\sigma}.
\label{eq: SLS bis}
\end{equation}

Instead of Eq.~(\ref{eq: TRS bis}),
we seek a representation of the TRS
in terms of 8-component Grassmann-valued spinors obeying
the Majorana constraint 
\begin{equation}
\bar{\psi}=
\psi^{T}
(-{i}\rho^{\ }_{0}\otimes\tau^{\ }_{0}\otimes\sigma^{\ }_{y}).
\label{eq: majorana constraint}
\end{equation}
This can be achieved by observing that the ``square root'' of
$\tau^{\ }_{x}$ is given by
\begin{equation}
\tau^{\ }_{x}=
-{i}\,
\tau^{T }_{z-y}
\tau^{\ }_{z-y},
\qquad
\tau^{\ }_{z-y}:=
\frac{\tau^{\ }_{z}-\tau^{\ }_{y}}{\sqrt{2}}.
\end{equation}
Now, we take advantage of the fact that
the kernel~(\ref{eq: def kernel in rho, tau, sigma rep})
commutes with 
\begin{equation}
T^{\ }_{z-y}:=
\rho^{\ }_{0}\otimes\tau^{\ }_{z-y}\otimes\sigma^{\ }_{0}
\end{equation}
so that
\begin{subequations}
\begin{equation}
\begin{split}
\mathcal{L}=&\,
-{i}
\bar{\Psi}
({i}\eta -\mathcal{H})^{\ }_{\rho,\tau,\sigma}
\Psi
\\
=&\,
{\Psi}^{T}
T^{T }_{z-y}
({i}\rho^{\ }_{0}\otimes\tau^{\ }_{0}\otimes\sigma^{\ }_{y})
({i}\eta -\mathcal{H})^{\ }_{\rho,\tau,\sigma}
T^{\ }_{z-y}
\Psi
\\
\equiv&\,
-
\bar{\psi}
({i}\eta -\mathcal{H})^{\ }_{\rho,\tau,\sigma}
\psi
\end{split}
\end{equation}
where
\begin{equation}
\psi:=
T^{\ }_{z-y}
\Psi
\end{equation}
\end{subequations}
determines $\bar{\psi}$ through the Majorana constraint 
Eq.~(\ref{eq: majorana constraint})
that follows because of the isospin-1/2 TRS.
In view of the Majorana constraint~(\ref{eq: majorana constraint}),
the isospin-1/2
TRS is now equivalent to the global $\mathrm{O}(2)$ invariance 
under the transformation
\begin{equation}
\bar{\psi} \to 
\bar{\psi}\, (\rho^{\ }_{0}\otimes\sigma^{\ }_{0}\otimes O^{T }_{\tau}),
\qquad
\psi \to
(\rho^{\ }_{0}\otimes\sigma^{\ }_{0}\otimes O^{\ }_{\tau})\, \psi
\label{eq: invariance under O(2)}
\end{equation}
for any $2\times 2$ orthogonal matrix
$O^{\ }_{\tau}$
acting in the $\tau$ subspace.

Finally, it is time to make use of the chS~(\ref{eq: SLS}).
By making the flavor subspace explicit,
\begin{subequations}
\label{eq: making SLS explicit}
\begin{equation}
\begin{split}
\mathcal{L}=&
\bar{\psi}^{\ }_{1} D^{\   }_{\tau,\sigma} \psi^{\ }_{2}
+
\bar{\psi}^{\ }_{2} D^{\dag}_{\tau,\sigma} \psi^{\ }_{1}
\\
&\,
-{i}\eta
\left(
\bar{\psi}^{\ }_{1} \psi^{\ }_{1}
+
\bar{\psi}^{\ }_{2} \psi^{\ }_{2}
\right),
\end{split}
\end{equation}
where ($a^{\prime}_{\mu},m^{\prime}_{z},m^{\prime}_{0}\in\mathbb{R}$)
\begin{equation}
\begin{split}
&
D^{\ }_{\tau,\sigma}:= 
\tau^{\ }_{0}\otimes
\left(
-{i}\sigma^{\ }_{\mu}\partial^{\ }_{\mu} 
+ V
\right),
\\
&
V:= 
{i}\sigma^{\ }_{\mu}a^{\prime}_{\mu}
-
{i}\sigma^{\ }_{z  }m^{\prime}_{z  }
+ 
   \sigma^{\ }_{0  }m^{\prime}_{0  },
\end{split}
\end{equation}
acts on the two independent 4-component Grassmann-valued
spinors $\psi^{\ }_{1}$ and $\psi^{\ }_{2}$ while the spinors
$\bar{\psi}^{\ }_{1}$ and $\bar{\psi}^{\ }_{2}$ 
obey the Majorana condition
($\Sigma^{\ }_{y}:=\tau^{\ }_{0}\otimes\sigma^{\ }_{y}$)
\begin{equation}
\bar{\psi}^{\ }_{1}=
\psi^{T}_{1}(-{i}\Sigma^{\ }_{y}),
\qquad
\bar{\psi}^{\ }_{2}=
\psi^{T}_{2}(-{i}\Sigma^{\ }_{y}).
\label{eq: final Majorana cond}
\end{equation}
\end{subequations}\noindent
With the help of the identity
\begin{equation}
\bar{\psi}^{\ }_{2}D^{\dag}_{\tau,\sigma}\psi^{\ }_{1}=
-\psi^T_1 D^*_{\tau,\sigma}\bar\psi^T_2=
\bar{\psi}^{\ }_{1}
D^{\ }_{\tau,\sigma}
\psi^{\ }_{2},
\end{equation}
we arrive at
\begin{equation}
\mathcal{L} =
2\,
\bar{\psi}^{\ }_{1} D^{\ }_{\tau,\sigma}\psi^{\ }_{2}
-{i}\eta
\left(
\bar{\psi}^{\ }_{1} \psi^{\ }_{1}
+
\bar{\psi}^{\ }_{2} \psi^{\ }_{2}
\right).
\label{eq: final rep cal L}
\end{equation}

This presentation of the Lagrangian
reveals that, upon quantization,
$\bar{\psi}^{\ }_{1}$ and $\psi^{\ }_{2}$
form a canonical pair of fermionic operators.
In other words, because of the chS,  the kinetic part 
$\bar{\psi}^{\ }_{1} D^{\ }_{\tau,\sigma}\psi^{\ }_{2}$
of the Lagrangian is invariant under any global 
$\mathrm{U}(2)$ transformation 
\begin{equation}
\bar{\psi}^{\ }_{1}\to 
\bar{\psi}^{\ }_{1} (\sigma^{\ }_{0}\otimes U^{\dag }_{\tau}),
\qquad
\psi^{\ }_{2}\to
(\sigma^{\ }_{0}\otimes U^{\ }_{\tau})\, \psi^{\ }_{2}, 
\label{eq: U(2) global symmetry}
\end{equation}
where
$U^{\ }_{\tau}$ is a $2\times 2$ unitary matrix
acting in the $\tau$ subspace.

\subsection{
Replicas and disorder averaging
           }

We now assume that
$a^{\prime}_{x}$,
$a^{\prime}_{y}$,
$m^{\prime}_z$,
and
$m^{\prime}_0$ 
from Eq.~(\ref{eq: making SLS explicit})
are all white-noise distributed 
with the \textit{same} variance $g$.
In doing so, we limit ourselves to the 
nearly-critical line of region CII
in Fig.~\ref{fig: phase diagram}.

We replicate the Lagrangian $\mathsf{N}^{\ }_{\mathrm{r}}$ times,
\begin{equation}
\begin{split}
\mathcal{L}^{\ }_{\mathsf{N}^{\ }_{\mathrm{r}}}=&\,
\sum_{\mathsf{a}=1}^{2\mathsf{N}^{\ }_{\mathrm{r}}}
\big[
2
\bar{\psi}^{\ }_{\mathsf{a}1}
(-i\sigma_\mu\partial_\mu+V)
\psi^{\ }_{\mathsf{a}2}
\\
&\,
-
{i}\eta
\left(
\bar{\psi}^{\ }_{\mathsf{a}1} \psi^{\ }_{\mathsf{a}1}
+
\bar{\psi}^{\ }_{\mathsf{a}2} \psi^{\ }_{\mathsf{a}2}
\right)
\big].
\end{split}
\end{equation}
This Lagrangian is invariant under any global
$\mathrm{O}(2\mathsf{N}^{\ }_{\mathrm{r}})$ rotation in the 
$\tau$ and replica subspaces.
After disorder averaging has been performed,
we arrive at the interacting Lagrangian
\begin{subequations}
\begin{equation}
\begin{split}
\mathcal{L}^{\ }_{\mathsf{N}^{\ }_{\mathrm{r}}}=&\,
2\sum_{\mathsf{a}=1}^{2\mathsf{N}^{\ }_{\mathrm{r}}}
\boldsymbol{d}^{\dag}_{\mathsf{a}}
\left(
-{i}\sigma^{\ }_{\mu}\partial^{\ }_{\mu}
\right)
\boldsymbol{d}^{\ }_{\mathsf{a}}
\\
&\,
+{i}\eta
\sum_{\mathsf{a}=1}^{2\mathsf{N}^{\ }_{\mathrm{r}}}
\left[
\boldsymbol{d}^{\dag}_{\mathsf{a}}
{i}\sigma^{\ }_{y}
(
\boldsymbol{d}^{\dag}_{\mathsf{a}}
)^{T}
+
\boldsymbol{d}^{T }_{\mathsf{a}} 
{i}\sigma^{\ }_{y} 
\boldsymbol{d}^{\ }_{\mathsf{a}}
\right]
\\
&\,
+8g
\sum_{\mathsf{a},\mathsf{b}=1}^{2\mathsf{N}^{\ }_{\mathrm{r}}}
\left(
\vec{S}^{\ }_{\mathsf{a}} 
\cdot
\vec{S}^{\ }_{\mathsf{b}}
-
\frac{1}{4}
n^{\ }_{\mathsf{a}} 
n^{\ }_{\mathsf{b}}
\right).
\end{split}
\label{eq: t-J Langrangian}
\end{equation}
Here,
since $\bar{\psi}^{\ }_{1}$ and $\psi^{\ }_{2}$ are canonically conjugate,
we have introduced the following notation for any 
$\mathsf{a}=1,\cdots,2\mathsf{N}^{\ }_{\mathrm{r}}$,
\begin{equation}
\begin{split}
& 
\boldsymbol{d}^{\dag}_{\mathsf{a}}:=
\bar{\psi}^{\ }_{ \mathsf{a}1},
\qquad
\boldsymbol{d}^{\ }_{\mathsf{a}}:=
\psi^{\ }_{ \mathsf{a}2},
\\
&
\vec{S}^{\ }_{\mathsf{a}}:=
\frac{1}{2}
\boldsymbol{d}^{\dag}_{\mathsf{a}}
\vec{\sigma}
\boldsymbol{d}^{\ }_{\mathsf{a}},
\qquad
n^{\ }_{\mathsf{a}}:=
\boldsymbol{d}^{\dag}_{\mathsf{a}}
\boldsymbol{d}^{\ }_{\mathsf{a}}.
\end{split}
\end{equation}
\end{subequations}
It is worth remembering that the replicated ``spin''
$\vec{S}^{\ }_{\mathsf{a}}$
in the $t$-$J$-like Lagrangian
(\ref{eq: t-J Langrangian})
originates from the $\sigma$ subspace
and not the true electronic spin-1/2. 
When $\eta=0$ and in accordance with the
global symmetry~(\ref{eq: U(2) global symmetry}), 
the action is invariant under 
any global $\mathrm{U}(2\mathsf{N}^{\ }_{\mathrm{r}})$ rotation 
\begin{equation}
\boldsymbol{d}^{\dag}_{\mathsf{a}}
\to 
\boldsymbol{d}^{\dag}_{\mathsf{b}}
U^{*}_{\mathsf{a}\mathsf{b}},
\qquad
\boldsymbol{d}^{\ }_{\mathsf{a}} 
\to 
U^{\ }_{\mathsf{a}\mathsf{c}} \boldsymbol{d}^{\ }_{\mathsf{c}},
\qquad
U^{*}_{\mathsf{a}\mathsf{b}} 
U^{\ }_{\mathsf{a}\mathsf{c}} =\delta_{\mathsf{b}\mathsf{c}},
\label{eq: U(2N) sym}
\end{equation}
while,
in accordance with the
global symmetry~(\ref{eq: invariance under O(2)}),
any non-zero $\eta$ breaks this symmetry down to 
the global $\mathrm{O}(2\mathsf{N}^{\ }_{\mathrm{r}})$ rotation 
\begin{equation}
\boldsymbol{d}^{T}_{\mathsf{a}}
\to 
\boldsymbol{d}^{T}_{\mathsf{b}}
O^{\ }_{\mathsf{a}\mathsf{b}},
\qquad
\boldsymbol{d}^{\ }_{\mathsf{a}} 
\to 
O^{\ }_{\mathsf{a}\mathsf{c}} \boldsymbol{d}^{\ }_{\mathsf{c}},
\qquad
O^{\ }_{\mathsf{a}\mathsf{b}} 
O^{\ }_{\mathsf{a}\mathsf{c}}=\delta_{\mathsf{b}\mathsf{c}},
\label{eq: O(2N) sym}
\end{equation}
where summation over repeated indices is assumed.

\subsection{Hubbard-Stratonovich transformation}

It is time to introduce auxiliary (Hubbard-Stratonovich) 
fields that decouple the interactions among replicas.
A possible channel for decoupling is singlet superconductivity
as it is favored by the symmetry breaking term $\eta$.
Hence for any $\mathsf{a},\mathsf{b}=1,\cdots,2\mathsf{N}^{\ }_{\mathrm{r}}$, 
we introduce the order parameters
\begin{equation}
\begin{split}
\mathcal{O}^{\dag}_{\mathsf{a}\mathsf{b}}:=&\,
\frac{-1}{\sqrt{2}}
\boldsymbol{d}^{\dag}_{\mathsf{a}}
{i}\sigma^{\ }_y
(\boldsymbol{d}^{\dag}_{\mathsf{b}})^{T}
\\
=&\,
\frac{-1}{\sqrt{2}}
\left(
d^{\dag}_{\mathsf{a}\uparrow}
d^{\dag}_{\mathsf{b}\downarrow}
-
d^{\dag}_{\mathsf{a}\downarrow}
d^{\dag}_{\mathsf{b}\uparrow}
\right),
\\
\mathcal{O}^{\ }_{\mathsf{a}\mathsf{b}}:=&\,
\frac{1}{\sqrt{2}}
\boldsymbol{d}^T_{\mathsf{a}}
{i}\sigma^{\ }_y
\boldsymbol{d}^{\ }_{\mathsf{b}}
\\
=&\,
\frac{1}{\sqrt{2}}
\left(
d^{\ }_{\mathsf{a}\uparrow}
d^{\ }_{\mathsf{b}\downarrow}
-
d^{\ }_{\mathsf{a}\downarrow}
d^{\ }_{\mathsf{b}\uparrow}
\right),
\end{split}
\end{equation}
in terms of which the ``exchange term'' becomes
\begin{eqnarray}
\vec{S}^{\ }_{\mathsf{a}}\cdot \vec{S}^{\ }_{\mathsf{b}}
-\frac{1}{4} n^{\ }_{\mathsf{a}} n^{\ }_{\mathsf{b}}
=
-
\mathcal{O}^{\dag}_{\mathsf{a}\mathsf{b}}
\mathcal{O}^{\ }_{\mathsf{a}\mathsf{b}},
\end{eqnarray}
and, in turn, the Lagrangian becomes
\begin{equation}
\begin{split}
\mathcal{L}^{\ }_{\mathsf{N}^{\ }_{\mathrm{r}}}=&\,
2
\sum_{\mathsf{a}=1}^{2\mathsf{N}^{\ }_{\mathrm{r}}}
\boldsymbol{d}^{\dag}_{\mathsf{a}}
\left(
-{i}\sigma^{\ }_{\mu}\partial^{\ }_{\mu}
\right)
\boldsymbol{d}^{\ }_{\mathsf{a}}
\\
&\,
+
{i}\sqrt{2}\,\eta 
\sum_{\mathsf{a}=1}^{2\mathsf{N}^{\ }_{\mathrm{r}}}
\left(
\mathcal{O}^{\ }_{\mathsf{a}\mathsf{a}}
-
\mathcal{O}^{\dag}_{\mathsf{a}\mathsf{a}}
\right)
\\
&\,
-8g \sum_{\mathsf{a},\mathsf{b}=1}^{2\mathsf{N}^{\ }_{\mathrm{r}}}
\mathcal{O}^{\dag}_{\mathsf{a}\mathsf{b}}
\mathcal{O}^{\ }_{\mathsf{a}\mathsf{b}}.
\end{split}
\end{equation}
The interacting term is then decoupled by
the $2\mathsf{N}^{\ }_{\mathrm{r}}\times 2\mathsf{N}^{\ }_{\mathrm{r}}$ 
Hubbard-Stratonovich field 
$\Delta^{\ }_{\mathsf{a}\mathsf{b}}$ 
and its complex conjugate
$\Delta^{* }_{\mathsf{a}\mathsf{b}}$,
\begin{equation}
\begin{split}
\mathcal{L}^{\ }_{\mathsf{N}^{\ }_{\mathrm{r}}}=&\,
2\sum_{\mathsf{a}=1}^{2\mathsf{N}^{\ }_{\mathrm{r}}}
\boldsymbol{d}^{\dag}_{\mathsf{a}}
\left(
-{i}\sigma^{\ }_{\mu}\partial^{\ }_{\mu}
\right)
\boldsymbol{d}^{\ }_{\mathsf{a}}
\\
&\,
+
{i}\sqrt{2}\,\eta 
\sum_{\mathsf{a}=1}^{2\mathsf{N}^{\ }_{\mathrm{r}}}
\left(
\mathcal{O}^{\ }_{\mathsf{a}\mathsf{a}}
-
\mathcal{O}^{\dag}_{\mathsf{a}\mathsf{a}}
\right)
\\
&\,
+
\sum_{\mathsf{a},\mathsf{b}=1}^{2\mathsf{N}^{\ }_{\mathrm{r}}}\left(
\frac{1}{8g}
\Delta^*_{\mathsf{b}\mathsf{a}}
\Delta^{\ }_{\mathsf{b}\mathsf{a}}
-
\mathcal{O}^{\dag}_{\mathsf{a}\mathsf{b}} 
\Delta^{\ }_{\mathsf{b}\mathsf{a}}
-
\mathcal{O}^{\ }_{\mathsf{a}\mathsf{b}} 
\Delta^{*}_{\mathsf{b}\mathsf{a}}
\right).
\end{split}
\label{eq: lagrantian with HS fields}
\end{equation}

No approximation has yet been invoked. 
As the interacting Lagrangian 
(\ref{eq: lagrantian with HS fields})
is not readily tractable,
we shall restrict the path integral to
slowly varying bosonic degrees of freedom 
(Nambu-Goldstone bosons). 
We first look for a diffusive saddle point
of the Lagrangian (\ref{eq: lagrantian with HS fields}).
In the diffusive regime, 
the auxiliary field $\Delta^{\ }_{\mathsf{a}\mathsf{b}}$ (spontaneously)
breaks the global $\mathrm{U}(2\mathsf{N}^{\ }_{\mathrm{r}})$ symmetry, 
along the symmetry breaking ``direction''
controlled by the symmetry breaking term $\eta$.
Thus, the spatially homogeneous configuration
\begin{eqnarray}
\Delta^{\ }_{0\mathsf{a}\mathsf{b}}=
-{i}
\left|\Delta^{\ }_{0}\right| \delta^{\ }_{\mathsf{a}\mathsf{b}}
\end{eqnarray}
should be a representative diffusive saddle point,
where $\left|\Delta^{\ }_{0}\right| \in \mathbb{R}$ is determined from the 
self-consistent equation 
\begin{equation}
\ln  
\left[
1
+ 
\left(\frac{\Lambda}{\left|\Delta^{\ }_{0}\right|}\right)^2 
\right]
=
\frac{\pi}{2g}
\end{equation}
and $\Lambda$ is an ultra-violet cutoff.

This choice of a saddle point is not exhaustive. 
Generic saddle points can be constructed by making use of 
the global $\mathrm{U}(2\mathsf{N}^{\ }_{\mathrm{r}})$ symmetry%
~(\ref{eq: U(2N) sym})
of the kinetic energy:
\begin{equation}
\Delta^{\ }_{\mathsf{a}\mathsf{b}}=
\sum^{2\mathsf{N}^{\ }_{\mathrm{r}}}_{\mathsf{p},\mathsf{q}=1}
U^{\ }_{\mathsf{a}\mathsf{p}}
\Delta^{\ }_{0\mathsf{p}\mathsf{q}}
U^{\ }_{\mathsf{b}\mathsf{q}}
=
-{i}\left|\Delta^{\ }_{0}\right|
\sum^{2\mathsf{N}^{\ }_{\mathrm{r}}}_{\mathsf{p}=1}
U^{\ }_{\mathsf{a}\mathsf{p}}
U^{\ }_{\mathsf{b}\mathsf{p}}
\label{eq: generic saddle points}
\end{equation}
where $U\in \mathrm{U}(2\mathsf{N}^{\ }_{\mathrm{r}})$.
However, 
not all $U \in \mathrm{U}(2\mathsf{N}^{\ }_{\mathrm{r}})$ 
generate a new saddle point configuration.
If $U\in \mathrm{O}(2\mathsf{N}^{\ }_{\mathrm{r}})$, 
$\Delta^{\ }_{\mathsf{a}\mathsf{b}}$ 
coincides with the reference configuration 
$\Delta^{\ }_{0\mathsf{a}\mathsf{b}}$
owing to the global 
$\mathrm{O}(2\mathsf{N}^{\ }_{\mathrm{r}})$ symmetry~(\ref{eq: O(2N) sym}).
This means that the set of saddle points $\Delta^{\ }_{\mathsf{a}\mathsf{b}}$ 
is the coset manifold 
\begin{subequations}
\begin{equation}
G/H = \mathrm{U}(2\mathsf{N}^{\ }_{\mathrm{r}})/\mathrm{O}(2\mathsf{N}^{\ }_{\mathrm{r}}),
\end{equation} 
whose elements can be parametrized by 
\begin{equation}
U U^T,
\qquad
U\in \mathrm{U}(2\mathsf{N}^{\ }_{\mathrm{r}}).
\end{equation}
\end{subequations}
Note that since $U U^T$ is symmetric and unitary,
$\mathrm{U}(2\mathsf{N}^{\ }_{\mathrm{r}})/\mathrm{O}(2\mathsf{N}^{\ }_{\mathrm{r}})$ 
is a set of symmetric and unitary matrices.

We now include fluctuations around 
the saddle points~(\ref{eq: generic saddle points}).
Since the longitudinal fluctuations (i.e., fluctuations
that changes $\left|\Delta^{\ }_{0}\right|$) are gaped, 
we shall freeze them and only consider
the transverse fluctuations
\begin{equation}
\Delta^{\ }_{\mathsf{a}\mathsf{b}}(\boldsymbol{r})
=
-{i}\left|\Delta^{\ }_{0}\right|
\sum^{2\mathsf{N}^{\ }_{\mathrm{r}}}_{\mathsf{p}=1}
U^{\ }_{\mathsf{a}\mathsf{p}}(\boldsymbol{r})
U^{\ }_{\mathsf{b}\mathsf{p}}(\boldsymbol{r})
\label{eq: generic background}
\end{equation}
with $U(\boldsymbol{r})\in \mathrm{U}(2\mathsf{N}^{\ }_{\mathrm{r}})$.
With the help of the Nambu representation,
the effective Lagrangian becomes
\begin{subequations}
\label{eq: L eff and majorana cond}
\begin{equation}
\mathcal{L}^{\ }_{\mathrm{eff}}
=
\sum_{\mathsf{a},\mathsf{b}=1}^{2\mathsf{N}^{\ }_{\mathrm{r}}}
\bar{\gamma}^{\ }_{\mathsf{a}}
D^{\ }_{\mathsf{a}\mathsf{b}}[\Delta]
\gamma^{\ }_{\mathsf{b}},
\end{equation}
where 
\begin{equation}
\bar{\gamma}^{\ }_{\mathsf{a}}
=
\left(
 \begin{array}{cc}
 \boldsymbol{d}^{\dag}
 & 
 \boldsymbol{d}^T
 (-{i}\sigma^{\ }_{y})
 \end{array}
 \right)^{\ }_{\mathsf{a}},
\quad
\gamma^{\ }_{\mathsf{a}}=
\left(
 \begin{array}{c}
 \boldsymbol{d} \\
 {i}\sigma^{\ }_y (\boldsymbol{d}^{\dag})^T \\
 \end{array}
 \right)^{\ }_{\mathsf{\mathsf{a}}}
\end{equation}
are related by the Majorana condition
\begin{equation}
\bar{\gamma}^{\ }_{\mathsf{a}}=
(
-{i}\sigma^{\ }_y \otimes i\tau^{\ }_{y}
\gamma^{\ }_{\mathsf{a}}
)^T
\label{eq: final Majorana}
\end{equation}
with $\tau^{\ }_{y}$ acting in the Nambu space, and
the kernel is
\begin{equation}
D^{\ }_{\mathsf{a}\mathsf{b}}[\Delta]=
\left(
\begin{array}{cc}
\displaystyle
-{i}
\delta^{\ }_{\mathsf{a}\mathsf{b}}
\sigma^{\ }_{\mu}\partial^{\ }_{\mu} & 
\displaystyle
\Delta^{\ }_{\mathsf{b}\mathsf{a}}(\boldsymbol{r}) \\
\displaystyle
\Delta^{*}_{\mathsf{b}\mathsf{a}}(\boldsymbol{r}) &
\displaystyle
+{i}
\delta^{\ }_{\mathsf{a}\mathsf{b}}
\sigma^{\ }_{\mu}\partial^{\ }_{\mu} 
\end{array}
\right).
\label{eq: Dirac Kernel D}
\end{equation}
\end{subequations}
(We have absorbed
$i\sqrt{2}\eta$ in a rescaling of $\Delta$.)
Because of Eq.~(\ref{eq: generic background})
$\Delta^{\ }_{\mathsf{a}\mathsf{b}}=
\Delta^{\ }_{\mathsf{b}\mathsf{a}}$
and thus
$D^{\ }_{\mathsf{a}\mathsf{b}}[\Delta]$
is Hermitian.
Observe that the eigenvalues of the kernel~(\ref{eq: Dirac Kernel D})
are real-valued 
and the nonvanishing ones come in pairs of opposite sign.
Indeed, we could have equally well presented the effective Lagrangian
(\ref{eq: L eff and majorana cond})
as
\begin{equation}
\mathcal{L}^{\ }_{\mathrm{eff}}
=
\left(
 \begin{array}{cc}
 \boldsymbol{d}^{\dag}
 & 
 \boldsymbol{d}^T
 \end{array}
 \right)
\left(
\begin{array}{cc}
\displaystyle
K
& 
\displaystyle
{i}\sigma^{\ }_{y}
\Delta(\boldsymbol{r}) 
\\
\displaystyle
-
{i}\sigma^{\ }_{y}
\Delta^{\dag}(\boldsymbol{r}) 
&
\displaystyle
-K^{T}
\end{array}
\right)
\left(
 \begin{array}{c}
 \boldsymbol{d} 
\\
(\boldsymbol{d}^{\dag})^T 
\\
 \end{array}
 \right)
\label{eq: BdG is there}
\end{equation}
where the kinetic energy $K$
was defined in Eq.~(\ref{eq: def K})
and we use a matrix convention to make
explicit the 
Bogoliubov-de-Gennes particle-hole symmetry
responsible for the aforementioned
pairing of eigenvalues.

The effective field theory 
$S^{\ }_{\mathrm{eff}}[\Delta]$
describing the dynamics of the slowly varying bosonic field 
$\Delta^{\ }_{\mathsf{a}\mathsf{b}}$
follows from integrating out the fermionic fields 
$\boldsymbol{d}^{\dag}$ and $\boldsymbol{d}$
in the partition function,
\begin{equation}
\begin{split}
e^{-S^{\ }_{\mathrm{eff}}[\Delta]}
\equiv&\,
\int 
\mathcal{D}
\left[
\boldsymbol{d}^{\dag},\boldsymbol{d}
\right]
\exp
\left(
-
\int \mathrm{d}^2\,r\,
\mathcal{L}^{\ }_{\mathrm{eff}}
\right)
\\
=&\,
(\pm)
\sqrt{
\mathrm{Det}\, D[\Delta]
     }
\\
\equiv&\,
\mathrm{Pf}\, D[\Delta].
\end{split}
\label{eq: def eff action}
\end{equation}
Here, the Pfaffian
$\mathrm{Pf}\, D[\Delta]$
implements the isospin-1/2 TRS through
the Majorana condition~(\ref{eq: final Majorana})
[see also Eq.~(\ref{eq: final Majorana cond})].
A gradient expansion of the exponentiated
Pfaffian gives the standard kinetic energy
of the NL$\sigma$M on 
the target space
$\mathrm{U}(2\mathsf{N}^{\ }_{\mathrm{r}})/\mathrm{O}(2\mathsf{N}^{\ }_{\mathrm{r}})$. 
However, since the second homotopy group of 
$G/H=\mathrm{U}(2\mathsf{N}^{\ }_{\mathrm{r}})/\mathrm{O}(2\mathsf{N}^{\ }_{\mathrm{r}})$ is non-trivial,
\begin{eqnarray}
\pi^{\ }_{2}[\mathrm{SU}(M)/\mathrm{O}(M)]= 
\mathbb{Z}^{\ }_{2}, 
\hbox{ for $M>2$},
\label{eq: homotopy group}
\end{eqnarray}
the NL$\sigma$M is allowed to have a topological term of 
the $\mathbb{Z}^{\ }_2$ type. 
In other words, 
Eq.~(\ref{eq: homotopy group}) tells us that the space of
all field configurations is divided into two sectors
that are not smoothly connected. Consequently, 
these two sectors can be weighted differently in the 
effective partition function. This possibility
is encoded in the ambiguity in defining the sign
of the Pfaffian~(\ref{eq: def eff action}), 
a global property of the
target manifold $G/H=\mathrm{U}(2\mathsf{N}^{\ }_{\mathrm{r}})/\mathrm{O}(2\mathsf{N}^{\ }_{\mathrm{r}})$.
In the following, 
we use the same approach as in Ref.\ \onlinecite{Ryu07b},
to show that the ambiguity in defining the sign of
the Pfaffian can be interpreted as the presence of a
$\mathbb{Z}^{\ }_2$-topological term.

\subsection{
$\mathbb{Z}^{\ }_{2}$ configurations of the $\Delta$-field
           }

In this section, we construct 
representative $\Delta$-field configurations
that belong to the two complementary 
$\mathbb{Z}^{\ }_{2}$-topological sectors
as defined by the second homotopy group%
~(\ref{eq: homotopy group}).
To this end, we introduce the generator 
\begin{eqnarray}
\lambda^{\ }_2:=
\left(
\begin{array}{ccc}
0 &-{i} &   \\
{i} &0 &    \\
 &  &   0_{2\mathsf{N}^{\ }_{\mathrm{r}}-2}\\ 
\end{array}
\right)
\in\mathrm{o}(2\mathsf{N}^{\ }_{\mathrm{r}})
\end{eqnarray}
of the symmetry-broken group $\mathrm{U}(2\mathsf{N}^{\ }_{\mathrm{r}})$ 
that leaves the saddle-points~(\ref{eq: generic saddle points}) 
invariant. We also define the generators 
$\lambda^{\ }_{1}$ 
and
$\lambda^{\ }_{3}$ 
through
\begin{equation}
\begin{split}
&
\lambda^{\ }_{1}:=
\left(
\begin{array}{ccc}
0 & 1 &   \\
1 &0 &   \\
 &  &     0^{\ }_{2\mathsf{N}^{\ }_{\mathrm{r}}-2}\\ 
\end{array}
\right),
\\
&
\lambda^{\ }_{3}:=
\left(
\begin{array}{ccc}
1 & 0 &   \\
0 &-1 &    \\
 &    &   0^{\ }_{2\mathsf{N}^{\ }_{\mathrm{r}}-2}\\ 
\end{array}
\right).
\end{split}
\end{equation}
Here,
$0^{\ }_{2\mathsf{N}^{\ }_{\mathrm{r}}-2}$
is the $(\mathsf{N}^{\ }_{\mathrm{r}}-2)\times(\mathsf{N}^{\ }_{\mathrm{r}}-2)$
matrix with 0 in all entries. 
The three matrices
$\lambda^{\ }_{1}$, 
$\lambda^{\ }_{2}$,
and 
$\lambda^{\ }_{3}$ 
generate an SU(2) algebra.
Unlike $\lambda^{\ }_2$,
neither
$\lambda^{\ }_{1}$
nor
$\lambda^{\ }_{3}$ 
leave the saddle-points~(\ref{eq: generic saddle points})
invariant. Hence, neither 
$\lambda^{\ }_{1}$
nor
$\lambda^{\ }_{3}$
belong to 
the unbroken symmetry group $H=\mathrm{O}(2\mathsf{N}^{\ }_{\mathrm{r}})$.

Let $S^2$ denote the two-sphere
and choose the polar $-\pi/2\leq\theta\leq+\pi/2$ 
and azimuthal $0\leq\phi<2\pi$ angles
as spherical coordinates of $S^2$.
Following Weinberg et al.\ in Ref.~\onlinecite{Weinberg84}
we define on $S^2$ the unitary matrices
\begin{equation}
U^{\ }_{l}(\theta,\phi):=
e^{{i}l\lambda^{\ }_2 \phi/2}\,
e^{{i}\lambda^{\ }_3 \theta/2}\,
e^{-{i}l\lambda^{\ }_2 \phi/2}
\in\mathrm{U}(2\mathsf{N}^{\ }_{\mathrm{r}})
\end{equation}
that we label by the integer $l \in\mathbb{Z}$.
Finally, we define the family 
\begin{subequations}
\label{eq: family label k over S2}
\begin{equation}
\begin{split}
\Delta^{\ }_{l}(\theta,\phi):=&\,
-{i}\left|\Delta^{\ }_{0}\right|
U^{\ }_{l} U^T_{l}
\\
=&\,
-{i}\left|\Delta^{\ }_{0}\right|
\left(
\begin{array}{cc}
R^{\ }_{l}(\theta,\phi) 
& 
0 
\\
0 
& 
I^{\ }_{2\mathsf{N}^{\ }_{\mathrm{r}}-2}
\end{array}
\right)
\end{split}
\label{eq: family label k over S2 a}
\end{equation}
in
$\mathrm{U}(2\mathsf{N}^{\ }_{\mathrm{r}})/\mathrm{O}(2\mathsf{N}^{\ }_{\mathrm{r}})$
where
\begin{equation}
R^{\ }_{l}(\theta,\phi)=
\left(
\begin{array}{cc}
\cos\theta+{i}\sin\theta\cos l\phi  
&
-{i}\sin\theta \sin l\phi  
\\
-{i}\sin\theta \sin l\phi 
&
\cos\theta-{i}\sin\theta\cos l\phi 
\\
\end{array}
\right)
\label{eq: def R of k}
\end{equation}
\end{subequations}
is labeled by the integer $l\in\mathbb{Z}$
\textit{but is independent of the replica number $\mathsf{N}^{\ }_{\mathrm{r}}$}.

The $\mathbb{Z}^{\ }_{2}$ configurations of the $\Delta$-field
on the two-dimensional torus $T^{2}$ 
with the coordinates  $0\le x,y \le L$
($L$ is serving as an infrared cutoff)
can be obtained from the parametrization
of the unit sphere $S^{2}$ in terms of the 
$L$-periodic unit vector
\begin{subequations}
\begin{equation}
\begin{split}
&
\boldsymbol{n}(x, y):=
\frac{
\boldsymbol{r}(x,y)
     }
     {
|\boldsymbol{r}(x,y)|
     },
\end{split}
\end{equation}
itself given by the $L$-periodic vector 
\begin{equation}
\begin{split}
\boldsymbol{r}(x, y):=
\begin{pmatrix}
-\sin(2\pi y/L)
\\
-\sin(2\pi x/L)
\\
\cos(2\pi x/L)
+ 
\cos(2\pi y/L)
- 
1
\end{pmatrix}.
\end{split}
\end{equation}
\end{subequations}
For example, the $L$-periodic 
$\Delta^{\ }_{l=1}(x,y)$ is obtained from
Eq.~(\ref{eq: family label k over S2})
by replacing $R^{\ }_{l=1}(\theta,\phi)$
with
\begin{equation}
R^{\ }_{l=1}(x,y)=
\left(
\begin{array}{cc}
n_z +{i} n_x
&
-{i}n_y
\\
-{i}n_y
&
n_z-{i}n_x
\\
\end{array}
\right).
\end{equation}

\subsection{
Spectral flow
           }
\label{appsubsec: Spectral flow}

\begin{figure}
  \begin{center}
  \includegraphics[width=7cm,clip]{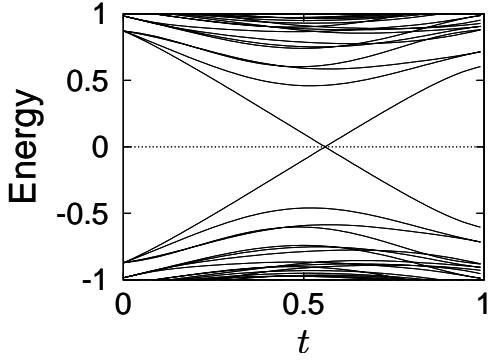} 
\caption{
\label{fig: spec_flow}
The energy eigenvalue spectrum 
in the vicinity of the band center 
for the kernel $D[\Delta(t)]$
from Eqs.~(\ref{eq: Dirac Kernel D})
and~(\ref{eq: def Qt}),
is computed numerically as a function of the 
parameter $0\leq t\leq 1$
for
$\left|\Delta^{\ }_{0}\right|/\Lambda=1$.
The field $\Delta(t)$
interpolates between 
$\Delta^{\ }_{i}$ when $t=0$
and
$\Delta^{\ }_{f}$ when $t=1$.
Without loss of generality,
the replica number $\mathsf{N}^{\ }_{\mathrm{r}}=1$
was chosen.
        }
\end{center}
\end{figure}

We are going to argue numerically that
\begin{equation}
\mathrm{sgn}\,
\mathrm{Pf}
\left( D[\Delta^{\ }_{l  }]\right)=-
\mathrm{sgn}\,
\mathrm{Pf}
\left( D[\Delta^{\ }_{l+1}]\right),
\quad
l\in\mathbb{Z},
\label{eq: Pf for Qt}
\end{equation}
by looking at the spectral flow of the kernel
\begin{equation}
\Delta(t):= 
(1-t)\Delta^{\ }_i 
+ 
t    \Delta^{\ }_f
\label{eq: def Qt}
\end{equation}
as a function of $0\leq t\leq 1$.
Here, the initial, $\Delta^{\ }_{i}$, and final, $\Delta^{\ }_{f}$, 
configurations belong to $G/H=\mathrm{U}(2\mathsf{N}^{\ }_{\mathrm{r}})/\mathrm{O}(2\mathsf{N}^{\ }_{\mathrm{r}})$,
while
$\Delta(t)$ 
is not a member of $G/H=\mathrm{U}(2\mathsf{N}^{\ }_{\mathrm{r}})/\mathrm{O}(2\mathsf{N}^{\ }_{\mathrm{r}})$ for $0<t<1$. 
According to Eq.~(\ref{eq: BdG is there}),
the spectrum $\lambda^{\ }_{\iota}(t)$
of $D[\Delta(t)]$ is symmetric about the band center at the energy zero.
Configurations $\Delta^{\ }_i$ and $\Delta^{\ }_f$
have Pfaffians of opposite signs whenever an odd number
of level crossing occurs at the band center (``spectral flow'')
during the $t$-evolution of the kernel $D[\Delta(t)]$. 
This is accompanied by
the closing of the spectral gap 
of $D[\Delta(t)]$
by an odd number of pairs 
$\big(-\lambda^{\ }_{\iota}(t),+\lambda^{\ }_{\iota}(t)\big)$
as $t$ interpolates between $0$ and $1$.
The spectral $t$-evolution
is obtained numerically using the regularization 
of the kernel $D[\Delta(t)]$
by choosing the family
on the torus $T^{2}$. In this way, the index $\iota$
takes discrete values.
In Fig.~\ref{fig: spec_flow},
we show the evolution of the eigenvalues 
for $\Delta(t)$ interpolating between 
$\Delta^{\ }_{l=0}$ and $\Delta^{\ }_{l=1}$.
Observe that in $\Delta^{\ }_{l}$ the part responsible 
for the winding configuration $R_l(\theta,\phi)$ 
is entirely localized in the sector of the first replica.
Thus, when computing the spectral flow,
we can focus on this sector alone.
Since level crossing at the band center 
takes place for a single pair of levels,
we conclude that
$\mathrm{Pf}({D}[\Delta^{\ }_{l=0}])$
and 
$\mathrm{Pf}({D}[\Delta^{\ }_{l=1}])$
differ by their sign.
This supports numerically
Eq.\ (\ref{eq: Pf for Qt}).

\subsection{
Summary
           }

In summary, after integration over the Majorana spinors
along the nearly-critical line of region CII
in Fig.~\ref{fig: phase diagram}, 
the effective action for the Nambu-Goldstone field 
$\Delta$, a symmetric and unitary matrix,
is given by
\begin{subequations}
\label{appeq: final NLSM CII with topological term}
\begin{equation}
Z^{\mathrm{topolo}}_{\mathrm{NL}\sigma\mathrm{M}}=
\int \mathcal{D}[\Delta]\,
(-1)^{n[\Delta]}\,
e^{-S[\Delta]}
\label{appeq: NLSM action with Z2 topology}
\end{equation}
where $S[\Delta]$ is the 
(fermionic replica version of the)
action for the NL$\sigma$M on 
$G/H=
\mathrm{U}(2\mathsf{N}^{\ }_{\mathrm{r}})/
\mathrm{O}(2\mathsf{N}^{\ }_{\mathrm{r}})$,
i.e.,%
~\cite{Gade91-93}
\begin{equation}
\begin{split}
S[\Delta]=&\,
\frac{1}{t^{\ }_{\mathrm{M}'}}
\int \mathrm{d}^{2}\,r\,
\mathrm{tr}
\left[
\left(
\partial^{\ }_{\mu}
\Delta^{\dag}
\right)
\left(
\partial^{\ }_{\mu}
\Delta
\right)
\right]
\\
&\,
+
\frac{1}{t^{\ }_{a'}}
\int \mathrm{d}^{2}\,r\,
\mathrm{tr}
\left(
\Delta^{\dag}
\partial^{\ }_{\mu}
\Delta
\right)
\mathrm{tr}
\left(
\Delta
\partial^{\ }_{\mu}
\Delta^{\dag}
\right),
\end{split}
\end{equation}
while
\begin{equation}
n[\Delta]=0,1
\end{equation}
\end{subequations}
-- the $\mathbb{Z}^{\ }_{2}$-topological quantum number 
of $\Delta$ --
reflects the ambiguity in defining globally the
sign of the Pfaffian of Majorana spinors.
Because of the block structure~(\ref{eq: def R of k}),
the topological quantum number $n[\Delta]=0,1$
is expected to survive the replica limit $\mathsf{N}^{\ }_{\mathrm{r}}\to0$.

\section{
Patterns of symmetry breaking and supermanifolds
        }
\label{subsec: Patterns of symmetry breaking and supermanifolds}

There are 10 target spaces
$
{G}/{H}
$
for the NL$\sigma$Ms of relevance to the 10 symmetry classes of
Anderson localization.~\cite{Zirnbauer96,Altland97,Heinzner05}
They encode 10 distinct
patterns of symmetry breaking.
These patterns have been exhaustively 
classified within a supersymmetric approach
by Zirnbauer in Ref.~\onlinecite{Zirnbauer96}.
Each target superspace
$
{G}/{H}
$
is a Riemannian symmetric supermanifold that
can be parametrized in its bosonic sector by
the Riemannian symmetric manifold 
\begin{equation}
{M}^{\ }_{B}=
{M}^{\ }_{BB}
\times
{M}^{\ }_{FF}.
\end{equation}
Here, ${M}^{\ }_{B}$
is the direct product between a non-compact 
Riemannian symmetric manifold
${M}^{\ }_{BB}$
originating from the boson-boson sector of the Riemannian 
symmetric supermanifold
and a compact Riemannian symmetric manifold
${M}^{\ }_{FF}$
originating from the fermion-fermion sector of the Riemannian 
symmetric supermanifold.
The target superspaces
${G}/{H}$
and Riemannian symmetric manifolds
$
{M}^{\ }_{B}=
{M}^{\ }_{BB}
\times
{M}^{\ }_{FF}
$
relevant to this paper are:
\begin{itemize}

\item
$
{G}/{H}=
\mathrm{GL}(n|n)
\times
\mathrm{GL}(n|n)/
\mathrm{GL}(n|n)=
\mathrm{GL}(n|n)
$
with the Riemannian symmetric  manifolds
\begin{equation}
\begin{split}
& 
{M}^{\ }_{BB}=
\mathrm{GL}(n,\mathbb{C})
/
\mathrm{U}(n),
\\
&
{M}^{\ }_{FF}=
\mathrm{U}(n),
\end{split}
\label{appeq: G/H and MBxMF for AIII}
\end{equation}
for the chiral symmetry class AIII,

\item
$
{G}/{H}=
\mathrm{GL}(2n|2n)
/
\mathrm{OSp}(2n|2n)
$
with the Riemannian symmetric manifolds
\begin{equation}
\begin{split}
&
{M}^{\ }_{BB}=
\mathrm{GL}(2n,\mathbb{R})
/
\mathrm{O}(2n),
\\
&
{M}^{\ }_{FF}=
\mathrm{U}(2n)
/
\mathrm{Sp}(2n),
\end{split}
\label{appeq: G/H and MBxMF for BDI}
\end{equation}
for the chiral symmetry class BDI,

\item
$
{G}/{H}=
\mathrm{GL}(2n|2n)
/
\mathrm{OSp}(2n|2n)
$
with the Riemannian symmetric manifolds
\begin{equation}
\begin{split}
&
{M}^{\ }_{BB}=
\mathrm{U}^{*}(2n)
/
\mathrm{Sp}(2n),
\\
&
{M}^{\ }_{FF}=
\mathrm{U}(2n)
/
\mathrm{O}(2n),
\end{split}
\label{appeq: G/H and MBxMF for CII}
\end{equation}
for the chiral symmetry class CII,

\item
$
{G}/{H}=
\mathrm{OSp}(2n|2n)
/
\mathrm{GL}(n|n)
$
with the Riemannian symmetric manifolds
\begin{equation}
\begin{split}
&
{M}^{\ }_{BB}=
\mathrm{Sp}(4n,\mathbb{R})
/
\mathrm{U}(n),
\\
&
{M}^{\ }_{FF}=
\mathrm{O}(2n)
/
\mathrm{U}(n),
\end{split}
\label{appeq: G/H and MBxMF for D}
\end{equation}
for the BdG symmetry class D, 

\item
$
{G}/{H}=
\mathrm{OSp}(4n|4n)
/
\mathrm{OSp}(2n|2n)
\times
\mathrm{OSp}(2n|2n)
$
with the Riemannian symmetric manifolds
\begin{equation}
\begin{split}
&
{M}^{\ }_{BB}=
\mathrm{Sp}(2n,2n)
/
\mathrm{Sp}(2n)
\times
\mathrm{Sp}(2n),
\\
&
{M}^{\ }_{FF}=
\mathrm{SO}(4n)
/
\mathrm{SO}(2n)
\times
\mathrm{SO}(2n),
\end{split}
\label{appeq: G/H and MBxMF for AII}
\end{equation}
for the symplectic symmetry class AII.

\end{itemize}

The compatibility of these target superspaces
with the addition of a topological term 
in the corresponding NL$\sigma$M
is solely determined by the compact Riemannian symmetric
manifold ${M}^{\ }_{FF}$:
A topological term requires a non-trivial
second homotopy group of
${M}^{\ }_{FF}$
(e.g., CII and AII).
In this context, observe that the Riemannian symmetric supermanifolds for
symmetry classes BDI and CII merely differ by the exchange
of the boson-boson and fermion-fermion
stabilizers $\mathrm{O}(2n)$ and $\mathrm{Sp}(2n)$ 
[in that regard, it is convenient to view
$\mathrm{U}^{*}(2n)$ as a 
non-compact real subgroup of
$\mathrm{GL}(2n,\mathbb{C})$].
This small difference is of great consequences
since the NL$\sigma$M for the symmetric class BDI
cannot be augmented by a topological term.

For simplicity, we consider symmetry class
CII with $n=1$. According to 
Eq.~(\ref{appeq: G/H and MBxMF for CII}),
the target superspace is
\begin{equation}
\mathrm{GL}(2|2)/\mathrm{OSp}(2|2)\approx
\mathrm{GL}(2|2)/\mathrm{SL}(1|2)
\label{appeq: iso OSp(2|2) Sl(1|2)}
\end{equation}
whereby we used the isomorphism
$\mathrm{OSp}(2|2)\approx\mathrm{SL}(1|2)$.
The projected CII target superspace 
obtained by quotienting out the two diagonal
generators of $\mathrm{GL}(2|2)$ is
\begin{subequations}
\label{appeq: projected CII if n=1}
\begin{equation}
\mathrm{PSL}(2|2)/\mathrm{OSp}(2|2)\approx
\mathrm{PSL}(2|2)/\mathrm{SL}(1|2).
\label{appeq: projected CII if n=1 a}
\end{equation}
One must carry this projection on the non-compact and compact
Riemannian symmetric manifolds%
~(\ref{appeq: G/H and MBxMF for CII}).
This is done by quotienting out their $\mathbb{R}^{\ }_{+}$
and $\mathrm{U}(1)$ factors, respectively.
Hence, the projected boson-boson Riemannian symmetric manifold is
\begin{equation}
\mathrm{SU}^*(2)/\mathrm{Sp}(2)
\approx
\mathrm{SU}^*(2)/\mathrm{SU}(2)
\label{appeq: projected CII if n=1 b}
\end{equation} 
while the projected fermion-fermion Riemannian symmetric manifold is
\begin{equation}
\mathrm{SU}(2)/\mathrm{O}(2)\sim 
S^2.
\label{appeq: projected CII if n=1 c}
\end{equation} 
\end{subequations}

\section{
The supergroup $\mathrm{PSL}(n|n)$
        }
\label{app sec: The group mathrm{PSL}(2|2)}

Let $n\in\mathbb{N}$ be an integer and denote with
$\mathrm{smat}(n|n)$
the set of all \textit{real} $(n|n)$ supermatrices 
${M}$, i.e., matrices of the form
\begin{equation}
{M}:=
\begin{pmatrix}
{M}^{\ }_{BB}
&
{M}^{\ }_{BF}
\\
{M}^{\ }_{FB}
&
{M}^{\ }_{FF}
\end{pmatrix}
\label{eq: def smat}
\end{equation}
where ${M}^{\ }_{BB}$ and ${M}^{\ }_{FF}$
are $n\times n$ \textit{real}-valued matrices
while ${M}^{\ }_{BF}$ and ${M}^{\ }_{FB}$
are $n\times n$ Grassmann-valued matrices.
We shall denote with $I$ and $J$ 
the $(n|n)$ diagonal supermatrices
\begin{equation}
\begin{split}
&
I:=
\mathrm{diag}
\begin{pmatrix}
1&
\cdots&
1&
1&
\cdots&
1
\end{pmatrix},
\\
&
J:=
\mathrm{diag}
\begin{pmatrix}
1&
\cdots&
1&
-1&
\cdots&
-1
\end{pmatrix},
\end{split}
\end{equation}
respectively. 
For any element ${M}\in\mathrm{smat}(n|n)$,
the supertrace is defined by
\begin{equation}
\mathrm{str}\,{M}:=
\mathrm{tr}\,{M}^{\ }_{BB}
-
\mathrm{tr}\,{M}^{\ }_{FF}.
\end{equation}
Observe that
\begin{equation}
\begin{split}
&
\mathrm{tr}\,{M}^{\ }_{BB}=
\frac{1}{2}
\left(
\mathrm{str}\,{MJ}
+
\mathrm{str}\,{MI}
\right),
\\
&
\mathrm{tr}\,{M}^{\ }_{BF}=
\frac{1}{2}
\left(
\mathrm{str}\,{MJ}
-
\mathrm{str}\,{MI}
\right),
\end{split}
\end{equation}
i.e., demanding that
$\mathrm{tr}\,{M}^{\ }_{BB}$
and
$\mathrm{tr}\,{M}^{\ }_{BF}$
both vanish is equivalent to demanding that
$\mathrm{str}\,{MJ}$
and
$\mathrm{str}\,{MI}$
both vanish.  
For any element ${M}\in\mathrm{smat}(n|n)$
with 
$\mathrm{det}\,{M}^{-1}_{FF}\neq0$
or
$\mathrm{det}\,{M}^{-1}_{BB}\neq0$
the superdeterminant is defined by
\begin{equation}
\mathrm{sdet}\,{M}:=
\frac{
\mathrm{det}\,
\left(
{M}^{\ }_{BB}
-
{M}^{\ }_{BF}
{M}^{-1}_{FF}
{M}^{\ }_{FB}
\right)
     }
     {
\mathrm{det}\,
{M}^{\ }_{FF}
     },
\end{equation}
or
\begin{equation}
\mathrm{sdet}\,{M}:=
\frac{
\mathrm{det}\,
\left(
{M}^{\ }_{FF}
-
{M}^{\ }_{FB}
{M}^{-1}_{BB}
{M}^{\ }_{BF}
\right)
     }
     {
\mathrm{det}\,
{M}^{-1}_{BB}
     },
\end{equation}
respectively. 

An obvious generalization of $\mathrm{smat}(n|n)$
is achieved through the complexification
\begin{equation}
{M}\to
{M}
+
{i}
{M}',
\qquad
{M},{M}'\in\mathrm{smat}(n|n).
\end{equation}
Another one follows from the substitution
\begin{equation}
\mathrm{smat}(n|n)\to
\mathrm{smat}(m|n)
\end{equation}
where the supermatrices from the set $\mathrm{smat}(m|n)$
are of the form~(\ref{eq: def smat}) 
with the entries of the 
$m\times m$ matrix ${M}^{\ }_{BB}$  
and the $n\times n$ matrix ${M}^{\ }_{FF}$
commuting numbers while the entries of the 
the $m\times n$ matrix ${M}^{\ }_{BF}$  
and the $n\times m$ matrix ${M}^{\ }_{FB}$  
are anticommuting numbers.

The following definitions apply to both real and complex 
supermatrices. The supergroup $\mathrm{PSL}(n|n)$ 
is constructed from the supergroup 
$\mathrm{GL}(n|n)\subset\mathrm{smat}(n|n)$ as follows. 
The supergroup $\mathrm{GL}(n|n)$ consists of all $(n|n)$
supermatrices for which both
${M}^{\ }_{BB}$
and
${M}^{\ }_{FF}$
are non-singular (i.e., have nonvanishing determinants)
and with the matrix multiplication as the group operation.
The supergroup $\mathrm{GL}(n|n)$ is not semisimple.
It possesses the matrix subsupergroup 
$\mathrm{SL}(n|n)$
that follows from restricting the superdeterminants 
in $\mathrm{GL}(n|n)$ to one.
The supergroup $\mathrm{SL}(n|n)$ is also not semisimple,
for it contains the $(n|n)$ unit supermatrix $I$
that commutes with all $(n|n)$
supermatrices.
All elements of $\mathrm{SL}(n|n)$
are generated through exponentiation of elements of
the Lie superalgebra 
$\mathrm{sl}(n|n)$
whereby any element of $\mathrm{sl}(n|n)$ 
is a $(n|n)$ supermatrix of the form 
\begin{equation}
{X}:=
\begin{pmatrix}
{X}^{\ }_{BB}
&
{X}^{\ }_{BF}
\\
{X}^{\ }_{FB}
&
{X}^{\ }_{FF}
\end{pmatrix}
\end{equation}
with ${X}^{\ }_{BB}$ and ${X}^{\ }_{FF}$
$n\times n$ real-valued matrices
while ${M}^{\ }_{BF}$ and ${M}^{\ }_{FB}$
are $n\times n$ Grassmann-valued matrices
with the vanishing supertrace
\begin{equation}
\mathrm{str}\,{X}=0.
\end{equation}
The supergroup $\mathrm{PSL}(n|n)$ 
is defined to be the factor group
$\mathrm{SL}(n|n)/\mathbb{R}^{\ }_{+}$
($\mathbb{R}^{\ }_{+}$ the set of positive real numbers)
by which any two elements in $\mathrm{sl}(n|n)$
that differ by a multiple of the unit element $I$
generate upon exponentiation the very same element of
$\mathrm{PSL}(n|n)$. 
The supergroup $\mathrm{PSL}(n|n)$ is semisimple,
an element of $\mathrm{PSL}(n|n)$ 
cannot be written as a supermatrix.

\medskip

\section{
The Lie group U$^{*}(2)$
        }
\label{app sec: The Lie group U{*}(2)}

Let $n\in\mathbb{N}$.
The Lie group U$^{*}(2n)$ is the set of matrices in
GL$(2n,\mathbb{C})$ that commutes with the linear transformation
\begin{equation}
\psi:\mathbb{C}^{2n}\to\mathbb{C}^{2n},
\begin{pmatrix}
z^{\ }_{1}
\\
\vdots
\\
z^{\ }_{n}
\\
z^{\ }_{n+1}
\\
\vdots
\\
z^{\ }_{2n}
\end{pmatrix}
\to
\begin{pmatrix}
z^{* }_{n+1}
\\
\vdots
\\
z^{* }_{2n}
\\
-
z^{* }_{1}
\\
\vdots
\\
-
z^{\* }_{n}
\end{pmatrix}
\end{equation}
where complex conjugation is denoted by $^*$.
It follows that the Lie algebra u$^{*}(2n)$ is
the set of matrices in GL$(2n,\mathbb{C})$ of the form
\begin{equation}
\begin{pmatrix}
Z^{\ }_{1}
&
Z^{\ }_{2}
\\
-
Z^{* }_{2}
&
Z^{* }_{1}
\end{pmatrix}
\end{equation}
where $Z^{\ }_{1}$ and $Z^{\ }_{2}$ are any complex-valued
$n\times n$ matrices.

We now specialize to the Lie group U$^{*}(2)$ with the Lie algebra
u$^{*}(2)$. Let $X\in$u$^{*}(2)$. There exist the complex numbers
$z^{\ }_{1}$ and $z^{\ }_{2}$ such that
\begin{equation}
\begin{split}
X=&\,
\begin{pmatrix}
z^{\ }_{1}
&
z^{\ }_{2}
\\
-
z^{* }_{2}
&
z^{* }_{1}
\end{pmatrix}
\\
=&\,
\mathrm{Re}\, z^{\ }_{1}\,\sigma^{\ }_{0}
+
{i}
\mathrm{Im}\, z^{\ }_{2}\,\sigma^{\ }_{1}
+
{i}
\mathrm{Re}\, z^{\ }_{2}\,\sigma^{\ }_{2}
+
{i}
\mathrm{Im}\, z^{\ }_{1}\,\sigma^{\ }_{3}.
\end{split}
\end{equation}
Here, we have introduced the unit $2\times2$ matrix
$\sigma^{\ }_{0}$ and the three Pauli matrices
$(\sigma^{\ }_{1},\sigma^{\ }_{2},\sigma^{\ }_{3})$.
Evidently, U$^{*}(2)$ and 
GL$(1,\mathbb{R})\times$SU$(2)$ share the same Lie algebra,
\begin{equation}
\mathrm{u}^{*}(2)\approx
\mathbb{R}\oplus\mathrm{su}(2).
\end{equation}
Locally, one thus have the isomorphism
\begin{equation}
\mathrm{U}^{*}(2)\approx
\mathbb{R}^{\ }_{+}\otimes\mathrm{SU}(2)
\end{equation}
where $\mathbb{R}^{\ }_{+}$ is the set of positive real numbers.

\section{
NL$\sigma$M on homogeneous target spaces versus
NL$\sigma$M on symmetric target spaces
        }
\label{appsec: homogeneous vs symmetric}

In this Appendix we briefly summarize a number of facts
about NL$\sigma$M on target spaces which are homogeneous
spaces $G/H$.  (For more details we refer the reader, for example,
to the appendix of \onlinecite{footnote: homogeneous space},
and references therein.)

The essential properties of a NL$\sigma$M 
whose target space is of the form of a coset space $G/H$, where $G$ is
a Lie group and $H$ a Lie-subgroup, are as follows.

If $H$ is a maximal subgroup of $G$, then the NL$\sigma$M has precisely one 
coupling constant (the target space $G/H$ is then what
is known as a `symmetric space').  If, on the other hand,  
there exists precisely one intervening subgroup 
$H'$, i.e., 
\begin{equation}
H\subset H'\subset G,
\label{appeq: nested sequence subgroups H, H', G}
\end{equation} 
then the NL$\sigma$M  with target space $G/H$ 
turns out to have precisely two independent coupling constants. 
If,  in the latter case, we were to run the RG into the 
infrared (i.e., to large length scales), then one of the two coupling 
constants will disappear. In this case one ends up,
asymptotically at long scales, 
with a NL$\sigma$M  with one coupling constant, whose target space will 
be either $G/H'$ or $H'/H$ 
(both are, by assumption, symmetric 
spaces in the sense of maximal subgroups).  The number of
coupling constants for a NL$\sigma$M on a homogeneous 
space with more intervening subgroups increases accordingly.

The relevance of the symmetric space NL$\sigma$M  for the universal 
physics appearing at large length scales is thus simply a consequence of 
the RG. If we begin with a NL$\sigma$M  on a general homogeneous space, 
the RG will select, asymptotically at long length scales, a target space 
which is a symmetric space $G/H$ where $H$ is a maximal subgroup
of $G$. So, all NL$\sigma$M on coset spaces become NL$\sigma$M on 
symmetric spaces, asymptotically at long length scales. It is for this 
reason that they are the ``stable'' large-scale limits, and appear 
naturally, without fine-tuning.

The appearance of a  NL$\sigma$M  
(on a, in general, possibly homogeneous and not necessarily symmetric space)
arises in a physical context from the  general principle of symmetry 
breaking. The group $G$ is the global symmetry group of the problem. The 
subgroup $H$ (not necessarily maximal) characterizes the symmetries which 
are preserved when the global symmetry $G$ is broken.

In the situation discussed in this article, the global symmetry group is
\begin{equation} 
G=\mathrm{GL}(2m|2m),
\qquad
m=1,2,\cdots.
\end{equation}

When a global density of states (DOS) is generated in 
the Dirac fermion formulation of the theory, the resulting expectation 
value breaks the symmetry $G$ but preserves the symmetry $H\subset G$.
It follows from Ref.~\onlinecite{Guruswamy00}  
that on the ``dotted line''
in our global phase diagram in Fig.~\ref{fig: phase diagram} 
this subgroup is
${H'}=\mathrm{OSp}(2m|2m)$.
The manifold 
$G/{H'}= \mathrm{GL}(2m|2m)/\mathrm{OSp}(2m|2m)$ 
is the symmetric space corresponding to symmetry class CII.
On the boundary of the CII region the symmetry class is AII,
corresponding to a target manifold
${H'}/H = 
\mathrm{OSp}(2m|2m)/
[\mathrm{OSp}(m|m)\times\mathrm{OSp}(m|m)]$.
Thus, the entire lower half of 
our global phase diagram in Fig.~\ref{fig: phase diagram}
can be described by the NL$\sigma$M on the homogeneous space
\begin{equation}
G/H=
\mathrm{GL}(2m|2m)/[\mathrm{OSp}(m|m)\times\mathrm{OSp}(m|m)].
\end{equation}
which has two coupling constants (besides the coupling of the Gade term).
This is analogous to the discussion in 
\onlinecite{footnote: homogeneous space}.


\begin{thebibliography}{99}
\bibitem{footnote on AII}
The analysis made in Ref.~\onlinecite{Ryu07b}
applies to 
the case
(\ref{eq: def AII boundary})
or 
(\ref{eq: def AII boundary bis})
although 
the corresponding Hamiltonian in this paper is in 
twice as large a representation
of the Dirac equation than in Ref.\ \onlinecite{Ryu07b}.
Indeed, the chiral symmetry at the band center here
implies that there is no need to 
distinguish retarded from advanced
Green's 
functions and thus to add another two-dimensional
subspace for that purpose. There is no chiral symmetry
in Ref.\ \onlinecite{Ryu07b} so that the representation
of the Dirac Hamiltonian in Ref.\ \onlinecite{Ryu07b} 
must be doubled to account for the product of a pair of
retarded and advanced Green's functions.
\bibitem{footnote on AII B}
Such a type of RG flow from a free Dirac state
to a metallic state had been conjectured by
M. Bocquet, D. Serban, and M. R. Zirnbauer
in Ref.\ \onlinecite{Bocquet00}.
However, in that work
the symmetry class in which this
flow takes place was misidentified as 
class D, as opposed to the correct class AII.
Moreover, the presence of a $\mathbb{Z}^{\ }_{2}$
topological term on the sigma model side of the flow was overlooked. 
The latter is absolutely crucial, and was correctly identified in
the recent works, 
Refs.\
\onlinecite{Ostrovsky07}
and
\onlinecite{Ryu07b}.
\bibitem{Ryu06b}
See, for our conventions of graded objects, 
Ref.~\onlinecite{Ryu06a} 
and references therein.
\bibitem{footnote-erratum}
Here, we are using the conventions from
Ref.\ \onlinecite{Mudry03}
for the normalization of the variances
and for the normalization of
the integration variables in the
path integral~(\ref{eq: final partition fct HWK}). 
Be aware that the beta function
$\beta^{\ }_{g^{\ }_{A}}$ 
in Eq.\ (5.2) from Ref.\ \onlinecite{Mudry03},
contains a typographical error: The right-hand side
of Eq.\ (5.2) from Ref.\ \onlinecite{Mudry03}
should be multiplied by $2\pi$.
(The same beta function is denoted by
$\beta^{\ }_{g^{\ }_{a'}}$ in the present paper.)
\bibitem{footnote-slope flow away boundaries}
Observe that $g^{\ }_{a'}>0$ does not flow along the 
BDI boundaries $g^{2}_{+}=g^{2}_{-}$
or the CII boundaries $g^{\prime 2}_{+}=g^{\prime 2}_{-}$.
The projected slope in region BDI is
\begin{equation}
\frac{
\beta^{\ }_{g^{\ }_{\mathrm{Im}\,m}}
     }
     {
\beta^{\ }_{g^{\ }_{\mathrm{Re}\,m}}
     }=
-
\frac{
2\,g^{\ }_{\mathrm{Im}\,m}
+
g^{\ }_{a'}
     }
     {
2\,g^{\ }_{\mathrm{Re}\,m}
+
g^{\ }_{a'}
     }
\nonumber
\end{equation}
in Fig.~\ref{fig: phase diagram}(b).
The projected slope in region CII is
\begin{equation}
\frac{
\beta^{\ }_{g^{\ }_{\mathrm{Im}\,m'}}
     }
     {
\beta^{\ }_{g^{\ }_{\mathrm{Re}\,m'}}
     }=
-
\frac{
2\,g^{\ }_{\mathrm{Im}\,m'}
-
g^{\ }_{a'}
     }
     {
2\,g^{\ }_{\mathrm{Re}\,m'}
-
g^{\ }_{a'}
     }
\nonumber
\end{equation}
in Fig.~\ref{fig: phase diagram}(b).
Given these finite slopes of the flows out of
the boundaries of regions BDI and CII, the reader might
conclude that it is possible to flow from the region
AIII to the regions BDI or CII. This is not so, however,
as any pair of couplings
$g^{\ }_{\mathrm{Re}\,m}>0$ and $g^{\ }_{\mathrm{Im}\,m}<0$
($g^{\ }_{\mathrm{Re}\,m'}<0$ and $g^{\ }_{\mathrm{Im}\,m'}>0$)
or
$g^{\ }_{\mathrm{Re}\,m}<0$ and $g^{\ }_{\mathrm{Im}\,m}>0$
($g^{\ }_{\mathrm{Re}\,m'}>0$ and $g^{\ }_{\mathrm{Im}\,m'}<0$)
in region AIII always generates a flow to strong coupling of
the variance $g^{\ }_{a}$ associated to the vector gauge potentials
that break time-reversal symmetry, 
see Eq.~(4.84) from Ref.~\onlinecite{Guruswamy00}.
These flows thus escape the regions BDI and CII for which
$g^{\ }_{a}=0$.
\bibitem{appendix_potential}
Such a symmetry-breaking potential can be 
chosen to be 
\begin{align}
V(G):= 
-
\lambda 
\left[
\mathrm{str}(G^{2}) 
+ 
\left(\mathrm{str}(G^{2} )\right)^{*}
\right],
\nonumber 
\end{align}
where $\lambda>0$
and 
$G$ is 
the fundamental (super) matrix field of the
$\mathrm{PSL}(2|2)^{\ }_{1}$-WZNW model (and principal chiral model),
which is a $4\times 4$ (super) matrix, arranged in four blocks of
$2\times 2$ matrices as
\begin{align}
G:=
\left(
\begin{array}{cc}
G^{\ }_{BB} 
& 
G^{\ }_{BF} 
\\
G^{\ }_{FB} 
& 
G^{\ }_{FF}
\end{array}
\right).
\nonumber
\end{align}
Here, matrix elements are all bosonic for 
$G^{\ }_{BB}$ and $G^{\ }_{FF}$, 
and all fermionic for $G^{\ }_{BF}$ and $G^{\ }_{FB}$.
This potential breaks the 
$\mathrm{PSL}(2|2)^{\ }_{L}\times\mathrm{PSL}(2|2)^{\ }_{R}$ 
symmetry of the
$\mathrm{PSL}(2|2)^{\ }_{1}$-WZNW model down to the diagonal 
$\mathrm{PSL}(2|2)$
symmetry group. 
Moreover, for any value of the coupling constant of the 
principal chiral model, 
this potential is a relevant perturbation. 
In particular, at weak coupling it is
strongly relevant since the scaling dimension of the fundamental 
matrix field $G$ is close to zero. 
Consider the weakly coupled 
$\mathrm{PSL}(2|2)$-principal chiral model, 
where the coupling constant $\lambda$ of
the potential grows large very rapidly.
For very large values of $\lambda$ 
only the configurations of the matrix field $G$
that minimize the potential
\begin{footnotesize}
\begin{equation}
\begin{split}
\mathrm{str}(G^2) 
+ 
\left(\mathrm{str}(G^{2})\right)^{*}
=&\,
\mathrm{tr} (G^{\ }_{BB})^{2} 
-
\mathrm{tr} (G^{\ }_{FF})^{2}
\\
&\,
+
\mathrm{tr} (G^{\ }_{BF} G^{\ }_{FB}) 
-
\mathrm{tr} (G^{\ }_{FB} G^{\ }_{BF}) 
\nonumber 
\end{split}
\nonumber
\end{equation}
\end{footnotesize}

\noindent
will contribute to the functional integral. 
The matrix elements of $G^{\ }_{BF}$ and 
$G^{\ }_{FB}$ are fermionic, and vanish at
the minimum of the potential. In terms of the eigenvalues 
$e^{\pm {i}\varphi'}$ of 
$G^{\ }_{BB}$ and the eigenvalues 
$e^{\pm {i}\varphi}$ of 
$G^{\ }_{FF}$,
the potential in the absence of fermionic
coordinates reads
\begin{align}
V(G)\propto 
2\left[\cos(2\varphi')-\cos(2\varphi)\right].
\nonumber 
\end{align}
Then the minimum occurs at 
$\varphi'=0$, and $\varphi=\pi$. 
The matrix field
\begin{align}
G^{\ }_{0}=
\mathrm{diag}\,(0, 0, 1, 1)
\nonumber 
\end{align}
that minimizes the potential is thus invariant under
conjugation by the 
$\mathrm{Sp}(2)\times\mathrm{SO}(2)$ subgroup of the bosonic
part 
$\mathrm{SU}(2)\times\mathrm{SU}(2)$ of the 
$\mathrm{PSL}(2|2)$ symmetry. 
Therefore, the addition of the relevant potential perturbation
to the action reduces the target space 
$\mathrm{PSL}(2|2)$ of the
principle chiral model to the target space 
$\mathrm{PSL}(2|2)/\mathrm{OSp}(2|2)
\sim \mathbb{C}P^{2|1}$ 
of the NL$\sigma$M in symmetry class CII. At the same
time, the WZ term reduces to a theta term at
$\theta=\pi$ in the NL$\sigma$M.
\end{thebibliography}
\end{document}